\documentstyle[aps,prd,psfig]{revtex}

\newcommand{\bm}[1]{{\mbox{\boldmath$#1$}}}
\newcommand{\bs}[1]{{\mbox{\scriptsize\boldmath$#1$}}}
\def\Quadrat#1#2{{\vcenter{\hrule height #2
  \hbox{\vrule width #2 height #1 \kern#1
    \vrule width #2}
  \hrule height #2}}}
\def\dAl{\mathop{\kern 1pt\hbox{$\Quadrat{8pt}{0.4pt}$} \kern1pt}}
\begin{document}
 
\author{Sergei M. Kopeikin\thanks{%
On leave from: ASC FIAN, Leninskii Prospect, 53, Moscow 117924, Russia}
and Gerhard Sch\"afer}
\address{FSU Jena, TPI, Max-Wien-Platz 1, 
D - 07743, Jena, 
Germany}
\title{Lorentz Covariant Theory of Light Propagation  
in Gravitational Fields of Arbitrary-Moving Bodies}
\maketitle
\tableofcontents
\date{\today}
\baselineskip=20pt
\newpage
\begin{abstract}
The Lorentz covariant theory of propagation of light in the (weak)
gravitational fields of N-body systems consisting of arbitrarily moving 
point-like bodies with constant masses $m_a$ ($a=1,2,...,N$) is constructed. 
The theory is based on the {\it Li\'enard-Wiechert} representation of the 
metric tensor which describes a retarded type solution of the
gravitational field equations. A new approach  for integrating the 
equations of motion of light particles (photons) depending on the 
retarded time argument is invented. Its application in 
the first post-Minkowskian 
approximation, which is linear with respect to the universal
gravitational constant, G, makes it evident that the equations of light 
propagation admit to be integrated straightforwardly by quadratures.
Explicit expressions for 
the trajectory of light ray and its tangent vector are obtained in
algebraically closed form in terms
of functionals of the retarded time. 
General expressions for the 
relativistic time delay, the angle of light deflection, and the 
gravitational shift of electromagnetic frequency are derived in the form
of instantaneous functions of the retarded time. 
They generalize previously known results for the case of 
static or uniformly moving bodies. The most important
applications of the theory to relativistic astrophysics and astrometry 
are given. They include a discussion of 
the velocity dependent terms in the gravitational lens equation, the Shapiro 
time delay in binary pulsars, gravitational Doppler shift, and a precise 
theoretical formulation of the general relativistic algorithms
of data processing of radio and optical astrometric measurements
made in the non-stationary gravitational field of the solar system.
Finally, proposals for future theoretical work being important for 
astrophysical applications are formulated.    
\end{abstract}
\newpage
\section{Introduction and Summary}

The exact solution of the problem of propagation of electromagnetic 
waves in 
non-stationary gravitational fields is extremely important for modern 
relativistic astrophysics and fundamental astrometry. Until now it are 
electromagnetic signals coming from various astronomical objects which deliver 
the most exhaustive and accurate physical information about numerous intriguing
phenomena going on in the surrounding universe. Present day technology has
achieved a level at which the extremely high precision of current ground-based 
radio interferometric astronomical observations approaches 1 $\mu$arcsec. This 
requires a better theoretical treatment of secondary effects in
the propagation of electromagnetic signals in variable gravitational fields of
oscillating and precessing stars, stationary and coalescing binary systems, and
colliding galaxies \cite{1}. 
Future space astrometric 
missions like GAIA \cite{2} or SIM \cite{3} 
will also 
have precision of about 1-10 $\mu$arcsec on positions and parallaxes of stars, 
and about 1-10 $\mu$arcsec per year for their proper motion. At this level of 
accuracy we are not allowed anymore to treat the 
gravitational field of the solar 
system as static and spherically symmetric. Rotation and oblateness of the Sun 
and large planets as well as time variability of the gravitational field should
be seriously taken into account \cite{4}. 

As far as we know, all approaches developed for integrating equations of
propagation of electromagnetic signals in gravitational fields were based 
on the usage of the post-Newtonian presentation of the metric tensor 
of the gravitational field. It is well-known
(see, for instance, the textbooks \cite{5},
\cite{6}) that the post-Newtonian approximation for the 
metric
tensor is valid only within the so-called "near zone". Hence, 
the post-Newtonian metric can be used for the calculation of light propagation 
only
from the sources lying inside the near zone of a gravitating system of bodies. 
The near zone is restricted by the distance comparable to the wavelength of the 
gravitational radiation emitted from the system. For example, Jupiter orbiting 
the 
Sun emits
gravitational waves with wavelength of about 0.3 parsecs, and the binary pulsar 
PSR
B1913+16 radiates gravitational waves with wavelength of around 4.4 astronomical
units. It is obvious that the majority of stars, quasars, and other sources of
electromagnetic radiation are usually far beyond the boundary of the near zone
of the gravitating system of bodies and another method of solving the problem
of propagation of light from these sources to the observer at the Earth should 
be 
applied. Unfortunately, such an advanced technique has not yet been developed 
and
researches relied upon the post-Newtonian approximation of the metric tensor
assuming implicitly that perturbations from the gravitational-wave part of the
metric are small and may be neglected in the data processing algorithms
\cite{5}, \cite{7} - \cite{9}. However,
neither this assumption was ever scrutinized nor the magnitude of the 
neglected residual terms was estimated. An attempt to clarify this question has
been undertaken in the paper \cite{4} where the matching of
asymptotics of the internal near zone and external Schwarzschild solutions of 
equations of light propagation in the gravitational field of the solar system 
has been employed. Nevertheless, a rigorous solution of the equations of light 
propagation being simultaneously valid both far outside and inside the solar 
system was not found.  

One additional problem to be enlightened relates to how to treat the motion
of gravitating bodies during the time of propagation of light from the point of
emission to the point of observation. The post-Newtonian metric of a
gravitating system of bodies is not static and the bodies move while light is
propagating. Usually, it was presupposed that the biggest influence on 
the light ray
the body exerts when the photon passes nearest to it. For this reason, 
coordinates of 
gravitating bodies in the post-Newtonian metric were assumed to be fixed at a 
specific instant of time $t_a$ (see, for instance, \cite{8} - \cite{10}) being 
close to that of the closest approach of the photon to the body. 
Nonetheless, it was never fully clear how to specify the moment $t_a$ 
precisely and what magnitude of error in the calculation of relativistic time 
delay
and/or the light deflection angle one makes if one choses a slightly different 
moment of time. 
Previous researches gave us different conceivable prescriptions for choosing
$t_a$ which might be used in practice. Perhaps, the most fruitful suggestion
was that given by Hellings \cite{10} and discussed later on in a paper \cite{4}
in more detail. This was just to accept $t_a$ as to be
exactly the time of the closest approach of the photon to the gravitating body 
deflecting the light. Klioner \& Kopeikin \cite{4} have shown that such a choice 
minimizes residual terms in the solution of equation of propagation of light 
rays obtained by the asymptotic matching technique. We note, however, that
neither Hellings \cite{10} nor Klioner \& Kopeikin \cite{4} have justified 
that the choice for $t_a$ they made is unique.    

Quite recently we have started the reconsideration of the problem of propagation 
of
light rays in variable gravitational fields of gravitating system of bodies.
First of all, a profound, systematic approach to integration of light 
geodesic equations 
in arbitrary time-dependent gravitational fields possessing a multipole
decomposition \cite{1}, \cite{11} has been worked out. 
A special technique of integration of the equation of light propagation with
retarded time argument has been developed which allowed to discover a rigorous 
solution of the equations everywhere outside a localized source
emitting gravitational waves. The present paper continues the 
elaboration of the
technique and makes it clear how to construct a Lorentz covariant solution of 
equations of propagation of light rays both outside and inside of a gravitating
system of massive point-like particles moving along arbitrary world lines. 
In finding the solution we used the {\it Li\'enard-Wiechert} 
presentation for the metric 
tensor which accounts for all possible effects in the description of the 
gravitational field and is valid everywhere outside the world lines of 
the bodies. The solution, we have found, allows to give an unambiguous 
theoretical
prescription for chosing the time $t_a$. In addition, by a straightforward
calculation we obtain the complete expressions for the angle of 
light deflection,
relativistic time delay (Shapiro effect), and gravitational shift of
observed
electromagnetic frequency of the emitted photons. These expressions 
are exact at the linear
approximation with respect to the universal gravitational constant, $G$, and 
at arbitrary order of magnitude with respect to the parameter $v_a/c$ where
$v_a$ is a characteristic velocity of the $a$-th light-deflecting body, 
and $c$ is the speed
of light \cite{12}. We devote a large part of the paper to 
the discussion of
practical applications of the new solution of the equations of light propagation
including moving gravitational lenses, 
timing of binary pulsars, the consensus model of very long 
baseline interferometry, and the relativistic reduction of astrometric 
observations
in the solar system.

The formalism of the present paper can be also used in astrometric 
experiments for testing alternative scalar-tensor theories of gravity 
after formal replacing in all subsequent formulas the universal
gravitational constant $G$ by the product $G(\gamma^\ast+1)/2$, where 
$\gamma^\ast$
is the effective light-deflection parameter which is slightly different from 
its weak-field limiting value $\gamma$ of  
the standard parameterized post-Newtonian (PPN) formalism \cite{13}. 
This statement 
is a direct consequence of a conformal invariance
of equations of light rays \cite{16} and can be
immediately proved by straightforward calculations. Solar system
experiments have not been sensitive enough to detect the difference
between the two parameters. However, it may play a role in the binary
pulsars analysis \cite{14}.

The paper is organized as follows. Section 2 presents a short description
of the energy-momentum tensor of the light-deflecting bodies and the metric
tensor given in the form of the  {\it Li\'enard-Wiechert} potential.
Section 3 is devoted to the development of a mathematical technique for
integrating equations of propagation of electromagnetic waves in the
geometric optics approximation. Solution of these equations and
relativistic perturbations of a photon trajectory are given in Section 4.
We briefly outline equations of motion for slowly moving observers and
sources of light in Section 5.
Section 6 deals with a general treatment of observable relativistic effects -
the integrated time delay, the deflection angle, and gravitational shift
of frequency. Particular cases are presented in Section 7. They include
the Shapiro time delay in binary pulsars, moving gravitational lenses,
and general relativistic astrometry in the solar system.    

\section{Energy-Momentum and Metric Tensors}       

The tensor of energy-momentum of a system of massive particles is given 
in covariant form, for example, by Landau \& Lifshitz \cite{17}
\begin{eqnarray}
\label{1}
T^{\alpha\beta}(t, {\bf x})&=&\sum_{a=1}^N 
\hat{T}_{a}^{\alpha\beta}(t)\delta\left({\bf x}-{\bf
x}_a(t)\right)\;,\\\nonumber\\
\hat{T}_{a}^{\alpha\beta}(t)&=&m_a \gamma_a^{-1}(t)u_a^\alpha(t)
u_a^\beta(t)\;,\\\nonumber
\end{eqnarray} 
where $t$ is coordinate time, ${\bf x}=x^i=(x^1,x^2,x^3)$ denotes spatial
coordinates of a current point in space, $m_a$ is the constant (relativistic) 
rest mass 
of the $a$-th particle, ${\bf x}_a(t)$ are spatial coordinates of the
$a$-th massive particle which depend on time $t$, ${\bf v}_a(t)=
d{\bf x}_a(t)/dt$ 
is velocity of the $a$-th particle, $\gamma_a(t)=[1-v_a^2(t)]^{-1/2}$ is 
the (time-dependent) 
Lorentz factor, $u_a^\alpha(t)=\{\gamma_a(t),\;\gamma_a(t){\bf
v}_a(t)\}$ is the
four-velocity of the $a$-th particle, $\delta({\bf x})$ is the usual 
3-dimensional Dirac
delta-function. In particular, we have
\begin{equation}
\label{tens}
\hat{T}_{a}^{00}(t)=\frac{m_a}{\sqrt{1-v_a^2(t)}}\;,\hspace{1.5 cm}
\hat{T}_{a}^{0i}(t)=\frac{m_a v_a^i(t)}{\sqrt{1-v_a^2(t)}}\;,\hspace{1.5 cm}
\hat{T}_{a}^{ij}(t)=\frac{m_a v_a^i(t)v_a^j(t)}{\sqrt{1-v_a^2(t)}}
\;.\hspace{1.5 cm}
\end{equation}
The metric tensor in the linear approximation reads
\begin{eqnarray}
\label{2}
g_{\alpha\beta}(t, {\bf x})&=&\eta_{\alpha\beta}+h_{\alpha\beta}(t, {\bf
x})\;,\\\nonumber
\end{eqnarray}
where $\eta_{\alpha\beta}={\rm diag}(-1,+1,+1,+1)$ is the Minkowski metric of
flat space-time and the metric perturbation 
$h_{\alpha\beta}(t, {\bf x})$ is a function of time and
spatial coordinates \cite{18}. 
It can be found by solving the Einstein field equations which read
in the first post-Minkowskian approximation and in the harmonic
gauge \cite{19}  
as follows
(\cite{20}, chapter 10)
\begin{eqnarray}
\label{3}
\dAl h_{\alpha\beta}(t, {\bf x})&=&-16\pi S_{\alpha\beta}(t,{\bf x})\;,
\end{eqnarray}
where
\begin{eqnarray}
\label{4}
S_{\alpha\beta}(t, {\bf x})&=&T_{\alpha\beta}(t, {\bf x})-
\frac{1}{2}\eta_{\alpha\beta}\;T_{\;\;\lambda}^{\lambda}(t, {\bf 
x})\;.\\\nonumber 
\end{eqnarray}
The solution of this equations has the form of the {\it Li\'enard-Wiechert} 
potential \cite{21}. In order to see how it looks 
like we represent the tensor of 
energy-momentum in a form where all time dependence is included in a 
one-dimensional delta-function
\begin{eqnarray}
\label{5}
T^{\alpha\beta}(t, {\bf x})&=&\int_{-\infty}^{+\infty}dt'\delta(t'-t)\
T^{\alpha\beta}(t', {\bf x})\;.
\end{eqnarray}
Here $t'$ is an independent parameter along the world lines of the particles 
which does not depend on time $t$. The solution of equation (\ref{3}) can be 
found 
using the retarded Green function \cite{22}, and after 
integration with respect to spatial coordinates, using the 3-dimensional 
delta-function, it is given in the form of an one-dimensional retarded-time 
integral
\begin{eqnarray}
\label{6}
h^{\alpha\beta}(t, {\bf
x})&=&\sum_{a=1}^N \int_{-\infty}^{+\infty}
\hat{h}^{\alpha\beta}_a(t',t, {\bf x})dt'\;,\\\nonumber\\
\label{hab}
\hat{h}^{\alpha\beta}_a(t',t, {\bf x})&=&
4\left[\hat{T}_{a}^{\alpha\beta}(t')-
\frac{1}{2}\eta^{\alpha\beta}\hat{T}_{a\lambda}^{\lambda}(t')\right]
\frac{\delta
\left[t'-t+r_a(t')\right]}{r_a(t')}\;,\\\nonumber
\end{eqnarray}
where ${\bf r}_a(t')={\bf x}-{\bf x}_a(t')$, and $r_a(t')=|{\bf
r}_a(t')|$ is the usual Euclidean length of the vector.

The integral (\ref{6})
can be performed explicitly as described in, e.g., (\cite{21}, 
section 14).
The result is the retarded {\it Li\'enard-Wiechert} tensor potential
\begin{eqnarray}
\label{lw}
h^{\alpha\beta}(t, {\bf x})&=&4\sum_{a=1}^N\;\frac{\hat{T}_{a}^{\alpha\beta}(s)-
\frac{1}{2}\eta^{\alpha\beta}\hat{T}_{a\lambda}^{\lambda}(s)}
{r_a(s)-{\bf v}_a(s)\cdot {\bf r}_a(s)}\;,
\end{eqnarray}
where the retarded time $s=s(t,{\bf x})$ for the $a$-th body 
is a solution of the light-cone equation \cite{26}  
\begin{eqnarray}
\label{rt}
s+|{\bf x}-{\bf x}_a(s)|&=&t\;.
\end{eqnarray}
Here it is assumed that the field is measured at time $t$ and at the 
point ${\bf x}$.  
We shall use this form of the metric perturbation $h_{\alpha\beta} (t,{\bf x})$ 
for the integration of the equations of light geodesics in the next section. It 
is
worth emphasizing that the expression for the metric tensor (\ref{lw}) is
Lorentz-covariant and is valid in any harmonic coordinate system admitting
a smooth transition to the asymptotically flat space-time at infinity and
relating to each other by the Lorentz transformations of theory of
special relativity \cite{6}, \cite{27}. A treatment of 
post-linear corrections to the {\it Li\'enard-Wiechert} potentials (\ref{lw})
is given, for example, in a series of papers by Kip Thorne and collaborators 
\cite{28}, \cite{31} - \cite{33}.    

\section{Mathematical Technique for Integrating Equations of Propagation of
Photons }

We consider the motion of a light particle (photon) in the background 
gravitational 
field described by the metric (\ref{6}). No back action of the 
photon on the gravitational field is assumed. Hence, we are allowed to use 
equations of light geodesics directly applying the metric tensor in 
question. 
Let the motion of the photon be defined by fixing the mixed 
initial-boundary conditions (see Fig \ref{covariant1}) 
\begin{equation}
{\bf x}(t_{0})={\bf x}_{0}\;, \hspace{2 cm} 
{\displaystyle {d{\bf x}(-\infty ) \over dt}}%
={\bf k}\;,
\label{7}
\end{equation}
where ${\bf k}^{2}=1$ and, henceforth, the spatial components of
vectors are denoted by bold letters. These conditions define the coordinates 
${\bf x}_{0}$ of the photon at the moment of emission of light, $t_{0}$, and its 
velocity at the infinite past and infinite distance from the origin of the 
spatial coordinates (that is, at the, so-called, past null infinity). 

The original equations of propagation of light rays are rather 
complicated \cite{1}. They can be simplified and
reduced to the form which will be shown
later in this section. In order to integrate them we shall have to resort to 
a special approximation method. In the Minkowskian
approximation of the flat space-time the unperturbed trajectory of the 
light ray is a straight line
\begin{eqnarray}
\label{8}
x^i(t)&=&x^i_N(t)=x^i_0+k^i\;\left(t-t_0\right)\;,\\\nonumber
\end{eqnarray}
where $t_0$, $x^i_0$, and $k^i={\bf k}$ have been defined in equation (\ref{7}). 
In this approximation, the coordinate speed of the photon is 
$\dot{x}^i=k^i$ and is considered as a constant in the expression for the
light-ray-perturbing force.

It is convenient to introduce a new independent parameter $\tau$ along the
photon's trajectory according to the rule \cite{1}, \cite{11} 
\begin{eqnarray}
\label{9}
\tau&=& {\bf k}\cdot{\bf x}_N(t)=t-t_0+{\bf k}\cdot{\bf x}_{0},
\end{eqnarray}
where here and in the following the dot symbol 
$"\cdot"$ between two spatial vectors 
denotes the Euclidean dot product. 
The time $t_0$ of the light signal's emission 
corresponds to the
numerical value of the parameter $\tau_0={\bf k}\cdot{\bf x}_{0}$, and
the numerical value of the parameter $\tau=0$ corresponds to the time 
\begin{eqnarray}
\label{uuu}
t^{\ast}&=&
t_0-{\bf k}\cdot{\bf x}_{0}
\;,
\end{eqnarray} 
which is the time of the closest approach of the unperturbed trajectory of 
the photon to the origin of an asymptotically flat harmonic coordinate system.
We emphasize that the numerical value of the moment $t^{\ast}$ is constant for
a chosen trajectory of light ray and
depends only on the space-time coordinates of the point of emission
of the photon and
the point of its observation. Thus, we find the relationships
\begin{eqnarray}
\label{10}
\tau&\equiv&t-t^{\ast},\hspace{2 cm}\tau_0=t_0-t^{\ast}\;,
\end{eqnarray}
which reveals that the variable $\tau$ is negative from the point of emission 
up to the point of the closest approach $x^i(t^\ast)=\hat{\xi}^i$, and is 
positive 
otherwise \cite{34}. The differential identity $dt=d\tau$ is valid and, for this 
reason, 
the integration along the light ray's path with respect to time $t$ can be 
always replaced by the integration with respect to variable $\tau$.

Making use of the parameter $\tau$, the equation of the unperturbed trajectory 
of 
the light ray can be represented as
\begin{eqnarray}
\label{11}
x^i(\tau)&=&x^i_N(\tau)=k^i \tau+\hat\xi^i\;,
\end{eqnarray}
and the distance, $r(\tau)=|{\bf x}_N(t)|$, of the photon from the origin of 
the coordinate system reads
\begin{eqnarray}
\label{12}
r(\tau)&=&\sqrt{\tau^2+d^2}\;.
\end{eqnarray}
The constant vector $\hat\xi^i={\hat\bm{\xi}}={\bf k}\times ({\bf x}_{0}
\times {\bf k})={\bf k}\times \left({\bf x}_N(t)\times {\bf k}\right)$ 
is called the impact parameter of the unperturbed trajectrory of
the light ray, $d=|{\hat\bm{\xi}}|$ is the length of the impact parameter, 
and the symbol $"\times"$ between two vectors denotes the usual Euclidean 
cross product of two vectors. We note that the vector ${\hat\bm{\xi}}$ is 
transverse
to the vector ${\bf k}$. It is worth emphasizing once again that the 
vector $\hat\xi^i$ 
is directed from the origin of the coordinate system towards the point of the 
closest approach of the unperturbed path of the light ray to the origin. This
vector plays an auxiliary role in our discussion and, in general, has no
essential physical meaning as it can be easily changed by the shift of
the origin of the coordinates \cite{35}.

Implementing the two new parameters $\tau$, ${\hat\bm {\xi}}$ and introducing
the four-dimensional isotropic vector $k^{\alpha}=(1,k^i)$ one can write 
the 
equations of light geodesics as follows (for more details see the paper
\cite{1} and reference \cite{35a})
\begin{eqnarray}
\label{14}
\ddot{x}^{i}(\tau)&=&
\frac{1}{2}k_{\alpha}k_{\beta}
{\hat{\partial}}_i h^{\alpha\beta}(\tau,{\hat\bm {\xi}})-
{\hat{\partial}}_{\tau}\left[
k_{\alpha}h^{\alpha i}(\tau,{\hat\bm {\xi}})+\frac{1}{2}k^i
h^{00}(\tau,{\hat\bm {\xi}})-\frac{1}{2}k^ik_p k_q
h^{pq}(\tau,{\hat\bm {\xi}})\right]
\;,\\\nonumber
\end{eqnarray}
where dots over the coordinates denote differentiation with respect to time, 
${\hat{\partial}}_{\tau}\equiv\partial/\partial\tau$, 
${\hat{\partial}}_i\equiv P_{ij}\partial/\partial \hat\xi^j$, and
$P_{ij}=\delta_{ij}-k_i k_j$ is the operator of projection onto the plane being
orthogonal to the vector $k^i$, and all quantities on the right hand side
of equation (\ref{14}) are 
taken along the light trajectory at the point corresponding to a  
numerical value of the running parameter $\tau$
while the parameter $\hat\bm {\xi}$ is assumed as constant. Hence, the equation
(\ref{14}) should be considered as an ordinary, second order differential 
equation in variable $\tau$ \cite{36}. The given form of 
equation 
(\ref{14}) already shows that only the first term on the right hand
side of it can contribute to the deflection of light if the observer and 
the source of 
light are at spatial infinity. Indeed, a first integration of the right hand 
side
of the equation (\ref{14}) with 
respect to time from $-\infty$ to $+\infty$ brings all terms 
showing time derivatives to zero due to the asymptotic flatness of the 
metric 
tensor which proves our statement (for more details see the next section).

However, if the observer and the source of light are located at finite distances 
from
the origin of coordinate system, we need to know how to perform the integrals 
from
the metric perturbations (\ref{6}) with respect to the parameter $\tau$ along 
the
unperturbed trajectory of light ray. Let us denote those integrals as
\begin{eqnarray}
\label{15}
B^{\alpha\beta}(\tau,{\hat\bm{\xi}})&=&
\int_{-\infty}^{\tau}
h^{\alpha\beta}[\sigma,{\bf x}(\sigma)]d\sigma\;,\\\nonumber\\
\label{16}
D^{\alpha\beta}(\tau,{\hat\bm{\xi}})&=&\int_{-\infty}^{\tau}B^{\alpha\beta}
(\sigma,{\hat\bm{\xi}})
d\sigma\;,\\\nonumber  
\end{eqnarray}
where the metric perturbation $h^{\alpha\beta}[\sigma,{\bf x}(\sigma)]$ is 
defined by the {\it Li\'enard-Wiechert} potential (\ref{6}) and $\sigma$
is a parameter along the light ray having the same meaning as the
parameter $\tau$ in equation (\ref{9}).
In order to calculate the integrals (\ref{15}), (\ref{16}) it is useful to 
change in the integrands
the time argument, $\sigma$, to the new one, $\zeta$, 
defined by the light-cone equation (\ref{rt}) which after substitution
for ${\bf x}$ the unperturbed light trajectory (\ref{11}) reads as follows
\cite{44}
\begin{equation}
\label{17}
\sigma+t^{\ast}=\zeta+|{\hat\bm {\xi}}+{\bf k}\sigma-{\bf 
x}_a(\zeta)|\;.\\\nonumber
\end{equation}
The differentiation of this equation  
yields a relationship between 
differentials of the time variables $\sigma$ and $\zeta$, and parameters
$t^\ast$, $\xi^i$, $k^i$
\begin{equation}
\label{18}
d\zeta\left(r_a-{\bf v}_a\cdot{\bf r}_a\right)=
d\sigma\left(r_a-{\bf k}\cdot{\bf r}_a\right)+r_a dt^\ast-{\bf r}_a\cdot 
d{\hat\bm{\xi}}-\sigma{\bf r}_a\cdot d{\bf k}\;,\\\nonumber
\end{equation}
where the coordinates, ${\bf x}_a$, and the velocity, ${\bf v}_a$, of the 
$a$-th body are 
taken at the retarded time $\zeta$, and coordinates of the photon, 
${\bf x}$, are taken at the time $\sigma(\zeta)$. From equation (\ref{18}) 
we immediately obtain the partial derivatives with respect to the parameters
\begin{equation}
\label{diff}
\frac{\partial \zeta}{\partial t^\ast}=\frac{r_a}{r_a-{\bf v}_a\cdot{\bf
r}_a}\;,\quad\quad\quad
\frac{\partial \zeta}{\partial \hat\xi^i}=-\frac{P_{ij}r_a^j}{r_a-{\bf v}_a\cdot{\bf
r}_a}\;,\quad\quad\quad
\frac{\partial \zeta}{\partial k^i}=-\frac{\sigma r_a^i}{r_a-{\bf v}_a\cdot{\bf
r}_a}\;,\\\nonumber
\end{equation}
and have the relationship between the time differentials along the world 
line of the photon which reads as follows
\begin{equation}
\label{newvar}
d\sigma=d\zeta\;\frac{r_a-{\bf v}_a\cdot{\bf r}_a}{r_a-{\bf k}\cdot{\bf
r}_a}\;.\\\nonumber
\end{equation}
If the parameter $\sigma$ runs from $-\infty$ to $+\infty$, the new 
parameter $\zeta$ runs from $\zeta_{-\infty}=-\infty$ to 
$\zeta_{+\infty}=t^{\ast}+{\bf k}\cdot{\bf x}_a(\zeta_{+\infty})$ 
provided the
motion of each body is restricted
inside a bounded domain of space, like in the case of a binary system. In case
the bodies move along straight lines with constant velocities, the parameter 
$\sigma$ runs from $-\infty$ to $+\infty$, and the parameter $\zeta$
runs from $-\infty$ to $+\infty$ as well. In addition, we note that when the 
numerical value of the parameter $\sigma$ is equal to the time of observation 
$\tau$, the numerical value of the parameter $\zeta$ equals to $s(\tau)$, which 
is 
found from the equation of the light cone (\ref{rt}) in which the
point ${\bf x}$ denotes spatial coordinates of observer.

After transforming time arguments the integrals (\ref{15}), (\ref{16}) take the
form
\begin{equation}
\label{19}
B^{\alpha\beta}(s)=\sum_{a=1}^N B^{\alpha\beta}_a(s)\;,\quad\quad\quad
B^{\alpha\beta}_a(s)=
4\int_{-\infty}^{s}
\frac{\hat{T}_{a}^{\alpha\beta}(\zeta)-
\frac{1}{2}\eta^{\alpha\beta}\hat{T}_{a\lambda}^{\lambda}(\zeta)}
{r_a(\sigma,\zeta)-{\bf k}\cdot{\bf r}_a(\sigma,\zeta)}d\zeta\;,
\end{equation}
\begin{eqnarray}
\label{20}
D^{\alpha\beta}(s)&=&\sum_{a=1}^N \int_{-\infty}^{\tau}
B^{\alpha\beta}_a[\zeta(\sigma)]d\sigma\;,\\\nonumber
\end{eqnarray}
where retarded times in the upper limits of integration depend on the index
of each
body as it has already been mentioned in the previous text.
Now we give a remarkable, exact relationship
\begin{equation}
\label{23}
r_a(\sigma,\zeta)-{\bf k}\cdot{\bf r}_a(\sigma,\zeta)=t^{\ast}+{\bf k}\cdot{\bf
x}_a(\zeta)-\zeta\;,\\\nonumber
\end{equation}
which can be proved by direct use of the light-cone equation (\ref{rt}) 
and the
expression (\ref{11}) for the unperturbed trajectory of light ray. 
It is important
to note that in the given relationship $t^{\ast}$ is a constant time 
corresponding to the moment of the closest 
approach of the photon to the
origin of coordinate system. The equation (\ref{23}) shows that the 
integrand on 
the left hand side of the second 
of equations (\ref{19}) does not depend on the parameter $\sigma$ at 
all, and the integration is performed only with respect to the retarded time
variable $\zeta$.
Thus, just as the law of motion of the bodies ${\bf x}_a(t)$ is known, the 
integral
(\ref{19}) 
can be calculated either analytically or numerically without solving the
complicated light-cone 
equation (\ref{rt}) to establish the relationship between the ordinary and
retarded time arguments. This statement is not  
applicable to the integral
(\ref{20}) because 
transformation to the new variable (\ref{newvar}) does not eliminate from the
integrand of this integral the explicit dependence on the argument of
time $\tau$. 
Fortunately, as it is evident from the structure of equation (\ref{14}), we do 
not 
need to calculate this integral. 

Instead of that,  
we need to know the first spatial derivative of 
$D^{\alpha\beta}(s)$ with respect to $\hat\xi^i$. In order to find it we
note that the integrand of $B^{\alpha\beta}(s)$ does not depend on the variable 
$\hat\xi^i$. This dependence manifests itself only indirectly through 
the upper limit $s(\tau,\xi)$ of
the integral because of the structure of the light-cone equation which
assumes at the point of observation the following form
\begin{eqnarray}
\label{obsp}
\tau+t^\ast&=&s+|\hat{\bm{\xi}}+{\bf k}\tau-{\bf x}_a(s)|\;.
\end{eqnarray}
For this reason, a straightforward differentiation of 
$B^{\alpha\beta}(s)$ with respect to the retarded time $s$ and the 
implementation
of formula (\ref{diff}) for the calculation of the derivative 
$\partial s/\partial \hat\xi^i$ at the point of observation yields \cite{44a}
\begin{eqnarray}
\label{21}
{\hat{\partial}}_i B^{\alpha\beta}(s)&=&-4\sum_{a=1}^N
\frac{\hat{T}_{a}^{\alpha\beta}(s)-
\frac{1}{2}\eta^{\alpha\beta}\hat{T}_{a\lambda}^{\lambda}(s)}
{r_a(s)-{\bf k}\cdot{\bf r}_a(s)}\;\frac{P^i_{\;j}\;r_a^j(s)}{r_a(s)-
{\bf v}_a(s)\cdot{\bf r}_a(s)}\;.
\end{eqnarray}
This result elucidates that ${\hat{\partial}}_i B^{\alpha\beta}(s)$ is not
an integral but instanteneous function of time and, 
that it can be calculated directly if the 
motion of the gravitating bodies
is given. 
While
calculating ${\hat{\partial}}_i D^{\alpha\beta}(s)$ we use, first, the
formula (\ref{21}) and, then, replacement of variables (\ref{newvar}). 
Proceeding in this
way we arrive at the result   
\begin{eqnarray}
\label{22}
{\hat{\partial}}_{i} D^{\alpha\beta}(s)&=&
\sum_{a=1}^N \int_{-\infty}^{\tau}
{\hat{\partial}}_{i}B^{\alpha\beta}_a[\zeta(\sigma)]d\sigma=
-4\sum_{a=1}^N
\int_{-\infty}^{s}
\frac{\hat{T}_{a}^{\alpha\beta}(\zeta)-
\frac{1}{2}\eta^{\alpha\beta}\hat{T}_{a\lambda}^{\lambda}(\zeta)}
{\left[r_a(\sigma,\zeta)-{\bf k}\cdot{\bf r}_a(\sigma,\zeta)\right]^2}
P^i_{\;j}\;r_a^j(\sigma,\zeta)d\zeta\\\nonumber\\\nonumber\mbox{}&=&
-4\sum_{a=1}^N\biggl\{\hat\xi^i\int_{-\infty}^{s}
\frac{\hat{T}_{a}^{\alpha\beta}(\zeta)-
\frac{1}{2}\eta^{\alpha\beta}\hat{T}_{a\lambda}^{\lambda}(\zeta)}
{\left[t^{\ast}+{\bf k}\cdot{\bf
x}_a(\zeta)-\zeta\right]^2}d\zeta-P^i_{\;j}
\int_{-\infty}^{s}
\frac{\hat{T}_{a}^{\alpha\beta}(\zeta)-
\frac{1}{2}\eta^{\alpha\beta}\hat{T}_{a\lambda}^{\lambda}(\zeta)}
{\left[t^{\ast}+{\bf k}\cdot{\bf%
x}_a(\zeta)-\zeta\right]^2}\;\;x^j_a(\zeta)\;d\zeta\biggr\}\;,
\end{eqnarray}
where the numerical value of the parameter $s$ in the upper limit of the 
integral 
is calculated by solving the light-cone equation (\ref{rt}). Going back 
to the
equation (\ref{23}) we find that the integrand of the integral (\ref{22}) 
depends 
only on the retarded time argument $\zeta$. Hence, again, as it has been proven
for $B^{\alpha\beta}(s)$, the integral (\ref{22}) admits a direct calculation as
soon as the motion of the gravitating bodies is prescribed \cite{45}.    

\section{Relativistic Perturbations of a Photon Trajectory}

Perturbations of the trajectory of the photon are found by straightforward 
integration
of the equations of light geodesics (\ref{14}) using the expressions (\ref{15}),
(\ref{16}). Performing the calculations we find 
\begin{eqnarray}
\label{25}
\dot{x}^i(\tau)&=&k^i+\dot{\Xi}^i(\tau)\;,\\			
\label{26}
x^i(\tau)&=&x^i_N(\tau)+\Xi^i(\tau)-\Xi^i(\tau_0)\;,
\end{eqnarray}
where $\tau$ and $\tau_0$ correspond, respectively, to the moment of observation
and emission of the photon. The functions $\dot{\Xi}^i(\tau)$ and 
$\Xi^i(\tau)$ are given as 
follows
\begin{eqnarray}
\label{27}
\dot{\Xi}^i(\tau)&=&\frac{1}{2}k_{\alpha}k_{\beta}
{\hat{\partial}}_i B^{\alpha\beta}(\tau)-
k_{\alpha}h^{\alpha i}(\tau)-\frac{1}{2}k^i
h^{00}(\tau)+\frac{1}{2}k^ik_p k_q
h^{pq}(\tau)\;,\\\nonumber\\
\label{28}
\Xi^i(\tau)&=&\frac{1}{2}k_{\alpha}k_{\beta}
{\hat{\partial}}_i D^{\alpha\beta}(\tau)-
k_{\alpha}B^{\alpha i}(\tau)-\frac{1}{2}k^i
B^{00}(\tau)+\frac{1}{2}k^ik_p k_q
B^{pq}(\tau)\;,\\\nonumber
\end{eqnarray}
where the functions $h^{\alpha\beta}(\tau)$, $B^{\alpha\beta}(\tau)$, 
${\hat{\partial}}_i B^{\alpha\beta}(\tau)$, and 
${\hat{\partial}}_i D^{\alpha\beta}(\tau)$ are defined by the relationships
(\ref{lw}), (\ref{19}), (\ref{21}), and (\ref{22}) respectively. 

The latter
equation can be used for the formulation of the boundary value problem 
for the equation of
light geodesics. In this case the initial position, 
${\bf x}_0={\bf x}(t_0)$, and final position,
${\bf x}={\bf x}(t)$, of the photon are given instead of the 
initial position ${\bf x}_0$ of the photon and the direction of light
propagation ${\bf k}$ given at past null infinity. All what we
need for the formulation of the boundary value problem is the relationship 
between the 
unit vector ${\bf k}$ and the unit vector 
\begin{eqnarray}
\label{vvv}
{\bf K}= -\;\frac{{\bf x}-{\bf x}_0}{|{\bf x}-{\bf x}_0|}\;,
\end{eqnarray}
which defines a geometric direction 
of the light propagation from observer to the source of light 
in flat space-time (see Fig \ref{covariant1}). The formulas (\ref{26}) and 
(\ref{28}) yield
\begin{eqnarray}
\label{29}
k^i&=&-K^i-\beta^i(\tau,{\hat\bm{\xi}})+\beta^i(\tau_0,{\hat\bm{\xi}})\;,
\end{eqnarray}
where relativistic corrections to the vector $K^i$ are defined as follows
\begin{eqnarray}
\label{30}
\beta^i(\tau,{\hat\bm{\xi}})&=&\frac{\frac{1}{2}k_{\alpha}k_{\beta}
{\hat{\partial}}_i D^{\alpha\beta}(\tau)-
k_{\alpha}P^i_{\;j}B^{\alpha j}(\tau)}{|{\bf x}-{\bf x}_0|}\;,\\\nonumber\\
\label{30a}
\beta^i(\tau_0,{\hat\bm{\xi}})&=&\frac{\frac{1}{2}k_{\alpha}k_{\beta}
{\hat{\partial}}_i D^{\alpha\beta}(\tau_0)-
k_{\alpha}P^i_{\;j}B^{\alpha j}(\tau_0)}{|{\bf x}-{\bf x}_0|}\;.
\end{eqnarray}
We emphasize that the vectors $\beta^i(\tau,{\hat\bm{\xi}})\equiv{\bm{\beta}}$
and $\beta^i(\tau_0,{\hat\bm{\xi}})\equiv{\bm{\beta}}_0$ are orthogonal to 
the vector
${\bf k}$ and are taken at the points of observation and emission of the
photon respectively. 
The relationships obtained in this section are used for the discussion of 
observable 
relativistic effects in the following section.

\section{Equations of Motion for Moving Observers and Sources of 
Light}

The knowledge of trajectory of motion of photons in the gravitational field
formed by a
N-body system of arbitrary-moving point masses
is necessary but not enough for
the unambiguous physical interpretation of observational effects. 
It also requires to know 
how observers and sources of light move in the gravitational field 
of this system. Let us assume that
observer and the source of light are point-like massless particles which 
move along
time-like geodesic world lines.
Then, in the post-Minkowskian approximation 
equations of motion of 
the particles, assuming no restriction on their velocities except for that
$v<c$ (see, however, discussion in \cite{30}), read
\begin{eqnarray}
\label{smp}
\ddot{x}^{i}(t)&=&\frac{1}{2}h_{00,i}-h_{0i,t}-\frac{1}{2}h_{00,t}\dot{x}^i-
h_{ik,t}\dot{x}^k-\left(h_{0i,k}-h_{0k,i}\right)\dot{x}^k\\\nonumber 
&&\mbox{} 
-h_{00,k}\dot{x}^k\dot{x}^i-\left(h_{ik,j}-\frac{1}{2}h_{kj,i}\right)
\dot{x}^k\dot{x}^j+
\left(\frac{1}{2}h_{kj,t}-h_{0k,j}\right)\dot{x}^k\dot{x}^j\dot{x}^i+
O(\frac{G^2}{c^4})\;.
\end{eqnarray}
In the given coordinate system for velocities much smaller than the speed
of light, the equation (\ref{smp}) reduces to
\begin{eqnarray}
\label{pnap}
\ddot{x}^{i}(t)&=&\frac{1}{2}h_{00,i}
+O(\frac{G}{c^{2}})+O(\frac{G^2}{c^4})\;.
\end{eqnarray}
Regarding specific physical conditions either the post-Minkowskian 
equation (\ref{smp}) or the post-Newtonian 
equation (\ref{pnap}) should be 
integrated with respect to time to give the 
coordinates of an observer, ${\bf x}(t)$, and a source of 
light, ${\bf x}_0(t_0)$, as a 
function of time of
observation, $t$, and of time of emission of light, 
$t_0$, respectively. We do not treat this problem in the
present paper as its solution has been developed with necessary accuracy
by a number of previous authors. In particular, the post-Minkowskian
approach for solving equations of motion of massive particles 
is thoroughly treated in \cite{42}, \cite{43}, \cite{46}, and references 
therein. The
post-Newtonian approach is outlined in details, for instance, 
in \cite{6}, \cite{8}, \cite{47} - \cite{49},  and references therein. 
In what follows, 
we assume the
motions of observer, ${\bf x}(t)$, and source of light, 
${\bf x}_0(t_0)$, to be known with the required precision.

\section{Observable Relativistic Effects}

\subsection{Shapiro Time Delay}

The relativistic time delay in propagation of electromagnetic signals 
passing through the {\it static, spherically-symmetric} gravitational 
field of the Sun was discovered by Irwin Shapiro \cite{50}. 
We shall give in this paragraph the generalization of his idea for the case
of the propagation of light through the {\it non-stationary} 
gravitational field formed by an ensemble of $N$ {\it arbitrary-moving} 
bodies. The result, which we shall obtain, is valid not only when the
light ray propagates outside the system of the bodies but also when light
goes through the system. In this sense we extend our calculations made in
a previous paper \cite{1} which treated relativistic
effects in propagation of light rays {\it only outside} the gravitating system
having a time-dependent quadrupole moment.

The total time of propagation of an electromagnetic signal from the point ${\bf 
x}_0$
to the point ${\bf x}$ is derived from equations (\ref{26}), (\ref{28}). 
First, we use the equation (\ref{26}) to express the difference ${\bf
x}-{\bf x}_0$ through the other terms of the equation. Then, we multiply
this difference by itself using the properties of the Euclidean dot
product. Finally, we find the total time of propagation of light, $t-t_0$, 
extracting the square root from the product, and using the expansion with 
respect 
to the relativistic parameter $(Gm_a)/(c^2 r_a)$ which is assumed to be small. 
It results in 
\begin{eqnarray}
\label{td}
t-t_0&=&|{\bf x}-{\bf x}_0|-{\bf{k}}\cdot{\bm{\Xi}}(\tau)+
{\bf{k}}\cdot{\bm{\Xi}}(\tau_0)\;,
\end{eqnarray}
or
\begin{eqnarray}
\label{qer}
t-t_0&=&|{\bf x}-{\bf x}_0|+\Delta(t,t_0)\;,
\end{eqnarray}
where $|{\bf x}-{\bf x}_0|$ is the usual Euclidean distance between the points
of emission, ${\bf x}_0$, and observation, ${\bf x}$, of the photon, and 
$\Delta(t,t_0)$ is the generalized Shapiro time delay produced by the 
gravitational field
of moving bodies 
\begin{eqnarray}
\label{shapd}
\Delta(t,t_0)&=&\frac{1}{2}k_{\alpha}k_{\beta}B^{\alpha\beta}(\tau)-
\frac{1}{2}k_{\alpha}k_{\beta}B^{\alpha\beta}(\tau_0)\;=\;2\sum_{a=1}^N
m_a\;B_a(s,s_0)\;.
\end{eqnarray}
In the integral
\begin{eqnarray}
\label{integral1}
B_a(s,s_0)&=&
\displaystyle{\int^{s}_{s_{0}}}\frac{[1-{\bf k}\cdot{\bf
v}_a(\zeta)]^2}{\sqrt{1-v^2_a(\zeta)}}
\frac{d\zeta}{t^{\ast}+{\bf k}\cdot{\bf x}_a(\zeta)-\zeta}\;,
\end{eqnarray} 
the retarded time $s$ is obtained by solving the equation (\ref{rt}) for 
the time of observation of the photon, and $s_0$ is found by solving the same 
equation written down for the time of emission of the photon \cite{51}
\begin{eqnarray}
\label{rt1}
s_0+|{\bf x}_0-{\bf x}_a(s_0)|&=&t_0\;.
\end{eqnarray} 
The relationships (\ref{qer}), (\ref{shapd}) for the time delay have 
been derived with respect to 
the coordinate time $t$. The transformation from the coordinate time 
to the proper time ${\cal T}$ of the observer is made by integrating the 
infinitesimal 
increment of the
proper time along the world line ${\bf x}(t)$ of the observer 
\cite{17}
\begin{eqnarray}
\label{prop}
{\cal T}&=&\int^t_{t_{\rm i}}\biggl\{1-{\bf v}^2(t)- h_{00}[t,{\bf x}(t)]-
2h_{0i}[t,{\bf x}(t)]v^i(t)-
h_{ij}[t,{\bf x}(t)]v^i(t)v^j(t)\biggr\}^{1/2}\;dt\;,
\end{eqnarray} 
where $t_{\rm i}$ is the initial epoch of observation, and $t$ is a 
time of observation.

The calculation of the integral (\ref{integral1}) is performed by means of using 
a new
variable
\begin{equation}
\label{new}
y=t^{\ast}+{\bf k}\cdot {\bf x}_a(\zeta)-\zeta\;,\quad\quad\quad\quad
dy=-[1-{\bf k}\cdot {\bf v}_a(\zeta)]d\zeta\;,
\end{equation}
so that the above integral (\ref{integral1}) reads
\begin{eqnarray}
\label{45}
B_a(s,s_0)&=&-\int^{s}_{s_{0}}\frac{1-{\bf k}\cdot{\bf v}_a(\zeta)}
{\sqrt{1-v^2_a(\zeta)}}\frac{d(\ln y)}{d\zeta}\;d\zeta\;.
\end{eqnarray}
Integration by parts results in
\begin{eqnarray}
\label{46}
B_a(s,s_0)&=&-\frac{1-{\bf k}\cdot{\bf v}_a(s)}
{\sqrt{1-v^2_a(s)}}\;\ln [r_a(s)-{\bf k}\cdot{\bf r}_a(s)]+
\frac{1-{\bf k}\cdot{\bf v}_a(s_0)}
{\sqrt{1-v^2_a(s_0)}}\;\ln [r_a(s_0)-{\bf k}\cdot{\bf
r}_a(s_0)]\\\nonumber\\\nonumber\mbox{}&&-\int^s_{s_{0}}\frac
{\ln(r_a-{\bf k}\cdot{\bf r}_a)}{\left(1-v^2_a\right)^{3/2}}
\left[{\bf k}-{\bf v}_a-{\bf v}_a\times ({\bf k}\times {\bf v}_a)\right]\cdot
\dot{\bf v}_a \;d\zeta\;.\\\nonumber 
\end{eqnarray}
The first and second terms describe the generalized form of the Shapiro time 
delay for the case of
arbitrary moving (weakly) gravitating bodies. The last term in the right
hand side of (\ref{46}) depends on the body's acceleration and is a
relativistic correction 
comparable, in general case, to the main terms of the
Shapiro time delay. This correction is identically zero if the bodies 
move along straight lines with constant velocities. Otherwise, we have to 
know the law of motion of the bodies for its calculation. Neglecting all 
terms of 
order $v_a^2/c^2$ for the Shapiro time delay we obtain the simplified
expression 
\begin{eqnarray}
\label{47}
\Delta(t,t_0)&=&-2\sum_{a=1}^N m_a\biggl\{\ln\frac{r_a-{\bf k}\cdot{\bf r}_a}
{r_{0a}-{\bf k}\cdot{\bf r}_{0a}}-
({\bf k}\cdot{\bf v}_a)\ln(r_a-{\bf k}\cdot{\bf r}_a)
\\\nonumber\\\nonumber\mbox{}&&+
({\bf k}\cdot{\bf v}_{a0})\ln(r_{0a}-{\bf k}\cdot{\bf r}_{0a})+
\int^s_{s_{0}}
\ln\left[t^{\ast}+{\bf k}\cdot {\bf x}_a(\zeta)-\zeta\right]\;
\left[{\bf k}\cdot\dot{\bf v}_a(\zeta)\right]\;d\zeta\biggr\}\;,
\end{eqnarray}
where ${\bf r}_a={\bf x}-{\bf x}_a(s)$, ${\bf r}_{0a}={\bf x}_0-{\bf
x}_{a}(s_0)$, $r_a=|{\bf r}_a|$, $r_{0a}=|{\bf r}_{0a}|$, ${\bf v}_a=\dot{\bf
x}_a(s)$, ${\bf v}_{a0}=\dot{\bf x}_{a}(s_0)$, and the retarded times
$s$ and $s_0$ should be calculated from the light-cone equations (\ref{rt}) and
(\ref{rt1}) respectively. The first term on the right hand side of 
the expression (\ref{47}) for the Shapiro delay was
already known long time ago (see. e.g., \cite{7} - \cite{9} and references 
therein). 
Our expressions (\ref{46}), (\ref{47}) vastly extends previously known
results for they are applicable to the case of arbitrary-moving bodies 
whereas the calculations of all previous authors were
severely restricted by the assumption that either 
the gravitating bodies are fixed
in space or move uniformly with constant velocities. 
In addition, there was no reasonable theoretical prescription 
for the definition
of the body's positions. The rigorous theoretical derivation of the 
formulas (\ref{46})
and (\ref{47}) has made 
a significant progress in clarifying this question and proved for the 
first time that in calculating the Shapiro delay the positions of the 
gravitating
bodies must be taken at the retarded times corresponding to the instants 
of emission and observation of electromagnetic signal. It is interesting
to note that in the right hand side of (\ref{47}) the terms being linearly 
dependent on velocities of bodies can be formally obtained in the
post-Newtonian approximate analysis as well under the assumption that 
gravitating
bodies move uniformly along straight lines \cite{52} -
\cite{54}. We emphasize once again that this assumption works well enough
only if the light travel time does not exceed the characteristic
Keplerian period of the gravitating system. Previous authors were never
able to prove that the assumption of uniform motion of bodies can be
applied, e.g., for treatment of the Shapiro time delay in binary pulsars.
We discuss this problem more deeply in the next sections of this paper.    

\subsection{Bending of Light and Deflection Angle}

The coordinate direction to the source of light measured at the point 
of observation ${\bf x}$ is
defined by the four-vector $p^\alpha=(1,p^i)$, where $p^i=-\dot{x}^i$, or
\vspace{0.3 cm}
\begin{eqnarray}
\label{coor}
p^i&=&-k^i-\dot{\Xi}^i(\tau,\hat{\bm{\xi}})\;,
\end{eqnarray} 
and where we have put the minus sign to make the vector $p^i$ 
directed from  
the observer to the source of light. However, the coordinate 
direction $p^i$ is not a directly observable quantity. A real observable vector 
towards the source of light, $s^\alpha=(1,s^i)$, is defined with respect to the
local
inertial frame of the observer. In this frame 
$s^i=-d{\cal X}^i/d{\cal T}$, where ${\cal T}$ is the
observer's proper time and ${\cal X}^i$ are spatial coordinates of the local 
inertial 
frame. We shall assume for simplicity that the observer is at rest \cite{55}
with respect to
the (global) harmonic coordinate system $(t,x^i)$. Then the infinitesimal
transformation from $(t,x^i)$ to $({\cal T},{\cal X}^i)$ is given by the formula 
\vspace{0.3 cm}
\begin{eqnarray}
\label{trans}
d{\cal T}=\Lambda^0_{\;0}\; dt+\Lambda^0_{\;j}\; dx^j\;\;\;&,&\;\;
d{\cal X}^i=\Lambda^i_{\;0}\; dt+\Lambda^i_{\;j}\; dx^j\;\;\;,
\end{eqnarray}
where the matrix of transformation $\Lambda^{\alpha}_{\;\beta}$ depends
on the space-time coordinates of the point of observation and 
is defined by the
requirement of orthonormality 
\vspace{0.3 cm}
\begin{eqnarray}
\label{ort}
g_{\alpha\beta}&=&\eta_{\mu\nu}\Lambda^{\mu}_{\alpha}\Lambda^{\nu}_{\beta}\;.
\end{eqnarray}
In particular, the orthonormality condition (\ref{ort}) pre-assumes that spatial 
angles and lengths at the point of observations are measured with the help of
the Euclidean metric $\delta_{ij}$. For this reason, as the vector $s^\alpha$ is 
isotropic, we conclude that the Euclidean length $|{\bf s}|$ of the vector $s^i$ 
is equal to 1. Indeed, one has 
\begin{eqnarray}
\label{unity}
\eta_{\alpha\beta}s^\alpha s^\beta&=&-1+{\bf s}^2=0\;.
\end{eqnarray}
Hence, $|{\bf s}|=1$, and the vector ${\bf s}$ points out the astrometric
position of the
source of light on the unit celestial sphere attached to the point of
observation. 

In linear approximation with respect to G the matrix of transformation is 
as follows \cite{1}  
\vspace{0.3 cm}
\begin{eqnarray}
\label{lambda}
\Lambda^0_{\;0}&=&1-\frac{1}{2}h_{00}(t,{\bf x})\;,\nonumber\\
\mbox{} \Lambda^0_{\;i}&=&-h_{0i}(t,{\bf x})\;,\nonumber\\ 
\mbox{} \Lambda^i_{\;0}&=&0\;,\nonumber\\\mbox{} 
 \Lambda^i_{\;j}&=&\delta_{ij}+\frac{1}{2}h_{ij}(t,{\bf x})\;.
\end{eqnarray}
Using the transformation (\ref{trans}) we obtain the relationship between the 
observable unit vector $s^i$ and the coordinate direction $p^i$
\vspace{0.3 cm}
\begin{eqnarray}
\label{rls}
s^i&=&\frac{\Lambda^i_{\;j}\; p^j-\Lambda^i_{\;0}}{\Lambda^0_{\;0}
-\Lambda^0_{\;j}\; p^j}\;.
\end{eqnarray}
In linear approximation it takes the form
\vspace{0.3 cm}
\begin{eqnarray}
\label{form}
s^i&=&
\left(1+\frac{1}{2}h_{00}-h_{0j}p^j\right)p^i+\frac{1}{2}h_{ij}p^j\;.
\end{eqnarray}
Remembering that $|{\bf s}|=1$, we obtain for the Euclidean norm of the 
vector 
$p^i$ 
\vspace{0.3 cm}
\begin{eqnarray}
\label{norma}
|{\bf p}|&=&1-\frac{1}{2}h_{00}+h_{0j}p^j-\frac{1}{2}h_{ij}p^i p^j\;,
\end{eqnarray}
which brings equation (\ref{form}) to the form \cite{56}
\begin{eqnarray}
\label{bnm}
s^i&=&m^i+\frac{1}{2}P^{ij}m^q h_{jq}(t,{\bf x})\;,
\end{eqnarray}
with the Euclidean unit vector $m^i=p^i/|{\bf p}|$.

Let now denote by $\alpha^i$ the dimensionless vector describing the angle of 
total deflection of the light ray measured at the point of observation and 
calculated with respect to vector $k^i$ given at past null infinity. It 
is defined according to the relationship \cite{1}
\vspace{0.3 cm}
\begin{eqnarray}
\alpha^i(\tau,\hat{\bm{\xi}})&=&k^i [{\bf k}\cdot 
\dot{\bm{\Xi}}(\tau,\hat{\bm{\xi}})]-\dot{\Xi}^i(\tau,\hat{\bm{\xi}})\;,
\end{eqnarray}
or
\begin{eqnarray}
\label{ang}
\alpha^i(\tau,\hat{\bm{\xi}})&=&-\;P^i_{\;j}\;\dot{\Xi}^j(\tau,\hat{\bm{\xi}})\;
.
\end{eqnarray}
As a consequence of the definitions (\ref{coor}) and (\ref{ang}) we conclude 
that
\begin{eqnarray}
\label{uio}
m^i&=&-k^i+\alpha^i(\tau,{\bm{\xi}})\;.
\end{eqnarray}
Taking into account expressions (\ref{rls}), (\ref{norma}), 
(\ref{ang}), and (\ref{29}) we obtain for the observed direction to the source 
of
light
\begin{eqnarray}
\label{dop}
s^i(\tau,\hat{\bm{\xi}})&=&K^i+\alpha^i(\tau,\hat{\bm{\xi}})+
\beta^i(\tau,\hat{\bm{\xi}})-
\beta^i(\tau_0,\hat{\bm{\xi}})+\gamma^i(\tau,\hat{\bm{\xi}})\;,
\end{eqnarray}
where the relativistic corrections $\beta^i$ are defined by the equation
(\ref{30}) and where
\vspace{0.3 cm} 
\begin{eqnarray}
\label{gamma}
\gamma^i(\tau,\hat{\bm{\xi}})&=&-\frac{1}{2}P^{ij}k^q h_{jq}(t,{\bf x})\;
\end{eqnarray}
describes the light deflection caused by the deformation of space at the
point of observations.
If two sources of light are observed along the directions $s_1^i$ and 
$s_2^i$, correspondingly, the measured angle $\psi$ between them is defined 
in the local inertial frame as follows  
\vspace{0.3 cm}
\begin{eqnarray}
\label{lkj}
\cos\psi&=&{\bf s}_1\cdot{\bf s}_2\;,
\end{eqnarray}
where the dot denotes the usual Euclidean scalar product. It is worth 
emphasizing
that the observed direction to the source of light (\ref{dop}) includes 
the relativistic
deflection of the light ray which depends not only on quantities taken 
at the point of observation but also on those 
$\beta^i(\tau_0,{\bm{\xi}})$ taken at the point of emission of light.
Usually this term is rather small and can be neglected. However, it
becomes important in the problem of propagation of light in the field of
gravitational waves \cite{1} or for a proper
treatment of high-precision astrometric observations of objects being
within the boundary of the solar system.

Without going into further 
details of the observational procedure we, first of all, give an explicit
expression for the angle $\alpha^i(\tau)$
\begin{eqnarray}
\label{31}
\alpha^i(\tau)&=&-\frac{1}{2}k_{\alpha}k_{\beta}
{\hat{\partial}}_i B^{\alpha\beta}(\tau)+
k_{\alpha}\;P^i_{\;j}\;h^{\alpha j}(\tau)\;.
\end{eqnarray} 
The relationships (\ref{lw}), (\ref{21}) along with the definition of the
tensor of 
energy-momentum (\ref{tens}) allow to recast the previous expression 
into the form
\begin{eqnarray}
\label{32}
\alpha^i(\tau)&=&2\sum_{a=1}^N \frac{m_a}{\sqrt{1-v^2_a}}\;
\frac{\left(1-{\bf k}\cdot {\bf v}_a\right)^2}
{r_a-{\bf k}\cdot{\bf r}_a}\;\frac{P^i_{\;j}\;r_a^j}{r_a-
{\bf v}_a\cdot{\bf r}_a}-4\sum_{a=1}^N\frac{m_a}{\sqrt{1-v^2_a}}
\frac{1-{\bf k}\cdot {\bf v}_a}
{r_a-{\bf v}_a\cdot {\bf r}_a}\;P^i_{\;j}\;v^j_a\;,
\end{eqnarray}
where all the quantities describing the motion of the $a$-th body have to be 
taken at the
retarded time $s$ which relates to $\tau=t-t^{\ast}$ by the 
light-cone equation (\ref{rt}). Neglecting all terms of the order $v_a/c$ we 
obtain a simplified form of the previous expression
\begin{eqnarray}
\label{48}
\alpha^i(\tau)&=&2\sum_{a=1}^N \frac{m_a}{r_a}
\frac{P^i_{\;j}\;r_a^j}
{\left(r_a-{\bf k}\cdot{\bf r}_a\right)}\;,
\end{eqnarray}
which may be compared to the analogous expression for the deflection angle
obtained previously by many other authors in the framework of the post-Newtonian
approximation (see \cite{8}, and references therein). We note that all
previous authors fixed the moment of time, at which 
the coordinates ${\bf x}_a$ of the gravitating bodies were to be calculated
rather arbitrarily, 
without 
having rigorous justification for their choice. 
Our approach gives a unique answer to this question and makes it obvious that 
the coordinates ${\bf x}_a$ should be fixed at the moment of retarded time $s$
relating to the time of observation $t$ by the light-cone equation (\ref{rt}).
 
The next step in finding the explicit expression for the observed coordinate 
direction $s^i$ is the computation of the quantity $\beta^i(\tau)$ given 
in (\ref{30}). We have from formulas (\ref{19}), 
(\ref{22})
the following result for the numerator of $\beta^i(\tau)$ 
\begin{eqnarray}
\label{49}
\frac{1}{2}k_{\alpha}k_{\beta}{\hat{\partial}}_i
D^{\alpha\beta}(\tau)-k_{\alpha}P^i_{\;j}B^{\alpha j}(\tau)
&=&-2\sum_{a=1}^N m_a\;
\left[
{\hat\xi}^i C_a(s)-
P^i_{\;j}
D^j_a(s)
\right]
+4\sum_{a=1}^N m_a\;P^i_{\;j} E_a^j(s)\;,
\end{eqnarray}
where the integrals 
$C_a(s)$, 
$D^j_a(s)$ and $E_a^j(s)$ read as follows
\begin{eqnarray}
\label{50}
C_a(s)&=&\int^s_{-\infty}\left[\frac{1-{\bf k}\cdot {\bf
v}_a(\zeta)}{t^{\ast}+{\bf k}\cdot {\bf x}_a(\zeta)-\zeta}\right]^2
\frac{d\zeta}{\sqrt{1-v_a^2(\zeta)}}
\;,\\\nonumber\\\nonumber\\
\label{51}
D^j_a(s)&=&\int^s_{-\infty}\left[\frac{1-{\bf k}\cdot {\bf
v}_a(\zeta)}{t^{\ast}+{\bf k}
\cdot {\bf x}_a(\zeta)-\zeta}\right]^2
\frac{x_a^j(\zeta)}{\sqrt{1-v_a^2(\zeta)}}\;d\zeta
\;,\\\nonumber\\\nonumber\\
\label{52}
E_a^j(s)&=&\int^s_{-\infty}\frac{1-{\bf k}\cdot {\bf v}_a(\zeta)}
{t^{\ast}+{\bf k}\cdot {\bf x}_a(\zeta)-\zeta}
\frac{v_a^j(\zeta)}{\sqrt{1-v^2_a(\zeta)}}\;d\zeta\;.\\\nonumber
\end{eqnarray}
Making use of the new variable $y$ introduced in (\ref{new}) and 
integrating 
by parts yields
\begin{eqnarray}
\label{53}
C_a(s)&=&\frac{1}{\sqrt{1-v^2_a}}
\frac{1-{\bf k}\cdot {\bf v}_a}{r_a-{\bf k}\cdot {\bf r}_a}+
\int^s_{-\infty}\frac{
\left[{\bf k}-{\bf v}_a-{\bf v}_a\times ({\bf k}\times {\bf v}_a)\right]\cdot
\dot{\bf v}_a}{r_a-{\bf k}\cdot {\bf r}_a}
\frac{d\zeta}{\left(1-v^2_a\right)^{3/2}}
\;,\\\nonumber\\
\label{57}
D_a^j(s)&=&\frac{1-{\bf k}\cdot {\bf v}_a}{\sqrt{1-v^2_a}}
\frac{x_a^j}{r_a-{\bf k}\cdot {\bf r}_a}+
\int^s_{-\infty}\frac{
\left[{\bf k}-{\bf v}_a-{\bf v}_a\times ({\bf k}\times {\bf v}_a)\right]\cdot
\dot{\bf v}_a}{r_a-{\bf k}\cdot {\bf r}_a}
\frac{x_a^j\;d\zeta}{\left(1-v^2_a\right)^{3/2}}-E_a^j(s)
\;,\\\nonumber\\
\label{54}
E_a^j(s)&=&-\frac{v_a^j}{\sqrt{1-v^2_a}}\ln(r_a-{\bf k}\cdot {\bf r}_a)+
\int^s_{-\infty}\ln(r_a-{\bf k}\cdot {\bf r}_a)\;
\Pi^j_k\;\dot{v}_a^k\;\frac{d\zeta}{\left(1-v^2_a\right)^{1/2}}\;,
\\\nonumber
\end{eqnarray}
where $\Pi^j_k(\zeta)=\delta^j_k+u^i(\zeta)u_k(\zeta)$ is the spatial part of 
the operator of projection onto the plane being perpendicular to the world line 
of 
the $a$-th body, and the bodies' coordinates and velocities in all terms, being 
outside the signs of integral, are taken at the moment of the retarded time $s$.
The equations (\ref{53})-(\ref{54}) will be used in section 7 for the discussion 
of 
the gravitational lens equation with taking into account the velocity of the 
body 
deflecting the light rays.

Finally, the quantity $\gamma^i(\tau)$ can be explicitly given by the following
expression
\begin{eqnarray}
\label{ggm}
\gamma^i(\tau)&=&-2\sum_{a=1}^N\frac{m_a}{\sqrt{1-v^2_a}}
\frac{({\bf k}\cdot {\bf v}_a)\;(P^i_{\;j}\;v^j_a)}
{r_a-{\bf v}_a\cdot {\bf r}_a}\;,
\end{eqnarray} 
where coordinates and velocities of the bodies must be taken at the retarded 
time
$s$ according to equation (\ref{rt}). We note that $\gamma^i$ is 
a very small quantity being proportional to the product
$(Gm_a/c^2 r_a)(v_a/c)$.
  
\subsection{Gravitational Shift of Frequency}

The exact calculation of the gravitational shift of electromagnetic 
frequency between
emitted and observed photons plays a crucial role for the adequate 
interpretation of measurements of radial velocities of astronomical 
objects, anisotropy of electromagnetic cosmic background radiation (CMB), 
and other spectral astronomical investigations. In the last several 
years, for instance,
radial velocity measuring technique has reached 
unprecedented accuracy and is approaching to the precision of about 
$10$ cm/sec \cite{57}. In the near future there is a hope to 
improve the
accuracy up to $1$ cm/sec \cite{58} when measurement of the post-Newtonian
relativistic effects in optical binary and/or multiple star systems
will be possible \cite{59}. 

Let a source of light move with respect to the harmonic coordinate system 
$(t,x^i)$ with velocity ${\bf v}_0(t_0)=d{\bf x}_0(t_0)/dt_0$ and emit 
electromagnetic
radiation  with frequency $\nu_0=1/(\delta{\cal T}_0)$, where $t_0$ and 
${\cal T}_0$ are 
coordinate
time and proper time of the source of light, respectively. We denote by
$\nu=1/(\delta{\cal T})$ the observed frequency of the electromagnetic radiation 
measured at 
the point of observation by an observer moving with velocity ${\bf v}(t)
=d{\bf x}/dt$
with respect to the harmonic coordinate system $(t,x^i)$. We can consider the
increments $\delta{\cal T}_0$ and $\delta{\cal T}$ as infinitesimally small. 
Therefore, the
observed gravitational shift of frequency $1+z=\nu/\nu_0$ can be defined 
through
the consecutive differentiation of the proper time of the source of
light, 
${\cal T}_0$, with respect to the proper time of the observer, ${\cal T}$,
\cite{corr.3}
- \cite{corr.2} 
\begin{eqnarray}
\label{58}
1+z&=&\frac{d{\cal T}_0}{d{\cal T}}=
\frac{d{\cal T}_0}{dt_0}\frac{dt_0}{dt}\frac{dt}{d{\cal T}}\;,
\end{eqnarray}
where the derivative 
\begin{eqnarray}
\label{59}
\frac{d{\cal T}_0}{dt_0}&=&
\left[1-v_0^2(t_0)-h_{00}(t_0,{\bf x}_0)-
2h_{0i}(t_0,{\bf x}_0)\;v^i_0(t_0)
-h_{ij}(t_0,{\bf x}_0)\;v^i_0(t_0)\;v^j_0(t_0)\right]^{1/2}\;,
\end{eqnarray}
is taken at the point of emission of light, and the derivative
\begin{eqnarray}
\label{60}
\frac{dt}{d{\cal T}}&=&
\left[1-v^2(t)-h_{00}(t,{\bf x})-
2h_{0i}(t,{\bf x})\;v^i(t)
-h_{ij}(t,{\bf x})\;v^i(t)\;v^j(t)\right]^{-1/2}\;,
\end{eqnarray}   
is calculated at the point of observation. 

The time derivative along the 
light-ray
trajectory is calculated from the equation (\ref{qer}) where we have to take 
into
account that the function $B_a(s,s_0)$ depends on times $t_0$ and $t$ not only
through the retarded times $s_0$ and $s$ in the upper and lower limits of the
integral (\ref{integral1}) but through the time $t^{\ast}$ and the 
vector ${\bf k}$ both being considered
in its integrand as time-dependent parameters. Indeed, the infinitesimal increment of times 
$t_0$ and/or $t$
causes variations in the positions of the source of light and/or
observer and, consequently, to the corresponding change in the trajectory of
light ray, that is in $t^\ast$ and ${\bf k}$. Hence, the derivative 
along the light ray reads as follows
\begin{eqnarray}
\label{61}
\frac{dt_0}{dt}&=&\frac{1+{\bf K}\cdot {\bf v}-2\displaystyle{
\sum_{a=1}^N\;m_a\left[\frac{\partial s}{\partial t}
\frac{\partial}{\partial s}+\frac{\partial s_0}{\partial t}
\frac{\partial}{\partial s_0}+\frac{\partial t^{\ast}}{\partial t}
\frac{\partial}{\partial t^{\ast}}+\frac{\partial k^i}{\partial t}
\frac{\partial}{\partial k^i}\right]B_a(s,s_0,t^{\ast},{\bf k})}}
{1+{\bf K}\cdot {\bf v}_0+2\displaystyle{
\sum_{a=1}^N\;m_a\left[\frac{\partial s}{\partial t_0}
\frac{\partial}{\partial s}+\frac{\partial s_0}{\partial t_0}
\frac{\partial}{\partial s_0}+\frac{\partial t^{\ast}}{\partial t_0}
\frac{\partial}{\partial t^{\ast}}+\frac{\partial k^i}{\partial t_0}
\frac{\partial}{\partial k^i}\right]B_a(s,s_0,t^{\ast},{\bf k})}
}\;,
\end{eqnarray}
where the unit vector ${\bf K}$ is defined in (\ref{vvv}) and where 
we explicitly
show the dependence of function $B_a$ on all parameters which implicitly 
depend on time \cite{60a}.

The time derivative of the vector ${\bf k}$ is calculated using the 
approximation ${\bf
k}=-{\bf K}$ and formula (\ref{vvv}) where the coordinates of the 
source of light, ${\bf
x}_0(t_0)$, and of the observer, ${\bf x}(t)$, are functions of time. It
holds
\begin{equation}
\label{vark}
\frac{\partial k^i}{\partial t}=\frac{({\bf k}\times({\bf v}\times{\bf
k}))^i}{R}\;,\quad\quad\quad
\frac{\partial k^i}{\partial t_0}=-\frac{({\bf k}\times({\bf v}_0\times{\bf
k}))^i}{R}\;,
\end{equation}
where $R=|{\bf x}-{\bf x}_0|$ is the distance between the observer and
the source of
light. 
The derivatives of retarded times $s$ and $s_0$ with respect to $t$ and
$t_0$ are calculated from the formulas
(\ref{rt}) and (\ref{rt1}) where we have to take into account that 
the spatial position of the point of observation is connected to the
point of emission of light by the unperturbed trajectory of light, 
${\bf x}(t)={\bf x}_0(t_0)+{\bf k}\;(t-t_0)$. 
More
explicitly, we use for the calculations 
the following relationships \cite{60b}
\begin{equation}
\label{ret}
s+|{\bf x}_0(t_0)+{\bf k}(t,t_0)\;(t-t_0)-{\bf x}_a(s)|=t\;,\quad\quad\quad\quad
\mbox{and}
\quad\quad\quad\quad s_0+|{\bf x}_0(t_0)-{\bf x}_a(s_0)|=t_0\;,
\end{equation}
where the unit vector ${\bf k}$ must be considered as 
a two-point function of times $t$, $t_0$ with
derivatives being taken from (\ref{vark}). 
The physical meaning of
relationships (\ref{ret}) and (\ref{rtuo}) is the preservation of the 
intersection at the
point of observation ${\bf x}(t)$ of two of the lines forming 
light cones which relate to propagation of the gravitational field and
electromagnetic signals, and having vertices at points ${\bf x}_a(s)$ and 
${\bf x}_0(t_0)$, respectively. 
Calculation of infinitesimal variations of equations (\ref{ret}) immediately gives
\begin{eqnarray}
\label{add1}
\frac{\partial s}{\partial t}&=&\frac{r_a-{\bf k}\cdot{\bf r}_a}
{r_a-{\bf v}_a\cdot{\bf r}_a}-\frac{({\bf k}\times{\bf v})\cdot({\bf 
k}\times{\bf
r}_a)}{r_a-{\bf v}_a\cdot{\bf r}_a}\;,\\\nonumber\\
\label{add2}
\frac{\partial s}{\partial t_0}&=&\frac{(1-{\bf k}\cdot{\bf v}_0)({\bf 
k}\cdot{\bf
r}_a)}{r_a-{\bf v}_a\cdot{\bf r}_a}\;,\\\nonumber\\
\label{add3}
\frac{\partial s_0}{\partial t_0}&=&\frac{r_{0a}-{\bf v}_0\cdot{\bf r}_{0a}}
{r_{0a}-{\bf v}_{a0}\cdot{\bf r}_{0a}}\;,\\\nonumber\\
\label{add4}
\frac{\partial s_0}{\partial t}&=&0\;.
\end{eqnarray}
Time derivatives of the parameter $t^{\ast}$ are calculated from its
original definition
$t^\ast=t_0-{\bf k}\cdot{\bf x}_0(t_0)$, which naturally appears in integrands
of all integrals, and read
\begin{equation}
\label{tstar}
\frac{\partial t^\ast}{\partial t_0}=1-{\bf k}\cdot{\bf
v}_0+\frac{{\bf v}_0\cdot\hat{{\bm{\xi}}}}{R}\;,
\quad\quad\quad\quad\frac{\partial t^\ast}{\partial t}=-
\frac{{\bf v}\cdot\hat{{\bm{\xi}}}}{R}\;,
\end{equation}
where the terms of order $\hat{\xi}/R$ in both formulas relate to the
time derivatives of the vector ${\bf k}$ 
.

Partial derivatives of the function $B_a(s,s_0,t^\ast,{\bf k})$ defined 
by the 
integral (\ref{integral1}) read as follows
\begin{eqnarray}
\label{62}
\frac{\partial B_a}{\partial s}&=&\frac{1}{\sqrt{1-v^2_a}}\;
\frac{(1-{\bf k}\cdot {\bf v}_a)^2}{r_a-{\bf k}\cdot {\bf r}_a}\;,
\\\nonumber\\
\label{63}
\frac{\partial B_a}{\partial s_0}&=&-\frac{1}{\sqrt{1-v^2_{a0}}}\;
\frac{(1-{\bf k}\cdot {\bf v}_{a0})^2}{r_{0a}-{\bf k}\cdot{\bf r}_{0a}}\;,
\\\nonumber\\
\label{63aa}
\frac{\partial B_a}{\partial t^\ast}&=&C_a(s_0)-C_a(s)\;,
\\\nonumber\\
\label{63aaa}
\frac{\partial B_a}{\partial k^i}&=&D_a^i(s_0)-D_a^i(s)+
2\left[E_a^i(s_0)-E_a^i(s)\right]\;.
\end{eqnarray}
The partial derivative $\partial B_a/\partial t^\ast$ is found with the
help of relationships (\ref{50}), (\ref{53}). Calculation of  
the partial derivative $\partial B_a/\partial k^i$ is realized by making
use of (\ref{51}), (\ref{52}) and (\ref{57}), (\ref{54}) respectively.
The integrals in (\ref{53})-(\ref{54}) are not calculable analytically 
in general. If we assume that the accelerations of gravitating
bodies are small so that the velocity of each body can be considered as a 
constant, the derivatives (\ref{63aa}), (\ref{63aaa}) are approximated by
simpler expressions 
\begin{eqnarray}
\label{perf}
\frac{\partial B_a}{\partial t^\ast}&=&
-\frac{1}{\sqrt{1-v^2_a}}
\frac{1-{\bf k}\cdot {\bf v}_a}{r_a-{\bf k}\cdot {\bf r}_a}+
\frac{1}{\sqrt{1-v^2_{a0}}}
\frac{1-{\bf k}\cdot {\bf v}_{a0}}{r_{0a}-{\bf k}\cdot {\bf
r}_{0a}}+...\;,
\end{eqnarray}
\begin{eqnarray}
\label{pe}
\frac{\partial B_a}{\partial k^i}&=&-
\frac{1-{\bf k}\cdot {\bf v}_a}{\sqrt{1-v^2_a}}
\frac{x_a^j(s)}{r_a-{\bf k}\cdot {\bf r}_a}+
\frac{1-{\bf k}\cdot {\bf v}_{a0}}{\sqrt{1-v^2_{a0}}}
\frac{x_a^j(s_0)}{r_{0a}-{\bf k}\cdot {\bf r}_{0a}}\\\nonumber
\\\nonumber&&
+\frac{2v_a^j}{\sqrt{1-v^2_a}}\ln(r_a-{\bf k}\cdot {\bf r}_a)-
\frac{2v_{a0}^j}{\sqrt{1-v^2_{a0}}}\ln(r_{0a}-{\bf k}\cdot {\bf
r}_{0a})+...\;.
\end{eqnarray}
Residual terms, denoted by ellipses, can be calculated from the
integrals in (\ref{53})-(\ref{54}) if one knows the explicit functional 
dependence of 
the bodies' velocities on time. One expects the magnitude of the residual term
to be so
small that it is unimportant for the following discussion \cite{62}. 
The expressions 
(\ref{perf}), (\ref{pe}) will be explicitly used in section VII.B for
discussion of the gravitational shift of frequency by a 
moving gravitational lens. 

\section{Applications to Relativistic Astrophysics and Astrometry}
\subsection{Shapiro Time Delay in Binary Pulsars}

\subsubsection{Approximation Scheme for Calculation of the Effect}

Timing of binary pulsars is one of the most important methods of testing
General Relativity in the strong gravitational field regime (\cite{63} -
\cite{67}, and 
references therein). Such an opportunity exists because of the possibility to 
measure in some binary pulsars the, so-called, post-Keplerian (PK)  
parameters of the pulsar's orbital motion. The PK parameters quantify
different relativistic effects and can be analyzed using
a theory-independent procedure in which the masses of the two stars are the
only dynamic unknowns \cite{68}. Each of the PK parameters
depends on the masses of orbiting stars in a different functional way. 
Consequently, if
three or more PK parameters can be measured, the overdetermined
system of the equations can be used to test the gravitational theory. 

Especially important for this test are binary pulsars on relativistic 
orbits visible nearly edge-on. In such systems observers can easily determine
masses of orbiting stars measuring the "range" and "shape" of the Shapiro time 
delay in the propagation of the radio pulses from the pulsar to the observer 
independently 
of other relativistic effects. 
Perhaps, the most famous examples of the nearly edge-on binary 
pulsars are PSR B1855+09 and PSR B1534+12. The sine of inclination angle, $i$, 
of the orbit of PSR B1855+09 to the line of sight makes up a value of about 
0.9992 and the range parameter of the Shapiro effect reaches 1.27 $\mu$s 
\cite{69}. The corresponding quantities for PSR B1534+12 are
0.982 and 6.7 $\mu$s \cite{70}. All binary pulsars emit
gravitational waves, a fact which was confirmed with the precision of 
about 
$0.3\%$ by Joe
Taylor and collaborators \cite{71}. New achievements in technological
development and continuous upgrading the largest radio telescopes extend our
potential to measure with a higher precision the static part of the 
gravitational
field of the binary system as well as the influence of the velocity-dependent 
terms in the metric tensor, generated by the moving stars, on propagation 
of radio signals from the pulsar to the observer. These terms produce an 
additional 
effect in timing observations which will reveal itself as a small excess to the 
range and shape of the known Shapiro delay making its representation more 
intricative. The effect under discussion can not be investigated thoroughly and
self-consistently within the
post-Newtonian approximation (PNA) scheme even if the velocity-dependent terms 
in the metric tensor are taken into account \cite{52} - \cite{54}. This is 
because the PNA scheme does not treat properly all retardation effects in 
the propagation of the gravitational field. 

In this section we present the exact Lorentz covariant theory of the Shapiro 
effect which includes, besides of the well known logarithm, all 
corrections for the 
velocities of the pulsar and its companion. However, later on we shall restrict 
ourselves to terms which are linear with respect to the velocities. The
matter is that due to the validity of the virial theorem in gravitational bound 
systems the terms being quadratic with respect to velocities are proportional 
to the gravitational potential of the system. It means that the proper treatment 
of
quadratic with respect to velocity terms can be achieved only within the second
post-Minkowskian approximation for the metric tensor which is not considered in 
the
present paper.

The original idea of the derivation of the relativistic time delay in the
{\it static} and 
{\it spherically symmetric} field of a self-gravitating body belongs to
Irwin 
Shapiro \cite{50}. Regarding binary pulsars the {\it static} part of the 
Shapiro time delay has been computed 
by Blandford \& Teukolsky \cite{72} under the assumption of everywhere weak and 
{\it static} gravitational fields. Nordtvedt \cite{52}, Klioner \cite{53},
and Wex \cite{54} calculated the
Shapiro time delay in the gravitational field of uniformly moving bodies 
but without
accounting for the retardation in the propagation of the gravitational 
field. The
mathematical technique of the present paper allows to treat the relativistic 
time delay rigorously and account for all effects caused by the {\it 
non-stationary} 
part of the gravitational field of a binary pulsar, that is to
find in the first post-Minkowskian approximation all special-relativistic 
corrections of order $v_a/c,\; v_a^2/c^2$, etc. to the static
part of the Shapiro effect where $v_a$ denotes characteristic velocity of
bodies in the binary pulsar. 

Let us assume that the origin of the coordinate system is at the 
barycenter of the 
binary pulsar. Radio pulses are emitted rather close to the
surface of the pulsar and the coordinates of the point of emission, 
${\bf x}_0$, can be given by the equation
\begin{eqnarray}
\label{point}
{\bf x}_0&=&{\bf x}_p(t_0)+{\bf X}(t_0)\;,
\end{eqnarray}
where ${\bf x}_p$ are the barycentric coordinates of the pulsar's center-of
-mass, and ${\bf X}$ are the barycentric coordinates of the point of emission
both taken at the moment of emission of the radio pulse, $t_0$. At the moment of
emission the spatial orientation of the pulsar's radio beam is almost 
the same with respect to observer at the Earth.
Hence, we are allowed to
assume that the vector ${\bf X}$ is constant at every ``time" when an 
emission of a radio
pulse takes place \cite{73}. 
In what follows the formula (\ref{47}) plays the key role. 
However, before performing the integral in this formula it is useful to derive
the relationship between retarded times $s$ and $s_0$ given by the expressions
(\ref{rt}) and (\ref{rt1}) respectively. Subtracting the 
equation (\ref{rt1}) from (\ref{rt}) and
taking into account the relationship (\ref{qer}), we obtain
\begin{eqnarray}
\label{s2}
s-s_0&=&R-r_a+r_{0a}+\Delta(t,t_0)\;,
\end{eqnarray}
where $R=|{\bf R}|$, ${\bf R}={\bf x}-{\bf x}_0$, $r_a=|{\bf x}-{\bf x}_a(s)|$, 
and 
$r_{0a}=|{\bf x}_0-{\bf x}_a(s_0)|$. We note that the point of observation, 
${\bf
x}$, is separated from the binary system by a very large distance 
approximately equal to $R$. On the other hand, the size of the binary
system can not exceed the distance $r_{0a}$. Thus,
the Taylor expansion of $r_a$ with respect to the small parameter 
$r_{0a}/R$ is admissible. It yields
\begin{eqnarray}
\label{s3}
r_a&=&|{\bf R}+{\bf x}_0-{\bf x}_a(s)|=R-{\bf K}\cdot [{\bf x}_0-{\bf
x}_a(s)]+O\left(\frac{r_{0a}}{R}\right)\;,
\end{eqnarray}
where the unit vector ${\bf K}$ is defined in (\ref{vvv}). Using the
approximation ${\bf K}=-{\bf k}+O(G)$, formula 
(\ref{s2}) is reduced to the form
\begin{eqnarray}
\label{s4}
s-s_0&=&r_{0a}-{\bf k}\cdot {\bf r}_{0a}+{\bf k}
\cdot [{\bf x}_a(s)-{\bf x}_a(s_0)]+
O\left(\frac{r_{0a}}{R}\right)+O(G)\;,
\end{eqnarray}
which explicitly shows that the difference between the retarded times $s$
and $s_0$ is of the order of time interval being required for light to 
cross the binary system. It is this interval which is characteristic in 
the problem of propagation of light rays from the binary (or any other 
gravitationally bound) system to the observer at the Earth. Therefore, 
the retarded time $s$ taken
along the light ray trajectory changes only a little during
the entire process of propagation of light from the pulsar to the observer while 
the coordinate time $t$ changes enormously. This remarkable fact was
never noted in any of previous works devoted to study of propagation of
electromagnetic signals from remote astronomical systems to observer at
the Earth.  

In addition to the expression (\ref{s4}) , we can
show that time differences $s_0-t_0$ and $s-t_0$ are also of the same order of
magnitude as $s-s_0$. Indeed, assuming that the velocities of pulsar and its
companion are small compared to the speed of light, we get 
from (\ref{rt1}) and (\ref{s4}) for these
increments
\begin{eqnarray}
\label{s5}
s_0-t_0&=&-|{\bf x}_0-{\bf x}_a(s_0)|=-\rho_{0a}-{\bm{\rho}}_{0a}\cdot 
{\bm{\upsilon}}_a+
O\left({\bm{\upsilon}}_a^2\right)+O(G)\;,\\
\label{s5a}
s-t_0&=&-({\bf k}\cdot {\bm{\rho}}_{0a})(1-{\bf k}\cdot {\bm{\upsilon}}_a)
+ O\left({\bm{\upsilon}}_a^2\right)+O(G)\;,
\end{eqnarray}
where ${\bm{\rho}}_{0a}={\bf x}_0-{\bf x}_a(t_0)$, 
$\rho_{0a}=|{\bm{\rho}}_{0a}|$, and ${\bm{\upsilon}}_a\equiv{\bf v}_a(t_0)$. 
The relationships (\ref{s5}), (\ref{s5a}) prove our previous statement and 
reveal
that coordinates of bodies comprising the
system and their time derivatives can be expanded in Taylor series around
the time of emission of the radio signal $t_0$ in powers of $s-t_0$ and/or 
$s_0-t_0$. Fig. \ref{shapiro} illustrates geometry of the mutual
positions of the binary pulsar and the observer and Fig. \ref{covariant4}
explains relationships between position of photon on the light trajectory
and retarded positions of pulsar and its companion.

In what follows we concentrate our efforts on the derivation of the 
linear with respect to
velocity of moving bodies corrections to the static part of the Shapiro delay. 
Calculations are realized using the expression (\ref{47}) where the integral is 
already proportional to the ratio $v_a/c$. Hence, in order to perform
the integration we take into account only first terms in the expansion of the
integrand with respect to $t_0$. Then, the integral reads as
\begin{eqnarray}
\label{s6}        
\int^s_{s_{0}}
\ln(r_a-{\bf k}\cdot{\bf r}_a)({\bf k}\cdot\dot{\bf v}_a)\;d\zeta&=&
{\bf k}\cdot \dot{\bf v}_a(t_0)\;
\int^s_{s_{0}}\ln\left[t^{\ast}+{\bf k}\cdot {\bf 
x}_a(t_0)-\zeta\right]d\zeta\;.
\end{eqnarray}
After this transformation the integral acquires table form and its 
calculation is rather trivial. Accounting for (\ref{s4})-(\ref{s5a}),
the result of integration yields 
\begin{eqnarray}
\label{s7} 
\int^s_{s_{0}}\ln\left[t^{\ast}+{\bf k}\cdot {\bf
x}_a(t_0)-\zeta\right]d\zeta&=&(r_{0a}-{\bf k}\cdot {\bf{r}}_{0a})\left[
\ln(r_{0a}-{\bf k}\cdot {\bf{r}}_{0a})-1\right]-
(r_{a}-{\bf k}\cdot {\bf{r}}_{a})
\ln(r_{a}-{\bf k}\cdot {\bf r}_{a})+O({\bm{\upsilon}}_a)
\;,
\end{eqnarray}
where $r_a$ and $r_{0a}$ have the same meaning as in (\ref{s2}).
The result (\ref{s7}) is multiplied by the radial acceleration of the
gravitating body according to (\ref{s6}). Terms forming such a product 
can reach in a binary pulsar the
maximal magnitude of order $(G m_a/c^3)(x/P_b)(\upsilon/c)\ln(1-\sin i)$,
where $x$ is
the projected semimajor axis of the binary system expressed in light 
seconds, 
$P_b$ is its orbital period, and $i$ is the angle of inclination of the
orbital plane of the binary system to the line of sight. For a binary 
pulsar like PSR B1534+12 
the terms under discussion are about $10^{-5}$ $\mu$s which is 
too small to be
measured. For this reason, all terms depending on the acceleration of the
pulsar and its
companion will be omitted from the following considerations.

Let us note that coordinates of the $a$-th body taken at the retarded time $s$
can be expanded in Taylor series in the neighborhood of time $s_0$ 
\begin{eqnarray}
\label{fff}
{\bf x}_a(s)&=&{\bf x}_a(s_0)+{\bf v}_a(s_0)(s-s_0)+O[(s-s_0)^2]\;,
\end{eqnarray}
or, accounting for (\ref{s4}), 
\begin{eqnarray}
\label{ff1}
{\bf x}_a(s)&=&{\bf x}_a(s_0)+{\bm{\upsilon}}_a\;(\rho_{0a}-{\bf k}\cdot 
{\bm{\rho}}_{0a})+O({\bm{\upsilon}}_a^2)\;.
\end{eqnarray} 
Making use of this expansion one can prove that the large distance, $r_a$, 
relates
to the small one, $r_{0a}$, by the important relationship 
\begin{eqnarray}
\label{s8}
r_a^2-\left({\bf k}\cdot {\bf r}_a\right)^2&=&
r^2_{0a}-\left({\bf k}\cdot {\bf r}_{0a}\right)^2-
2(\rho_{0a}-{\bf k}\cdot {\bm{\rho}}_{0a})\left[
{\bm{\upsilon}}_a\cdot{\bm{\rho}}_{0a}-({\bf k}\cdot {\bm{\rho}}_{0a})
({\bf k}\cdot{\bm{\upsilon}}_a)\right]+
O({\bm{\upsilon}}_a^2)\;.
\end{eqnarray}
Moreover,
\begin{eqnarray}
\label{mor}
r_{0a}+{\bf k}\cdot {\bf r}_{0a}&=&\rho_{0a}+{\bf k}\cdot {\bm{\rho}}_{0a}+
{\bm{\upsilon}}_a\cdot{\bm{\rho}}_{0a}+({\bf
k}\cdot{\bm{\upsilon}}_a)\rho_{0a}+O({\bm{\upsilon}}_a^2)\;.
\end{eqnarray}
As a concequence of simple algebra we obtain
\begin{eqnarray}
\label{s9}
\frac{r_a-{\bf k}\cdot{\bf r}_a}{r_{0a}-{\bf k}\cdot{\bf r}_{0a}}&&=
\frac{r_a^2-\left({\bf k}\cdot {\bf r}_a\right)^2}
{r^2_{0a}-\left({\bf k}\cdot {\bf r}_{0a}\right)^2}\;
\frac{r_{0a}+{\bf k}\cdot{\bf r}_{0a}}{r_{a}+{\bf k}\cdot{\bf r}_{a}}\;,
\end{eqnarray}
which gives after making use of (\ref{s8}), (\ref{mor}) the following result
\begin{eqnarray}
\label{s9a}
\frac{r_a-{\bf k}\cdot{\bf r}_a}{r_{0a}-{\bf k}\cdot{\bf r}_{0a}}&
=&\frac{1+{\bf k}\cdot{\bm{\upsilon}}_a}{r_{a}+{\bf k}\cdot{\bf r}_{a}}
\left[\rho_{0a}+{\bf k}\cdot {\bm{\rho}}_{0a}-
{\bm{\upsilon}}_a\cdot{\bm{\rho}}_{0a}+({\bf k}\cdot {\bm{\rho}}_{0a})
({\bf k}\cdot{\bm{\upsilon}}_a)\right]+
O\left({\bm{\upsilon}}_a^2\right)\;.
\end{eqnarray}
It is straightforward to prove that
\begin{eqnarray}
\label{s10}
r_{a}+{\bf k}\cdot{\bf r}_{a}&=&2\left(R+{\bf k}\cdot {\bf
r}_{0a}\right)+O\left(\frac{r_{0a}^2}{R}\right)\;,
\end{eqnarray}
where $R=|{\bf R}|$ is the distance from the point of emission to the point of
observation. This distance is expanded as
\begin{eqnarray}
\label{iop}
{\bf R}&=&{\bm{\cal R}}+{\bf x}_E+{\bf w}-{\bf x}_p-{\bf X}\;,
\end{eqnarray}
where ${\bm{\cal R}}$ is the distance between the barycenters of the binary 
pulsar
and the solar system, ${\bf x}_E$ is the distance from the barycenter of
the solar system to the center of mass of the
Earth, ${\bf w}$ is the geocentric position of the radio telescope, ${\bf x}_p$ 
are
coordinates of the center of mass of the pulsar with respect to the 
barycenter of the binary system,
and ${\bf X}$ are coordinates of the point of emission of radio pulses with
respect to the pulsar proper reference frame. The distance ${\bm{\cal R}}$ is
gradually changing because of the proper motion of the binary system in the
sky. It is well known that the proper motion of any star is small and,
hence, can be neglected in the time delay
relativistic corrections. All other distances in formula (\ref{iop}) are of
order of either diurnal, or annual, or pulsar's orbital parallax with 
respect to the distance ${\bm{\cal R}}$. Hence,
when considering relativistic corrections in the Shapiro time delay, 
the distance ${\bf R}$ can be taken 
as a constant. Such an approximation is more than enough to put
\begin{eqnarray}
\label{mlk}
\ln (r_{a}+{\bf k}\cdot{\bf r}_{a})&=&\ln (2{\cal R})+
O\left(\frac{r_{0a}}{R}\right)\simeq{\rm const.}\;,
\end{eqnarray}
where ${\cal R}=|{\bm{\cal R}}|$.
Constant terms are not directly observable in pulsar timing because they are 
absorbed in the initial rotational phase of the pulsar. For this reason, 
we shall omit for simplicity the 
term $\ln (r_{a}+{\bf k}\cdot{\bf r}_{a})$ from the final expression for the 
Shapiro time delay.

Accounting for all approximations having been developed in this section we
obtain from (\ref{47}), (\ref{s9a}), and (\ref{mlk}) 
\begin{eqnarray}
\label{s11}
\Delta(t,t_0)&=&-2\sum_{a=1}^N m_a\biggl\{(1-{\bf k}\cdot{\bm{\upsilon}_a})
\ln\left[\rho_{0a}+{\bf k}\cdot {\bm{\rho}}_{0a}-
{\bm{\upsilon}}_a\cdot{\bm{\rho}}_{0a}+({\bf k}\cdot {\bm{\rho}}_{0a})
({\bf k}\cdot{\bm{\upsilon}}_a)\right]+{\bf k}\cdot{\bm{\upsilon}_a}\biggr\}
\\\nonumber\\\nonumber\mbox{}&&
\hspace{1.5 cm}+O\left(\frac{G m_a}{c^3}\frac{\upsilon_a^2}{c^2}\right)+
O\left(\frac{G m_a}{c^3}\frac{\upsilon_a}{c}\frac{x}{P_b}\right)+
O\left(\frac{G m_a}{c^3}\frac{x}{\cal R}\right)\;.\\\nonumber
\end{eqnarray}
This formula completes our analytic derivation of the
velocity-dependent corrections to the Shapiro time 
delay in 
binary systems. It also includes residual terms which have not
been deduced by other authors \cite{79}.

\subsubsection{Post-Newtonian versus post-Minkowskian calculations 
of the Shapiro time delay in binary systems}  

Our approach clarifies the principal question why the post-Newtonian 
approximation was efficient for the correct calculation of the main
(velocity-independent) term in the 
formula (\ref{s11}) for the Shapiro time delay in binary systems. We 
recall that the
post-Newtonian theory operates with the instantaneous values of the 
gravitational
potentials in the near zone of the gravitating system. 
In the post-Newtonian scheme coordinates and velocities of gravitating
bodies, being arguments of the metric tensor, depend on the coordinate 
time $t$. Thus, if we expand these coordinates and velocities
around the time of emission of light, $t_0$, we get for the components of
metric tensor
$g_{\alpha\beta}[t,{\bf x}(t),{\bf x}_a(t),{\bf v}_a(t)]$ a Taylor 
expansion which reads as follows 
\begin{eqnarray}
\label{s12}
g_{\alpha\beta}[t,{\bf x}(t),{\bf x}_a(t),{\bf v}_a(t)]&=&
g_{\alpha\beta}[t,{\bf x}(t),{\bf x}_a(t_0),{\bf
v}_a(t_0)]+\\\nonumber\\\nonumber
&&\biggl\{
\frac{\partial 
g_{\alpha\beta}[t,{\bf x}(t),{\bf x}_a(t_0),{\bf v}_a(t_0)]}
{\partial x^i_a}\;v^i_a(t_0)+
\frac{\partial 
g_{\alpha\beta}[t,{\bf x}(t),{\bf x}_a(t_0),{\bf v}_a(t_0)]}
{\partial v^i_a}\;\dot{v}^i_a(t_0)\biggr\}(t-t_0)+...\;.
\end{eqnarray}
This expansion is divergent if the time interval $t-t_0$ exceeds the orbital 
period $P_b$ of the gravitating system. This is the reason why the 
post-Newtonian scheme
does not work if the time of integration of the equations of light propagation 
is bigger than the orbital period. 

On the other hand, the
post-Minkowskian scheme gives components of the metric tensor in terms of
the {\it Li\'enard-Wiechert} potentials being functions of retarded
time $s$. We have shown 
that in terms of the retarded time argument the characteristic time for the 
process of propagation of light rays from the pulsar to observer corresponds 
to the interval of time being required for light to cross the system. During 
this time gravitational potentials can not change their numerical values too 
much because of the slow motion of the gravitating bodies. 
Hence, if we expand coordinates of the bodies around $t_0$ we get for the 
metric tensor expressed in terms of the {\it Li\'enard-Wiechert} 
potentials the following expansion
\begin{eqnarray}
\label{s13}
g_{\alpha\beta}[t,{\bf x}(t),{\bf x}_a(s),{\bf v}_a(s)]&=&
g_{\alpha\beta}[t,{\bf x}(t),{\bf x}_a(t_0),{\bf
v}_a(t_0)]+\\\nonumber\\\nonumber &&
\biggl\{
\frac{\partial 
g_{\alpha\beta}[t,{\bf x}(t),{\bf x}_a(t_0),{\bf v}_a(t_0)]}
{\partial x^i_a}\;v^i_a(t_0)+
\frac{\partial 
g_{\alpha\beta}[t,{\bf x}(t),{\bf x}_a(t_0),{\bf v}_a(t_0)]}
{\partial v^i_a}\;\dot{v}^i_a(t_0)\biggr\}(s-t_0)+...\;,
\end{eqnarray}
which always converges because the time difference $s-t_0$ never exceeds
the orbital period (see equation (\ref{s5a})). 

Nevertheless, as one can easily see, the leading
terms in the expansions (\ref{s12}) and (\ref{s13}) coincide exactly which
indicates that the terms in the solution of the equations of light 
propagation depending only on the static part of gravitational field
should be identical independently on what kind of approximation 
scheme is used for finding the metric tensor.
Thus, the 
post-Newtonian approximation works fairly well for finding the 
{\it leading} part of the
solution of the equations of light geodesics. However, it can not be used for
taking into account perturbations of the light trajectory caused by the 
motion of 
massive bodies in the light-deflecting, gravitationally bounded 
astronomical systems \cite{81}. 

It is worth emphasizing once again
that our approach is based on the post-Minkowskian approximation scheme
for the calculation of gravitational potentials which properly 
accounts for all 
retardation
effects in the motion of bodies by means of the {\it Li\'enard-Wiechert} 
potentials. 

\subsubsection{Shapiro Effect in the Parametrized Post-Keplerian Formalism}

The parametrized post-Keplerian (PPK) formalism was introduced by Damour \&
Deruelle \cite{80} and partially improved by Damour \& Taylor \cite{68}. It
parametrizes the timing formula for binary pulsars in a general phenomenological
way \cite{82}. In
order to update the PPK presentation of the Shapiro delay we use expression
(\ref{s11}). A binary pulsar consists of two bodies - the pulsar (subindex 
"$p$")
and its companion (subindex "$c$").
The emission of a radio pulse takes place very near to the surface of 
the pulsar and,
according to (\ref{point}) and the related discussion, we can 
approximate ${\bf X}=X{\bf k}$ where $X$ is the distance from the center
of mass of the pulsar to the pulse-emitting point. In this approximation 
we get ${\bm{\rho}}_{0p}=X{\bf k} $ and, as a
consequence,
\begin{eqnarray}
\label{puls}
\ln\left[\rho_{0p}+{\bf k}\cdot {\bm{\rho}}_{0p}-
{\bm{\upsilon}}_p\cdot{\bm{\rho}}_{0p}+({\bf k}\cdot {\bm{\rho}}_{0p})
({\bf k}\cdot{\bm{\upsilon}}_p)\right]&=&\ln(2X)={\rm const.}\;.
\end{eqnarray}
Hence, the formula (\ref{s11}) for the Shapiro time delay can be displayed in 
the form
\begin{eqnarray}
\label{s11a}
\Delta(t,t_0)&=&-2m_c\biggl\{(1-{\bf k}\cdot{\bm{\upsilon}_c})
\ln\left[\rho_{0c}+{\bf k}\cdot {\bm{\rho}}_{0c}-
{\bm{\upsilon}}_c\cdot{\bm{\rho}}_{0c}+({\bf k}\cdot {\bm{\rho}}_{0c})
({\bf k}\cdot{\bm{\upsilon}}_c)\right]+{\bf
k}\cdot{\bm{\upsilon}_c}\biggr\}\\\nonumber\mbox{}&&
-2m_p\left[(1-{\bf k}\cdot{\bm{\upsilon}_p})
\ln(2X)+{\bf k}\cdot{\bm{\upsilon}_p}\right]\;,
\end{eqnarray}
where we have omitted residual terms for simplicity. It was shown in the paper
\cite{83a} that any constant term multiplied by the dot product 
${\bf k}\cdot{\bm{\upsilon}_p}$ or ${\bf k}\cdot{\bm{\upsilon}_c}$ is absorbed
into the epoch of the first pulsar's passage through the periastron. 
Thus, we conclude that terms relating to the pulsar in the formula (\ref{s11a})
and the very last term in the curl brackets are not directly observable. For 
this 
reason, we shall omit them in what follows
and consider only the logarithmic contribution to the Shapiro effect 
caused by the pulsar's companion.  
According to formula (\ref{point}) we have
\begin{equation}
\label{fgh}
{\bm\rho_{0c}}={\bf r}+X{\bf
k}\quad\quad,\quad\quad\rho_{0c}=r+\frac{X}{r}{\bf k}\cdot{\bf r}+...\;,
\end{equation}
where ${\bf r}={\bf x}_p(t_0)-{\bf x}_c(t_0)$ is the vector of relative
position of the pulsar with respect to its companion, $r=|{\bf r}|$, and 
dots
denote residual terms of higher order. Taking into account all previous
remarks and omitting directly unobservable terms we conclude that 
the Shapiro delay assumes the form
\begin{eqnarray}
\label{ert}
\Delta(t,t_0)&=&-2m_c(1-{\bf k}\cdot{\bm{\upsilon}_c})
\ln\left[\left((1+\frac{X}{r}\right)\left(r+{\bf k}\cdot {\bf r}\right)-
{\bm{\upsilon}}_c\cdot{\bf r}+({\bf k}\cdot {\bf r})
({\bf k}\cdot{\bm{\upsilon}}_c)\right]\;.
\end{eqnarray}
If the pulsar's orbit is not nearly edgewise and the ratio $X/r$ is
negligibly small the time delay can be
decomposed into three terms
\begin{eqnarray}
\label{wex}
\Delta(t,t_0)&=&-2m_c\ln\left(r+{\bf k}\cdot {\bf r}\right)+
2m_c({\bf k}\cdot{\bm{\upsilon}_c})\ln\left(r+{\bf k}\cdot {\bf
r}\right)+2m_c\frac{{\bm{\upsilon}}_c\cdot{\bf r}-({\bf k}\cdot {\bf r})
({\bf k}\cdot{\bm{\upsilon}}_c)}{r+{\bf k}\cdot {\bf r}}\;.
\end{eqnarray}
The first term on the right hand side of (\ref{wex}) is the standard 
expression for the Shapiro time delay. 
The second and third terms on the right hand side were
discovered by Nordtvedt \cite{52} and Wex \cite{54} under the
assumption of uniform and rectilinear motion of pulsar and companion in
the expression for the post-Newtonian metric tensor of the binary system.
One understands now that this assumption was equivalent to taking into
account primary terms of retardation effects in propagation of 
gravitational field of pulsar and its companion. 
Nevertheless, the
approximation used by Nordtvedt and Wex works fairly well only for terms linear
with respect to velocities of bodies. Had one tried to take into
account quadratic terms with respect to velocities using the
post-Newtonian approach an inconsistent result would have been obtained,
at least under certain circumstances \cite{83b}.     

In what follows only the case of the elliptic motion of the pulsar with 
respect to its 
companion is of importance. Moreover, we do not use the expansion
(\ref{wex}) keeping in mind the case of the nearly edgewise orbits for which
the magnitude of $r+{\bf k}\cdot {\bf r}$ term can be pretty small near
the event of the superior conjunction of pulsar and companion. 
The size and the shape of an elliptic orbit
of the pulsar with respect to its companion are
characterized by the semi-major axis $a_R$ and the eccentricity $e$ ($0\leq e <
1$). The orientation in space of the plane of the pulsar's motion is 
defined with respect to the plane of the sky by the
inclination angle $i$ and the longitude of the ascending node $\Omega$. For
orientation of the pulsar's position in the plane of motion one uses the 
argument of the
pericenter $\omega$. More precisely, the orientation of the orbit is defined by
three unit vectors $({\bf l}, {\bf m}, {\bf n})$ having coordinates \cite{8}, 
\cite{80}
\begin{eqnarray}
\label{triad}
{\bf l}&=&(\cos\Omega, \sin\Omega, 0)\;,\nonumber\\\
{\bf m}&=&(-\cos i \sin\Omega, \cos i \cos\Omega, \sin i)\;,\\\nonumber
{\bf n}&=&(\sin i \sin\Omega, -\sin i \cos\Omega, \cos i)\;.
\end{eqnarray}
In this coordinate system we have the unit vector ${\bf k}$ to be ${\bf k}=-{\bf
K}=(0,0,-1)$ \cite{84}.
The coordinates of the pulsar in the orbital plane are the
radius vector ${\bf r}$ and the true anomaly $f$. In terms of ${\bf r}$ and $f$
one has according to \cite{80} (see also \cite{8}, chapter 1) 
\begin{eqnarray}
\label{rv}
{\bf r}&=&r\left({\bf P}\cos f+{\bf Q}\sin f\right)\;,
\end{eqnarray}
where the unit vectors ${\bf P}$, ${\bf Q}$ are defined by
\begin{equation}
\label{uv}
{\bf P}={\bf l}\cos\omega+{\bf m}\sin\omega\quad\quad,\quad\quad
{\bf Q}=-{\bf l}\sin\omega+{\bf m}\cos\omega\;.
\end{equation}
The coordinate velocity of the pulsar's companion is given by
\begin{eqnarray}
\label{vel}
{\bm{\upsilon}}_c&=&-\frac{m_p}{M}\;\dot{\bf r}\;,\\
\dot{\bf r}&=&\left(\frac{GM}{p}\right)^{1/2}
\left[-{\bf P}\sin f+{\bf Q}(\cos f+e)\right]\;,
\end{eqnarray}
where $M=m_p+m_c;\;$ $p=a_R(1-e^2)^{1/2}\;$ is the focal parameter of the
elliptic orbit, and $m_p$ and 
$m_c$ are the masses of the pulsar and its companion. Accounting for 
relationships
\begin{equation}
\label{s14}     
r=a_R(1-e\cos u)\;,\quad\quad
r\cos f=a_R(\cos u-e)\;,\quad\quad
r\sin f=a_R(1-e^2)^{1/2}\sin u\;,
\end{equation}
where $u$ is the eccentric anomaly relating to the time of emission,
$t_0\equiv T$, and the moment of the first passage of the pulsar through the
periastron, $T_0$,
by the Kepler transcendental equation
\begin{eqnarray}
\label{s18}
u-e \sin u&=&n_b (T-T_0)\;,
\end{eqnarray}
we obtain
\begin{eqnarray}
\label{s15}
{\bf k}\cdot{\bf r}&=&-a_R\sin i\left[
(\cos u-e)\sin\omega+(1-e^2)^{1/2}
\cos\omega\sin u\right]\;,\\
\label{s16}
{\bf r}\cdot{\bm{\upsilon}}_{c}&=&-a_c\;a_R\;n_b\;e\sin u\;,\\
\label{s17}
{\bf k}\cdot{\bm{\upsilon}}_{c}&=&a_c\;n_b\;(1-e^2)^{-1/2}\sin i
\left[e \cos\omega+\frac{(\cos u-e)\cos\omega-(1-e^2)^{1/2}
\sin\omega\sin u}{1-e\cos u}\right]\;.
\end{eqnarray}
Here $a_c=a_R\;m_p/M$, and $n_b=(GM/a_R^3)^{1/2}$ is the orbital 
frequency related
to the orbital period $P_b$ by the equation $n_b=2\pi/P_b$. 

Ignoring all constant factors, the set of equations given in this 
section allows to write down the Shapiro
delay (\ref{ert}) in the form
\begin{eqnarray}
\label{s19}
\Delta_S(T)&=&-\frac{2G m_c}{c^3}\ln\biggl\{1-e\cos u-\sin i\left[
\sin\omega(\cos u-e)+(1-e^2)^{1/2}\cos\omega\sin u\right]
\\\nonumber\\\nonumber\mbox{}&&
+\frac{2\pi}{\sin i}
\frac{x}{P_b}\frac{m_p}{m_c}\;e\sin u
-\frac{2\pi\sin i}{(1-e^2)^{1/2}}
\frac{x}{P_b}\frac{m_p}{m_c}\left[
\sin\omega(\cos u-e)+(1-e^2)^{1/2}\cos\omega\sin u\right]\times
\\\nonumber\\\nonumber\mbox{}&&\times
\left[e \cos\omega+\frac{(\cos u-e)\cos\omega-(1-e^2)^{1/2}
\sin\omega\sin u}{1-e\cos u}\right]\biggr\}\;,     
\end{eqnarray}
where in front of the logarithmic function we have omitted the term of order 
$(G m/c^3)(\upsilon_c/c)$ which is small and hardly be detectable in
future. The term of order $X/r$ in
the argument of the logarithm is also too small and is omitted. The magnitude of 
the
velocity-dependent terms in the argument of the 
logarithm is of order $10^{-3}\div 10^{-4}$. These terms can be comparable with 
the
main terms in the argument of the logarithm when the pulsar is near the superior
conjunction with the companion and the orbit is nearly edge-on. The
velocity-dependent terms cause
a small surplus distortion in the shape of the Shapiro effect which may be 
measurable 
in future timing observations when better precision and time resolution will be
achieved. Unfortunately, existence of the, so-called, bending time delay
\cite{78} may make observation of the
velocity-dependent terms in the Shapiro time delay a rather hard problem. 
 
\subsection{Moving Gravitational Lenses}

The theoretical study of astrophysical phenomena caused by a moving 
gravitational 
lens certainly deserves a fixed attention. Though effects produced by the
motion of the lens are difficult to measure, they can give us an additional
valuable information on the lens parameters. In particular, a lensing object 
moving across the line of sight should
cause a red-shift difference between multiple images of a background object
like a quasar lensed by a galaxy, and a brightness anisotropy in the
microwave background radiation \cite{85}. Moreover, 
velocity-dependent terms in the equation of gravitational lens along with
proper motion of 
the deflector can distort the shape and the amplitude of magnification curve 
observed in a microlensing event. Slowly moving gravitational lenses
are `conventional' astrophysical objects and effects caused by their motion are
small and hardly detectable. However, a cosmic string, for example, 
may produce a noticeable observable effect
if it has sufficient mass per unit length. Gradually increasing precision
of spectral and photometric astronomical observations will make it possible
to measure all these and other effects in a foreseeable future. 

\subsubsection{Gravitational Lens Equation}

In this section we derive the 
equation of a moving gravitational lens for the case that the velocity 
${\bf v}_a$ of the $a$-th light-ray-deflecting mass is constant but without
any other restrictions on its magnitude. This assumption simplifies calculations
of all required integrals allowing to bring them to a manageable form.      
In what follows it is convenient to introduce two vectors ${\bm{\varsigma}}_a=
{\bf x}(s)-{\bf x}_a(s)$ and ${\bm{\varsigma}}_{0a}={\bf x}(2t_0-s_0)-{\bf x}_a
(s_0)$ (see Fig. \ref{movingGL} for more details on the geometry of 
lens). We also shall suppose that the length of vector 
${\bm{\varsigma}}_a$
is small compared to any 
of the distances: $R=|{\bf x}-{\bf x}_0|$, $r_a=|{\bf x}-{\bf x}_a(s)|$, or 
$r_{0a}=|{\bf x}_0-{\bf x}_a(s_0)|$.
It is not difficult to prove by straightforward calculations, taking
account of the light-cone equation, that
\begin{equation}
\label{g1}
{\bm{\varsigma}}_{a}={\bf r}_a-{\bf k} r_a\;,\quad\quad\quad\quad
{\bm{\varsigma}}_{0a}={\bf r}_{0a}+{\bf k} r_{0a}\;,
\end{equation}
where, as in the other parts of the present paper, we have 
${\bf r}_a={\bf x}-{\bf x}_a(s)$ and ${\bf r}_{0a}={\bf x}_0-{\bf x}_a(s_0)$.
From these equalities it follows that
\begin{equation}
\label{g2}
{\bf k}\cdot{\bm{\varsigma}}_{a}=-\frac{d^2_a}{2 r_a}\;,\quad\quad\quad\quad
{\bf k}\cdot{\bm{\varsigma}}_{0a}=\frac{d^2_{0a}}{2 r_{0a}}\;,
\end{equation}
and
\begin{equation}
\label{g3}
r_{0a}-{\bf k}\cdot{\bf r}_{0a}=2r_{0a}-\frac{d^2_{0a}}{2 r_{0a}}\;,
\end{equation}
where distances $d_a=|{\bm{\varsigma}}_{a}|$ and $d_{0a}=|{\bm{\varsigma}}_{0a}|
$ are Euclidean lengths of
corresponding vectors.
We can see as well
that making use of the relationships (\ref{g1}) yields
\begin{eqnarray}
\label{g4}
r_{a}-{\bf v}_a\cdot{\bf r}_{a}&=&r_a(1-{\bf k}\cdot{\bf
v}_a)-{\bm{\varsigma}}_{a}\cdot{\bf v}_a=
r_a(1-{\bf k}\cdot{\bf v}_a)+O\left(v_a d_a\right)\;,
\end{eqnarray}
and the residual term can be neglected because of its smallness compared to
the first one.

It is worth noting that the vector ${\bm{\varsigma}}_a$ 
is approximately equal to the impact
parameter of the light ray trajectory with respect to the position of the
deflector at the retarded time 
$s$. 
Indeed, let us
introduce the vectors ${\hat\xi}^i=P^i_{\;j} x^j$ and ${\hat\xi}^i_a=P^i_{\;j} 
x^j_a(s)$ 
which are lying in
the plane being orthogonal to the unperturbed trajectory of light ray. Then, 
from the definitions (\ref{g1}), (\ref{g2}) one immediately derives the exact 
relationship
\begin{eqnarray}
\label{g0}
{\hat\bm{\xi}}-{\hat\bm{\xi}}_a&=&{\bm{\varsigma}}_a+{\bf k}\;\frac{d_a^2}
{2r_a}\;,
\end{eqnarray}
from which follows
\begin{eqnarray}
\label{g0a}
{\bm{\varsigma}}_a&=&{\hat\bm{\xi}}-{\hat\bm{\xi}}_a-{\bf k}\;\frac{d_a^2}
{2r_a}\;,
\end{eqnarray}
and the similar relationships may be derived for ${\bm{\varsigma}}_{0a}$. It is
worthwhile to note that 
\begin{eqnarray}
\label{nnn}
P_j^i r_a^j&=&P_j^i{\varsigma}^j=\hat{\xi}^i-\hat{\xi}^i_a\;,
\end{eqnarray}
and
\begin{eqnarray}
\label{nnp}
r_a-{\bf k}\cdot{\bf
r}_a&=&\frac{d_a^2}{2r_a}=
\frac{|{\hat\bm{\xi}}-{\hat\bm{\xi}}_a|^2}{2r_a}+\frac{d_a^4}{8r_a^3}\;.
\end{eqnarray}

Let us denote the total angle of light deflection caused by the $a$-th
body as
\begin{eqnarray}
\label{defa}
\alpha^i_a(\tau)&=&4m_a\frac{1-{\bf k}\cdot {\bf v}_a}{\sqrt{1-v^2_a}}\;
\frac{{\hat\xi}^i-{\hat\xi}_a^i}{|{\hat\bm{\xi}}-{\hat\bm{\xi}}_a|^2}\;.
\end{eqnarray}
Thus, for the vectors $\alpha^i$ and $\beta^i$
introduced in (\ref{32}), (\ref{30}) and from the formulas (\ref{53})-(\ref{54})
one obtains \cite{86}
\begin{eqnarray}
\label{g5}
\alpha^i(\tau)&=&\sum_{a=1}^N \alpha^i_a(\tau)
+O\left(\frac{G m_a}{c^2 r_a}\frac{v_a}{c}\right)
+O\left(\frac{G m_a}{c^2 r_a} \frac{d_a}{r_a}\right)
\;,\\\nonumber\\
\label{g6}
\beta^i(\tau)&=&-\frac{1}{R}\sum_{a=1}^N r_a\alpha^i_a(\tau)
-\frac{2}{R}
\sum_{a=1}^N \frac{m_a v^i_{aT}}
{\sqrt{1-v^2_a}}\ln\left(\frac{|{\hat\bm{\xi}}-{\hat\bm{\xi}}_a|^2}
{2r_a}\right)
+O\left(\frac{G m_a}{c^2 r_a}\frac{v_a}{c}\right)
+O\left(\frac{G m_a}{c^2 r_a} \frac{d_a}{r_a}\right)\;,\\\nonumber\\
\label{g7}
\beta^i(\tau_0)&=&-\frac{2}{R}
\sum_{a=1}^N \frac{m_a v^i_{aT}}
{\sqrt{1-v^2_a}}\ln\left(2r_{0a}\right)
+O\left(\frac{G m_a}{c^2 r_a}\frac{v_a}{c}\right)
+O\left(\frac{G m_a}{c^2 r_a} \frac{d_a}{r_a}\right)\;,\\\nonumber\\
\label{g8a}
\gamma^i(\tau)&=&O\left(\frac{G m_a}{c^2 r_a}\frac{v_a^2}{c^2}\right)\;,
\end{eqnarray}
where (by definition) the transverse velocity $v^i_{aT}=P^i_{\;j} v^j_a$ is
the projection of the velocity
of the $a$-th body onto the plane being orthogonal to the unperturbed light 
trajectory.

Let us introduce the new operator of projection onto the plane which is 
orthogonal to the vector ${\bf K}$
\begin{eqnarray}
\label{g8}
{\cal P}^{ij}&=&\delta^{ij}-K^i K^j\;.
\end{eqnarray}
It is worth emphasizing that the operator ${\cal P}^{ij}$ differs from  
$P^{ij}=\delta^{ij}-k^i k^j$ by relativistic corrections because of 
the relation (\ref{29}) between the 
vectors ${\bf k}$ and ${\bf K}$. We define a new impact parameter
$\xi^i={\cal P}^i_j x^j={\cal P}^i_j x^j_0$ of the
unperturbed light trajectory with respect to the direction defined by the
vector
${\bf K}$. The old impact parameter $\hat{\bm{\xi}}$ differs from the new
one ${\bm{\xi}}$ by relativistic corrections. The direction of the 
perturbed light
trajectory at the point of observation is determined by the unit vector ${\bf
s}$ according to equation (\ref{dop}). We use that definition to draw a
straight line originating from the point of observation and directed 
along the vector 
${\bf s}$ up to the point of its intersection with the lens plane 
(see Fig \ref{movingGL2}). The line is parametrized through the 
parameter $\lambda$
and its equation is given by
\begin{eqnarray}
\label{g9}
x^i(\lambda)&=&x^i(t)+s^i\;(\lambda-t)\;,
\end{eqnarray}
where $\lambda$ should be understood as the running parameter, 
$t$ is the value of the parameter $\lambda$ fixed at the moment of 
observation, and
$x^i(t)$ are the spatial coordinates of the point of 
observation. On the other hand, the coordinates of the point 
$x^i(\lambda)$ at the instant of time $\lambda^\ast$ when the line 
(\ref{g9}) intersects the lens plane, can be
defined also as
\begin{eqnarray}
\label{g9a}
x^i(\lambda^\ast)&=&X^i(\lambda^\ast)+\eta^i-\xi^i_L\;,
\end{eqnarray}
where $\eta^i={\cal P}^i_j\;x^i(\lambda^\ast)$ is the perturbed value 
of the impact 
parameter $\xi^i$ caused by the influence of the combined gravitational 
fields of the 
(micro) lenses $m_a$, $X^i(\lambda^\ast)={\cal M}^{-1}\sum^N_{a=1}
m_a x^i_a(\lambda^\ast)$ are coordinates of the center of mass of
the lens at the moment $\lambda^\ast$. 
When the line (\ref{g9}) intersects the lens plane the 
numerical value of
$\lambda$ up to corrections of order $O(d/r)$ is equal to that of the 
retarded time $s$ defined by equation like (\ref{rt}) in which $r_a$ is
replaced by $r$ - the distance from observer to the lens. 
It means that at the lens
plane ${\lambda^\ast}-t\simeq -r$. 
Accounting for this note, and applying the operator of projection ${\cal
P}_{ij}$ to the equation (\ref{g9}), we obtain
\begin{eqnarray}
\label{g10}
\eta^i&=&\xi^i-\left[\alpha^i(\tau)+\beta^i(\tau)-\beta^i(\tau_0)+\gamma^i(\tau)
\right]\;r
\;.
\end{eqnarray}
Finally, making use of the relationships (\ref{g5})-(\ref{g7}) and
expanding distances $r_a$, $r_{0a}$ around the values $r$, $r_0$
respectively (see Fig. \ref{movingGL2} for explanation of meaning of
these distances), the equation
of gravitational lens in vectorial notations reads as follows
\begin{eqnarray}
\label{g11}
{\bm{\eta}}&=&{\bm{\xi}}-\frac{r\;r_{0}}{R}\;{\bm{\alpha}}({\bm{\xi}})+
\frac{r}{R}\;{\bm{\kappa}}({\bm{\xi}})\;,
\end{eqnarray}
where
\begin{eqnarray}
\label{g12}
{\bm{\alpha}}({\bm{\xi}})&=&
4\sum_{a=1}^N m_a\frac{1-{\bf k}\cdot {\bf v}_a}{\sqrt{1-v^2_a}}\;
\frac{{\xi}^i-{\xi}_a^i}{|{\bm{\xi}}-{\bm{\xi}}_a|^2}\;,\\\nonumber\\
\label{g13}
{\bm{\kappa}}({\bm{\xi}})&=&
2\sum_{a=1}^N \frac{m_a v^i_{aT}}
{\sqrt{1-v^2_a}}\ln\left(\frac{|{\bm{\xi}}-{\bm{\xi}}_a|^2}
{2r_a\;r_{0a}}\right)\;.
\end{eqnarray}
It is not difficult to realize that the third term on the right hand side of
the equation (\ref{g11}) is $(d_a/r_{0})(v_a/c)$ times smaller than the 
second one. For this reason we are allowed to neglect it and represent 
the equation of
gravitational lensing in its conventional form 
\cite{87}, \cite{88}
\begin{eqnarray}
\label{g11bb}
{\bm{\eta}}&=&{\bm{\xi}}-\frac{r\;r_{0}}{R}\;{\bm{\alpha}}({\bm{\xi}})\;,
\end{eqnarray}
where ${\bm{\alpha}}({\bm{\xi}})$ is given by (\ref{g12}). It is
worthwhile emphasizing that although the assumption of constant velocities
of particles ${\bf v}_a$ was made, the equation (\ref{g11bb}) is actually
valid for arbitrary velocities under the condition that the accelerations 
of the bodies are small and can be neglected.
  
It is useful to compare the expression for the angle of deflection $\alpha^i$
given in equation (\ref{g12}) with that derived one in our previous work 
\cite{1}. In that paper we have considered different aspects of 
astrometric and timing effects of gravitational waves from localized sources.
The gravitational field of the source was described in terms of static monopole,
spin dipole, and time-dependent quadrupole moments. Time delay and the angle 
of light
deflection $\alpha^i$ in case of gravitational lensing were obtained in the
following form \cite{1}
\begin{equation}
\label{wer}
t-t_0=|{\bf x}-{\bf x}_0|-4\psi+2{\cal M}\ln(4r
r_0)\;,\quad\quad\quad\quad
\alpha_i=4\;{\hat{\partial}}_i\psi\;,
\end{equation}
where the partial (`projective') derivative reads 
${\hat{\partial}}_i\equiv
P_{\;i}^j{\partial}/\partial\xi^j$, and $r$ and $r_0$ are distances from the
lens to observer and the source of light respectively.
The quantity $\psi$ is the, so-called, gravitational lens potential 
\cite{87}, \cite{88} having the form
\cite{1}
\begin{eqnarray}
\label{damour}
\psi&=&\left[{\cal M}+\epsilon_{jpq} k^p{\cal S}^q{\hat{\partial}}_j+
\frac{1}{2}\;{\cal I}^{pq}(t^{\ast})\;{\hat{\partial}}_{pq}
\right]\ln |{\bm{\xi}}|\;,
\end{eqnarray}
and $\epsilon_{jpq}$ is the fully antisymmetric Levi-Civita symbol. 
The expression (\ref{damour}) includes the explicit dependence on the static 
mass ${\cal
M}$, spin ${\cal S}^i$, and time-dependent quadrupole moment ${\cal I}^{ij}$ of
the deflector taken at the moment $t^{\ast}$ of the closest approach of
the light 
ray to the
origin of the coordinate system which was chosen at the center of mass 
of the
deflector emitting gravitational waves so that the dipole moment ${\cal
I}^i$ of the system equals to zero identically. It generalizes 
the result obtained independently in \cite{11} for the case of
a stationary gravitational field of the deflector
for the gravitational lens
potential which is a function of time.
In case of the isolated astronomical system of $N$ bodies the multipole moments
are defined in the Newtonian approximation as follows
\begin{equation}
\label{multi}
{\cal M}=\sum_{a=1}^N m_a\;,\quad\quad{\cal I}^i=\sum_{a=1}^N m_a x_a^i\;,
\quad\quad
{\cal S}^i=\sum_{a=1}^N m_a ({\bf x}_a\times{\bf v}_a)^i\;,\quad\quad
{\cal I}_{ij}=\sum_{a=1}^N m_a\left(x_a^i x_a^j-\frac{1}{3}{\bf
x}_a^2\;\delta^{ij}\right)\;,
\end{equation}
where the symbol `$\times$' denotes the usual Euclidean cross product and, what
is more important, coordinates and velocities of {\it all} bodies are
taken at  {\it{one and the same instant}} of time. In the rest of this
section we assume that velocity of light-ray-deflecting bodies are small
and the origin of coordinate frame is chosen at the
barycenter of the gravitational lens system. It means that
\begin{equation}
\label{dip}
{\cal I}^i(t)=\sum_{a=1}^N m_a x_a^i(t)=0\;,\quad\quad\mbox{and}\quad\quad
\dot{\cal I}^i(t)=\sum_{a=1}^N m_a v_a^i(t)=0\;.
\end{equation}

Now it is
worthwhile to note that coordinates of gravitating bodies in (\ref{g12}) are
taken at different instants of retarded time defined for each body by the
equation (\ref{rt}). In the case of gravitational lensing all these retarded
times are close to the moment of the closest approach $t^{\ast}$ and we are
allowed to use the Taylor expansion of the quantity
\begin{eqnarray}
\label{g13a}
\sum_{a=1}^N m_a x_a^i(s)&=&\sum_{a=1}^N m_a x_a^i(t^{\ast})+
\sum_{a=1}^N m_a v_a^i(t^\ast)(s-t^{\ast})+O(s-t^{\ast})^2\;.
\end{eqnarray} 
Remembering that retarded time $s$ is defined by equation (\ref{rt}) and the 
moment of the closest approach is given by the relationship
\begin{eqnarray}
\label{rem}
t^{\ast}&=&t-{\bf k}\cdot{\bf x}=t-{\bf k}\cdot{\bf r}_a-{\bf k}\cdot{\bf
x}_a(s)\;,
\end{eqnarray}
we obtain, accounting for (\ref{nnp}), 
\begin{eqnarray}
\label{st}
s-t^{\ast}&=&{\bf k}\cdot{\bf x}_a(s)-\frac{d_a^2}{2r_a}\simeq
{\bf k}\cdot{\bf x}_a(t^{\ast})+O\left(\frac{d_a^2}{r_a}\right)+
O\left(\frac{v_a}{c}\; x_a\right)\;.
\end{eqnarray}
Finally, we conclude that
\begin{eqnarray}
\label{fin}
\sum_{a=1}^N m_a x_a^i(s)&=&\sum_{a=1}^N m_a v_a^i(t^{\ast})
[{\bf k}\cdot{\bf x}_a(t^{\ast})]+...\;,
\end{eqnarray}
where ellipses denote terms of higher order of magnitude, and where the 
equation (\ref{dip}) has been used.

Let us assume that the impact parameter $\xi^i$ is always larger than the
distance $\xi_a^i$. Then making use of the Taylor expansion of the right hand
side of equation (\ref{g12}) with respect to $\xi_a^i$ and $v_a/c$ 
one can prove that 
the deflection angle $\alpha^i$ is
represented in the form
\begin{eqnarray}
\label{annn}   
\alpha_i&=&4\hat{\partial}_i\Psi\;,
\end{eqnarray}
where the potential $\Psi$ is given as follows
\begin{eqnarray}
\label{psii}
\Psi&=&\biggl\{\sum_{a=1}^N m_a-{\bf k}\cdot\sum_{a=1}^N m_a {\bf v}_a(s)-
\sum_{a=1}^N m_a x_a^j(s)\hat{\partial}_j+\\\nonumber\\\nonumber\mbox{}&&
{\bf k}\cdot\sum_{a=1}^N m_a{\bf v}_a(s)\;x_a^j(s)\;\hat{\partial}_j+
\frac{1}{2}\sum_{a=1}^N m_a x_a^p(s)\;x_a^q(s)\;\hat{\partial}_{pq}
\biggr\}\ln|{\bm{\xi}}|+...\;,
\end{eqnarray}
and ellipses again denote residual terms of higher order of magnitude. 
Expanding all terms depending on retarded time 
in this formula with respect to the time $t^{\ast}$, noting that the 
second `projective' derivative $\hat{\partial}_{pq}$ is traceless,
and taking into account the
relationship (\ref{fin}), the center-of-mass conditions (\ref{dip}), 
the definitions of multipole moments (\ref{multi}), and 
the vector equality 
\begin{eqnarray}
\label{veceq}
x_a^j({\bf k}\cdot{\bf v}_a)-v_a^j({\bf k}\cdot{\bf x}_a)&=&({\bf k}\times({\bf
x}_a\times{\bf v}_a))^j\;,
\end{eqnarray}
we find out that with necessary accuracy the gravitational lens 
potential is given by $\Psi=\psi$ \cite{89}.
Hence, the gravitational lens formalism elaborated in this paper gives the 
same result for the
angle of deflection of light as it is shown in formulas (\ref{wer}), 
(\ref{damour}).

\subsubsection{Gravitational Shift of Frequency by a Moving Gravitational
Lens}

We assume that the velocity ${\bf v}_a$ of each body comprising the lens 
is almost constant so that we can neglect the bodies' acceleration as it
was assumed in the previous section. 
The calculation of the gravitational shift of frequency by a moving 
gravitational lens
is performed by making use of a general equation (\ref{58}). As we are
primarily interested in gravitational
lensing, derivatives of proper times of the source of light, ${\cal
T}_0$, and observer, ${\cal T}$, with
respect to coordinate time, $t$, can be calculated neglecting contributions
from the metric tensor. It yields
\begin{eqnarray}
\label{tder1}
\frac{d{\cal T}_0}{dt_0}&=&\sqrt{1-v_0^2}\;,
\end{eqnarray}
\begin{eqnarray}
\label{tder2}
\frac{dt}{d{\cal T}}&=&\frac{1}{\sqrt{1-v^2}}\;.
\end{eqnarray}
Accounting for the identity (\ref{add4}), 
we obtain from (\ref{61}) 
\begin{eqnarray}
\label{neg}
\frac{dt_0}{dt}&=&
\frac{1+{\bf K}\cdot {\bf v}-2\displaystyle{
\sum_{a=1}^N\;m_a\left[\frac{\partial s}{\partial t}
\frac{\partial}{\partial s}+
\frac{\partial t^{\ast}}{\partial t}
\frac{\partial}{\partial t^{\ast}}+
\frac{\partial k^i}{\partial t}
\frac{\partial}{\partial k^i}
\right]B_a(s,s_0,t^{\ast},{\bf k})}}
{1+{\bf K}\cdot {\bf v}_0
+2\displaystyle{
\sum_{a=1}^N\;m_a\left[\frac{\partial s}{\partial t_0}
\frac{\partial}{\partial s}+
\frac{\partial s_0}{\partial t_0}\frac{\partial}{\partial s_0}+
\frac{\partial t^{\ast}}{\partial t_0}
\frac{\partial}{\partial t^{\ast}}+
\frac{\partial k^i}{\partial t_0}
\frac{\partial}{\partial k^i}
\right]B_a(s,s_0,t^{\ast},{\bf k})}}\;.
\end{eqnarray}
After taking partial derivatives with the
help of relationships (\ref{add1})-(\ref{perf}), using 
the expansions (\ref{g3}), 
(\ref{g4}), (\ref{nnp}), neglecting terms of order $d_a/r_a$, $m_a/r_a$,
$m_a/r_{0a}$, and reducing similar terms, one gets
\begin{eqnarray}
\label{hope}
1+z&=&
\left(\frac{1-v^2_0}{1-v^2}\right)^{1/2}
\frac{1+\left({\bf K}+{\bm{\beta}}-{\bm{\beta}}_0\right)\cdot{\bf v}+
4\displaystyle{
\sum_{a=1}^N\;m_a\frac{1-{\bf k}\cdot{\bf v}_a}{\sqrt{1-v_a^2}}
\frac{({\bf k}\times{\bf v})\cdot({\bf k}\times{\bf
r}_a)}{|{\bm{\xi}}-{\bm{\xi}}_a|^2}}}
{1+\left({\bf K}+{\bm{\beta}}-{\bm{\beta}}_0\right)\cdot {\bf v}_0+
4\displaystyle{
\sum_{a=1}^N\;m_a\frac{1-{\bf k}\cdot{\bf v}_0}{\sqrt{1-v_a^2}}
\frac{({\bf k}\times{\bf v}_a)\cdot({\bf k}\times{\bf
r}_a)}{|{\bm{\xi}}-{\bm{\xi}}_a|^2}}}\;,
\end{eqnarray}
where the relativistic corrections ${\bm{\beta}}={\bm{\beta}}(\tau,\hat{\bm{\xi}})$,
${\bm{\beta}}_0={\bm{\beta}}(\tau_0,\hat{\bm{\xi}})$ are given by means of
expressions (\ref{30}), (\ref{30a}), (\ref{49}) - (\ref{52}). Making use of
relationship (\ref{29}) between the unit vectors ${\bf K}$ and ${\bf k}$, the
previous formula can be displayed as follows
\begin{eqnarray}
\label{hop}
1+z&=&
\left(\frac{1-v^2_0}{1-v^2}\right)^{1/2}
\frac{1-{\bf k}\cdot{\bf v}+
4\displaystyle{
\sum_{a=1}^N\;m_a\frac{1-{\bf k}\cdot{\bf v}_a}{\sqrt{1-v_a^2}}
\frac{({\bf k}\times{\bf v})\cdot({\bf k}\times{\bf
r}_a)}{|{\bm{\xi}}-{\bm{\xi}}_a|^2}}}
{1-{\bf k}\cdot {\bf v}_0+
4\displaystyle{
\sum_{a=1}^N\;m_a\frac{1-{\bf k}\cdot{\bf v}_0}{\sqrt{1-v_a^2}}
\frac{({\bf k}\times{\bf v}_a)\cdot({\bf k}\times{\bf
r}_a)}{|{\bm{\xi}}-{\bm{\xi}}_a|^2}}}\;.
\end{eqnarray}
This formula is gauge-invariant with respect to small coordinate 
transformations in the first post-Minkowskian approximation which leave the 
coordinates asymptotically Minkowskian. Moreover, the formula (\ref{hop}) is invariant 
with respect to Lorentz transformations and can be
applied for arbitrary large velocities of observer, source of light, and
gravitational lens. In case of slow motion 
of the source of light,
the equation (\ref{hope}) can be further simplifed by expansion with 
respect to powers of $v_0/c$, $v/c$, and $v_a/c$. 
Neglecting terms of order $v^4/c^4$, $v^4_0/c^4$, $(m_a/d_a)
(v^2/c^2)$, etc.,
this yields for the frequency shift
\begin{eqnarray}
\label{vic}
\frac{\delta\nu}{\nu_0}&=&{\bf k}\cdot({\bf v}_0-{\bf
v})\left[1+{\bf k}\cdot{\bf v}_0+({\bf k}\cdot{\bf v}_0)^2
-\frac{v_0^2}{2}+\frac{v^2}{2}\right]
-\frac{v_0^2}{2}+\frac{v^2}{2}\\\nonumber\\
\nonumber&&
+\;4\displaystyle{
\sum_{a=1}^N\;m_a\frac{1-{\bf k}\cdot{\bf v}_a}{\sqrt{1-v_a^2}}
\frac{\left[{\bf k}\times({\bf v}-{\bf v}_a)\right]
\cdot({\bf k}\times{\bf
r}_a)}{|{\bm{\xi}}-{\bm{\xi}}_a|^2}}\;,
\end{eqnarray}
where $\delta\nu=\nu-\nu_0$.
The terms on the right hand side of this formula
depending only on the velocities of the source of light and observer 
are the part of the special relativistic Doppler
shift of frequency caused by the motion of the observer and the source of light. 
The last term on the right hand side of (\ref{vic}) describes the gravitational
shift of frequency caused by the time-dependent gravitational deflection 
of light rays due to relative motion of lens with respect to
observer \cite{93}. It shows that the static gravitational lens being at rest
with respect to observer does not
lead to the gravitational shift of frequency which appears only if there
is a relative transverse velocity of the lens with respect to the observer  
which brings in the dependence of the impact
parameter for the $a$-th body, ${\bm{\xi}}-{\bm{\xi}}_a$, on time \cite{corr.1}. 
By expanding the last term in the
expression (\ref{vic}) with respect to powers $\xi/\xi_a$ the gravitational
shift of frequency reads
\begin{eqnarray}
\label{rewr1}
\left(\frac{\delta\nu}{\nu_0}\right)_{gr}&=&4\;\frac{\partial\psi}
{\partial t^\ast}+{\bf v}\cdot{\bm{\alpha}}({\bm{\xi}})\;, 
\end{eqnarray}
where the deflection angle ${\bm{\alpha}}$ is displayed in (\ref{wer}).
It is remarkable that the formula (\ref{rewr1}) is a direct consequence of 
the equation 
(\ref{wer}) for the time delay in gravitational lensing. 

Indeed, let us assume that the lens is comprised of an ensemble of $N$ 
point-like
bodies each moving with (time-dependent) 
velocity ${\bf v}_a$ with respect to the origin of
the coordinate system chosen near the (moving) barycenter of the 
lensing object. The
velocity 
${\bf V}$ of the center-of-mass of the
gravitational lens is defined as the first time derivative of the dipole
moment ${\cal I}^i$  of the lens shown in (\ref{multi}), that is
\begin{eqnarray}
\label{vel1}
V^i&=&\frac{dX^i}{dt}=\frac{1}{\cal M}\sum_{a=1}^N m_a v_a^i=
\frac{\dot{\cal I}^i}{\cal M}\;.
\end{eqnarray} 
Using this definition and assuming that at the initial epoch the
barycenter of the lens is at the origin of the coordinate system, 
we find out that the time derivatives of the lens gravitational potential
(\ref{dipole}) 
reads as follows (see the remarks in \cite{corr.1}, \cite{94} which clarify 
calculation of the derivatives)
\begin{equation}
\label{rewr}
\frac{\partial\psi}{\partial t^\ast}=\left(
-{\cal M}V^i\;\hat{\partial}_i
+\frac{1}{2}\dot{\cal
I}^{ij}\hat{\partial}_{ij}\right)\ln|{\bm{\xi}}|\;,\quad\quad\quad\quad
\frac{\partial\psi}{\partial t}=0\;,\quad\quad\quad\quad
\frac{\partial\psi}{\partial t_0}=v_0^i\;\hat{\partial}_i\psi=\frac{1}{4}({\bf
v}_0\cdot{\bm{\alpha}})\;. 
\end{equation}
Formulas given in (\ref{rewr}) allow to find out the total differential of equation
(\ref{wer}) in terms of the increments of time $dt$ and $dt_0$. While finding the total
differential of the gravitational lens equation, it should be also kept in
mind that asymptotically in the limit $r_0\rightarrow +\infty$, $r=const.$, 
the following relationship (see formulas (\ref{29}), 
(\ref{g6}),(\ref{g7}))
\begin{eqnarray}
\label{asymp}
{\bf K}&=&-{\bf k}+{\bm{\alpha}}\;,
\end{eqnarray}
between vectors ${\bf K}$ and ${\bf k}$ holds. Taking differential of equation
(\ref{wer}) using results of (\ref{rewr}) and (\ref{asymp}) one confirms the 
validity of the
presentation (\ref{rewr1}) for the gravitational shift of frequency in gravitational
lensing. Formula (\ref{rewr1})  
reflects the fact that the gravitational shift of
frequency can be induced if and only if the gravitational lens potential is a
function of time. 

One sees that, in general, not only the translational 
motion of the lens with 
respect to observer generates the 
gravitational shift of frequency but also the
time-dependent part of the quadrupole moment of the lens. Besides this, 
we emphasize
that the motion of observer with respect to the solar system barycenter
should produce periodic annual changes in the observed spectra of images of
background sources in cosmological gravitational lenses. This is because
of the presence of the solar system time-periodic part of velocity 
${\bf v}$ 
of observer in equation (\ref{rewr1}). The effect of the frequency shift
may
reveal the small scale variations of the temperature of the CMB radiation in the 
sky caused by the time-dependent gravitational lens effect on clusters 
of galaxies having peculiar motion with respect to the cosmological
expansion. 
However,
it will be technically challenging to observe this effect because of its 
smallness. 

The simple relationship (\ref{rewr1}) can be compared to the 
result of the calculations by Birkinshaw \& Gull 
(\cite{95}, equation 9). We have checked that the derivation of the
corresponding formula for the gravitational shift of frequency given 
by Birkinshaw \& Gull \cite{95} on the ground of a
pure phenomenological approach and 
cited in \cite{85} is consistent, at least, in the first order approximation 
with respect
to the velocity of lens \cite{96}. Preliminary numerical simulations of the CMB 
anisotropies by moving gravitational lenses carried out in the paper 
\cite{98} on the premise of formula (\ref{rewr1}) under assumption ${\bf v}=0$, confirm a significance of the 
effect for future space experiments
being designed for detection of the small scale temperature fluctuations of 
the CMB. 

However, we would like to make it clear that in practice the gravitational 
shift of frequency caused by moving gravitational lens must be calculated on the 
basis of a different-from-equation (\ref{rewr1}) formulation. The matter is that 
the gravitational lens is not at infinity but at a finite distance. For 
this reason, calculation and subtraction of special relativistic Doppler shift 
of frequency in equation  (\ref{vic}) should be done using the unit vector ${\bf 
K}$ related to ${\bf k}$ by the transformation (\ref{29}). Using the given 
transformation for replacement of ${\bf k}$ by ${\bf K}$ in (\ref{rewr1}),  
remembering (see equations (\ref{g6}), (\ref{g7})) that 
${\bm{\beta}}(\tau)=-r/R\;{\bm{\alpha}}$, and the angle 
${\bm{\beta}}(\tau_0)$ is negligibly small, we obtain for the observable 
shift of frequency in gravitational lensing 
\begin{eqnarray}
\label{real} 
\left(\frac{\delta\nu}{\nu_0}\right)_{gr}^{obs}&=&4\;\frac{\partial\psi}
{\partial t^\ast}+\frac{r_0}{R}\left({\bf v}\cdot{\bm{\alpha}}\right)+
\frac{r}{R}\left({\bf v}_0\cdot{\bm{\alpha}}\right)\;, 
\end{eqnarray}
where $r$ and $r_0$ are distances from observer to the lens and from the lens to 
the source of light respectively, $R=|{\bf x}-{\bf x}_0|\simeq r+r_0$. 
In the limit $r_0\rightarrow +\infty$, $r=const.$, the equation (\ref{real})
goes over to equation (\ref{rewr1}). 

The last two terms in the right hand side of equation 
(\ref{real}) have been derived by Bertotti \& Giampieri \cite{corr.4} who used
a different mathematical technique assuming that the lens is static. Hence,
they missed the first term in the right hand side of (\ref{real}) 
discovered by Birkinshaw \& Gull 
\cite{95} who, in their own turn, neglected the contributions due to the
motion of source of light and observer. It is also useful to note that 
equation (22) in the paper 
\cite{corr.4} for Doppler shift in gravitational lensing contains a misprint of
algebraic sign
in front of the term depending on the velocity of observer. The error has been
corrected in the paper \cite{corr.5} by Iess {\it et al.} 
(see equation (8) in \cite{corr.5}) so that their result coincides precisely 
with the last two terms in the right hand side of our equation (\ref{real}).

\subsection{General Relativistic Astrometry in the Solar System}  

\subsubsection{Theoretical Background}

For a long time the basic theoretical principles of general relativistic 
astrometry
in the solar system were based on using the post-Newtoninan approximate
solution of the Einstein field 
equations \cite{7} - \cite{9}, \cite{60}, \cite{99}. 
The metric tensor of the post-Newtonian solution is 
an instantaneous 
function of coordinate time $t$. It depends on the field point, 
${\bf x}$, 
the coordinates, ${\bf x}_a(t)$, and velocities, ${\bf v}_a(t)$, of the 
gravitating 
bodies and is valid only inside the near zone of the solar system because
of the
expansion of retarded integrals with respect to the small parameter $v_a/c$
\cite{6}. This expansion restricts the domain of validity for which
the propagation 
of light rays can be considered from the mathematical point of view in a 
self-consistent manner by the boundary of the near zone. Finding a solution of 
the equations of light
propagation (\ref{14}) in the near zone of the, for instance, solar 
system 
can be achieved by means of expanding 
positions and velocities of the solar system bodies in Taylor series around 
some fixed instant of time, their substitution into the equations 
of motion of photons (\ref{14}), and their subsequent integration with 
respect to time. Such an
approach is theoretically well justified for a proper description of radar
\cite{100} and lunar laser \cite{101} 
ranging experiments, and the interpretation of the Doppler tracking of 
satellites \cite {corr.4}, \cite{corr.5},  
\cite{102}
- \cite{corr.6}. The only problem which arises in the approach under 
discussion
is how to determine that fiducial instant of time to which coordinates and
velocities of gravitating bodies should be anchored. Actually, 
the answer on 
this question is dimmed if one works in the framework of the
post-Newtonian approximation scheme which disguises the hyperbolic
character of the Einstein equations for the gravitational field and does not
admit us to distinguish between advanced and retarded solutions of the
field equations \cite{75}. 
For this reason, propagation of light rays, which always takes place along the 
isotropic characteristics of a light cone, is different in the post-Newtonian 
scheme from the gravitational-field propagation because the latter propagates in that
framework instanteneously and with infinite speed.
Thus, the true causal
relationship between the position of the light particle and location of 
the 
light-ray-deflecting bodies in the system is violated which leads to a
necessity to use some
artificial assumptions about the initial values of positions and velocities
of the bodies for integration of equations of light propagation 
(see Fig. \ref{covariant4} for more details).  
One of the reasonable choices is to fix
coordinates and velocities of the body at the moment of the closest approach 
of light ray to it. Such an assumption was used by Hellings \cite{10} and put on
more firm ground by Klioner \& Kopeikin \cite{4} revealing that it minimizes the
magnitude of residual terms of the post-Newtonian solution of the equations of
light propagation. 

As it has been explained above, 
the post-Newtonian approach has stringent limitations when applied to the
integration of equations of light propagation in the case when the 
light-ray-perturbing gravitating system is not in a steady state and 
the points of emission, ${\bf x}_0$, and observation, ${\bf x}$, of light are 
separated by the distance which is much larger than the characteristic 
(Keplerian) time of the system. The first limitation comes from the fact that
the general post-Newtonian expansion of the metric tensor diverges as the 
distance 
$r$ from the system increases (see, for instance, \cite{104} - \cite{106}). 
Usually this fact has
been ignored by previous researches who used for the integration of the 
equations of
light rays the following trancated form of the metric tensor
\begin{eqnarray}
\label{pn1}
g_{00}(t,{\bf x})&=&-1+\frac{2U(t,{\bf x})}{c^2}+O(c^{-4})\;,\\\nonumber\\
\label{pn2}
g_{0i}(t,{\bf x})&=&-\frac{4U^i(t,{\bf x})}{c^3}+O(c^{-5})\;,\\\nonumber\\
\label{pn3}
g_{ij}(t,{\bf x})&=&\delta_{ij}\left[1+\frac{2U(t,{\bf x})}{c^2}\right]
+O(c^{-4})\;,
\end{eqnarray}
where the instantaneous, Newtonian-like potentials are given by the
expressions
\begin{eqnarray}
\label{u}
U(t,{\bf x})&=&\sum_{a=1}^N\frac{m_a}{|{\bf x}-{\bf x}_a(t)|}\;,\\\nonumber\\
\label{ui}
U^i(t,{\bf x})&=&\sum_{a=1}^N\frac{m_a v^i_a(t)}{|{\bf x}-{\bf x}_a(t)|}\;,
\end{eqnarray}
and all terms describing high-order multipoles have been omitted \cite{107}.
From a purely formal point of view the expressions 
(\ref{pn1})-(\ref{pn3}) are not
divergent when the distance $r$ approaches infinity but residual terms in
the metric tensor are.
It means that the post-Newtonian metric can not be used for finding
solutions of equations of light propagation if the distance $r$ is larger some
specific value $r_0$. This spatial divergency of the metric tensor relates to
the fact that the expressions (\ref{pn1})-(\ref{pn3}) represent only 
the first terms in the
{\it near-zone expansion} of the metric and say nothing about the
behavior of the metric in the {\it far-zone} \cite{6}, \cite{106},
\cite{108}. The distance 
$r_0$ bounding the near-zone is about the characteristic wavelength 
$\lambda_{gr}$ of
the gravitational
radiation emitted by the system ($\lambda_{gr}\simeq (c P_b)/(4\pi)$, where
$P_b$ is the characteristic Keplerian time of the system). If we assume, for 
example, that the main bulk of the
gravitational radiation emitted by the solar system is produced by the orbital
motion of Jupiter, the distance $r_0$ does not exceed 0.3 pc. Almost all
extra-solar luminous objects visible in the sky lie far beyond this distance. 
From this point of view, the results of integration of the equations of light 
propagation from stars of 
our galaxy and extra-galactic objects having been performed previously by 
different authors on the premise of the implementation of metric tensor 
(\ref{pn1})-(\ref{pn3}) can not be considered as rigorous and conclusive
for residual terms of such an integration were never discussed.   

The second limitation for the application of the near-zone expansion of
the metric
tensor relates to the retarded character of the propagation of the gravitational
interaction. The expressions (\ref{pn1})-(\ref{pn3}) are instantaneous functions 
of
time and do not show this property of retardation 
at all. At the same time the
post-Newtonian metric (\ref{pn1})-(\ref{pn3}) can be still used for 
integration of equations of light
rays, at least from the formal point of view, because the integration
will give a convergent result. However, we may expect that the
trajectory of light ray obtained by solving the equations of propagation of
light using the instantaneous potentials will deviate from that obtained using 
the metric perturbations expressed as the {\it Li\'enard-Wiechert} 
potentials. Such
a deviation can be, in principle, so large that the error might be comparable
with the main term of relativistic deflection of light and/or time delay. None
of the methods of integration attempted so far contains error estimates in a
precise mathematical sense; at best, errors have been roughly estimated using
matched asymptotic technique \cite{4}. None of other previous
authors have ever tried to develop a self-consistent approach for calculation
of the errors. 

One more problem relates to the method of performing time
integration of the instantaneous potentials along the unperturbed trajectory of
the light ray. This is because coordinates and velocities of bodies are 
functions of
time. Even in the case of circular orbits we have a problem of solving
integrals of the type
\begin{eqnarray}
\label{inst1}
\int^t_{t_0}U(t,{\bf x})dt&=&\sum^N_{a=1}m_a\int^t_{t_0}\frac{dt}{|{\bf
x}_0+{\bf k}\;(t-t_0)-A_a\left[{\bf e}_1\sin(\omega_a t+\varphi_a)+
{\bf e}_2\cos(\omega_a t+\varphi_a)\right]|}\;,
\end{eqnarray}
where $A_a$, $\omega_a$, and $\varphi_a$ are the radius, 
the angular frequency, and the initial phase of the 
orbit of the $a$-th body respectively, and ${\bf e}_1$, ${\bf e}_2$ are the unit 
orthogonal
vectors lying in the orbital plane. The given integral can not be performed
analytically and requires the application of numerical methods. In case of 
elliptical motion calculations will be even more complicated. 
Implicitly, it was usually assumed that the main contribution to the integral
(\ref{inst1}) comes from that part of the trajectory of light ray which passes
nearby the body deflecting light rays so that one is allowed to fix the position 
of
the body at some instant of time which is close to the moment of the closest
approach of the light ray to the body. 
However, errors of such an approximation
usually were never disclosed except for the attempt made by Klioner \& Kopeikin
\cite{4}. Nevertheless, it is not obvious so far that the error analysis
fullfiled in \cite{4} is complete and that the use of
the Taylor expansion of coordinates and velocities
of the solar system bodies with respect to time, made in the neighborhood
of the instant of the closest approach of photon to the light-deflecting
body in order to perform the integration in (\ref{inst1}), minimizes
errors of calculations and allows to solve light ray equations with 
better precision. Moreover, such an expansion is allowed only if photon
moves near or inside of the gravitating system. Far outside the system
the other method of solving the integral (\ref{inst1}) is
required \cite{109}.

Regarding this difficulty Klioner \&
Kopeikin \cite{4} have used a matched asymptotic technique for finding the 
perturbed
trajectory of the light ray going to the solar system from a very remote source 
of
light like a pulsar or a quasar. The whole space-time was separated in two
domains - the near and far zones lying correspondingly inside and outside of 
the distance $r_0$ being approximately equal to the characteristic length of 
gravitational waves emitted by the solar system. The internal solution of 
the equations of light rays within the near zone have been obtained by expanding
coordinates and velocities of the bodies in the Taylor time series and then 
integrating
the equations. The external solution of
the equations has been found by decomposing the metric tensor in gravitational
multipoles and accounting only for the first monopole term which corresponds
mainly to the static, spherically symmetric field of the Sun. A global solution
was obtained by matching of the internal and external solutions at the buffer 
region in order to reach the required astrometric accuracy of 1 $\mu$arcsec. 
The approach we have used sounds
reasonable and may be used in theoretical calculations. However, it does not
help very much to give a final answer to the question at which
moment of time one has to fix positions and velocities of bodies when
integrating equations of light propagation inside the near zone. In addition,
the approach under consideration does not give any recipe how to integrate 
equations of light propagation in the external domain of space (beyond 
$r_0$) if the higher, time-dependent 
gravitational multipoles should be taken into account and what magnitude of the
perturbative effects one might expect. In any case, the global solution 
obtained
by the matched asymptotic technique consists of two pieces making the
visualization of the light ray trajectory obscure and the astrometric
implementation of the method impractical.

For these reasons we do not rely in this section upon the technique developed
in \cite{4} but resort to the method of integration of light
ray equations based on the usage of the {\it Li\'enard-Wiechert} potentials. 
This method allows to construct a smooth and unique global solution of the 
light propagation equations from arbitrary distant source of light to observer 
located in the solar system. We are able to handle the integration of the
equations more easily and can easily estimate the magnitude of all residual
terms. Proceeding in this way we also 
get a unique prediction for that moment of time at which coordinates and 
positions of
gravitating bodies should be fixed just as the location of observer is
known. We shall consider three kinds of observations - pulsar timing, very long
baseline interferometry (VLBI) of quasars, and optical astrometric 
observations of stars.

\subsubsection{Pulsar Timing}

The description of the timing formula is based on the usage of 
equations (\ref{qer}) and (\ref{shapd}). 
Taking in the equation (\ref{shapd}), which should be compared with its
post-Newtonian analogue (\ref{inst1}), terms up to the order $v_a/c$
inclusively, we obtain 
\begin{eqnarray}
\label{pt1}
B_a(s,s_0)&=&-\ln\left[\frac{r_a(s)-{\bf k}\cdot{\bf r}_a(s)}
{r_{a}(s_0)-{\bf k}\cdot{\bf r}_{a}(s_0)}\right]-\int^s_{s_0}\frac{
{\bf k}\cdot{\bf v}_a(\zeta)\;d\zeta}{t^{\ast}+
{\bf k}\cdot{\bf x}_a(\zeta)-\zeta}+O\left(\frac{v_a^2}{c^2}\right)\;,
\end{eqnarray}
where the retarded times $s$ and $s_0$ should be calculated from equations
(\ref{rt}) and (\ref{rt1}) respectively, ${\bf r}_a(s)={\bf x}-{\bf x}_a(s)$, 
${\bf r}_{a}(s_0)={\bf x}_0-{\bf x}_a(s_0)$, and we assume that the observation 
is made
at the point with spatial barycentric coordinates {\bf x} at the instant
of time $t$, and the pulsar's pulse is emitted at the moment $t_0$ from
the point ${\bf x}_0$ which is at the distance of the pulsar from the
solar system, typically more than 100 pc. 

In principle, the first term in
this formula is enough to treat the timing data for any pulsar with accuracy
required for practical purposes. The denominator in the argument of the 
logarithmic
function is $r_a(s_0)-{\bf k}\cdot{\bf r}_a(s_0)\simeq 2R$, where $R$ is the
distance between the barycenter of the solar system and the pulsar. 
The logarithm of $2R$ is a function which is nearly constant but may
have a secular change because of the slow relative motion of the pulsar with
respect to the solar system. All such terms are absorbed in the pulsar's
rotational phase and can not be observed directly. For this reason, in
what follows, we shall
omit the denominator in the logarithmic term of equation (\ref{pt1}). We
emphasize that positions of the solar system bodies in the numerator of the
logarithmic term are taken at the moment of retarded time which is found by
iterations of the equation $s=t-|{\bf x}-{\bf x}_a(s)|$. It makes
calculation of the Shapiro time delay in the solar system theoretically
consistent and practically more precise.

There is a difference
between the logarithmic term in (\ref{pt1}) and the corresponding 
logarithmic terms in timing formulas suggested by
Hellings \cite{10} and Doroshenko \& Kopeikin \cite{110} where the position of
the $a$-th 
body is fixed at the moment of the closest approach of the pulse to the body. 
It is, however, 
be not so important in practice as timing observations are not yet precise 
enough to
distinguish the Shapiro delay when positions of bodies are taken at the retarded
time or at any other one, being close to it. Indeed, the maximal 
difference
is expected to be of order of
$(4GM_{\odot}/c^3)(v_{\odot}/c)(1+\cos\theta)^{-1}$, where
$M_{\odot}$ and $v_{\odot}$ are mass of the Sun and its barycentric
velocity respectively, and $\theta$ is the angle between directions
towards the Sun and the pulsar. For $v_{\odot}$ is less than 20 m/s and
$\theta$ can not exceed $0.25$ degrees the error in timing formula
relating to various definitions of the Sun's barycentric coordinates
in the expression for the Shapiro time delay is less than 200
nanoseconds. Although this value is yet beoynd of observational limit it
would be desirable to update existing timing data processing programs 
like TEMPO
\cite{111} and TIMAPR \cite{78}, \cite{110}
to make their functional structure be in agreement with the latest
theoretical developments. 

The integral in equation (\ref{pt1}) can not be calculated
analytically if the trajectory of motion of the bodies is not simple. The matter 
is
that in the case when light propagates from the remote source to the
gravitating system the time interval between $s$ and $s_0$ is not small as it
was in the case of the derivation of timing formula for binary pulsars in 
section
7A (see equation (\ref{s2}) and related discussion as well as caption to
Fig. \ref{covariant4}). This was because light 
propagates from the binary
system in the same direction as gravitational waves emitted by it, so
that the gravitational field of the binary system is almost "frozen" as
seen by the outgoing photon. When we
consider propagation of light towards the solar system the infalling
photon moves in the 
direction
being opposite to that of propagation of gravitational field generated 
by the moving solar system bodies. For this reason, the difference 
$s-s_0\simeq 2R$, and 
is very large.
Thus, we are not allowed, as it was in the case of derivation of timing
formula for binary pulsars, to use the expansion of coordinates and 
velocities of the
solar system bodies in Taylor series with respect to time. Moreover,
integrals, like
that in (\ref{pt1}), should be also calculated without any expansion 
using the known law of motion of gravitating bodies, that is the solar 
system ephemerides like DE200, DE245, or an equivalent one.
Let us give an idea what kind of result we can get proceeding in this way.

First of all, we note that the orbital plane of any of the solar system bodies 
lies very
close to the ecliptic and can be approximated fairly good by circular 
motion up to the first order  
correction with respect to the orbital eccentricity which is usually 
small.
The motion of the Sun with respect to the barycenter of the solar system may be
described as a sum of harmonics corresponding to gravitational perturbations 
from
Jupiter, Saturn, and other smaller bodies. Thus, we assume that ${\bf x}_a$ is 
given in the ecliptic plane as follows
\begin{eqnarray}
\label{pt2}
{\bf x}_a(t)&=&A\left[\cos(n t)\;{\bf e}_1+\sin(n t)\;{\bf e}_2\right]\;,
\end{eqnarray}
where $A$ and $n$ are the amplitude and frequency of the corresponding harmonic 
in
the Fourier decomposition of the orbital motion of the $a$-th body, ${\bf e}_1$ 
is
directed to the point of the vernal equinox, ${\bf e}_2$ is orthogonal to ${\bf 
e}_1$
and lies in the ecliptic plane. 
The vector ${\bf k}$ is defined in ecliptic
coordinates as  
\begin{eqnarray}
\label{pt3}
{\bf k}&=&-\cos b\cos l\;{\bf e}_1-\cos b\sin l\;{\bf e}_2-\sin b\;{\bf e}_3\;,
\end{eqnarray}
where $b$ and $l$ are ecliptic spherical coordinates of the pulsar. Substituting 
these
definitions into the integral of equation (\ref{pt1}) and performing
calculations using approximate relationship $\zeta=t^{\ast}-y+{\bf k}
\cdot{\bf x}_a(y)$, where $y$ is the new variable defined in (\ref{new}), we get
\begin{eqnarray}
\label{pt4}
\int^s_{s_0}\frac{
{\bf k}\cdot{\bf v}_a(\zeta)\;d\zeta}{t^{\ast}+
{\bf k}\cdot{\bf x}_a(\zeta)-\zeta}&=&
-{\bf k}\cdot{\bf v}_a(t^{\ast})\biggl\{{\bf Ci}\left[n\left(
r_a-{\bf k}\cdot{\bf r}_a\right)\right]-{\bf Ci}\left[n\left(
r_{0a}-{\bf k}\cdot{\bf r}_{0a}\right)\right]\biggr\}
\\\nonumber\\\nonumber\mbox{}&&-
{\bf k}\cdot{\bf x}_a(t^{\ast})\biggl\{{\bf Si}\left[n\left(
r_a-{\bf k}\cdot{\bf r}_a\right)\right]-{\bf Si}\left[n\left(
r_{0a}-{\bf k}\cdot{\bf r}_{0a}\right)\right]\biggr\}\;,
\end{eqnarray}
where $t^{\ast}=t-{\bf k}\cdot{\bf x}$ is the time of the closest approach of
the light ray to the barycenter of the solar system. Taking into account the
asymptotic behaviour of sine and cosine integrals for large and small 
values of their
arguments in relationship (\ref{pt4}) we arrive at the approximate formula for
the Shapiro delay
\begin{eqnarray}
\label{pt5}
\Delta(t)&=&-\sum_{a=1}^N m_a\left[1-{\bf k}\cdot{\bf v}_a(t^{\ast})\right]
\ln\left[r_a(s)-{\bf k}\cdot{\bf r}_a(s)\right]+O\left(\frac{G m_a}{c^3}
\frac{v_a}{c}\right)\;,
\end{eqnarray}     
where the residual term denotes all contributions
which are simple products of the gravitational radius $Gm_a/c^3$ of the 
$a$-th body, expressed in time units, by the
ratio $v_a/c$ up to a constant factor. If one takes numerical values of 
masses and velocities of the solar system bodies one finds that such residual 
terms are extremely much smaller than the level of errors in timing 
measurements. We
conclude that these residual terms can not be detected by the present
day pulsar timing 
techniques. 

\subsubsection{Very Long Baseline Interferometry} 

VLBI measures the time differences in the arrival of microwave signals from 
extragalactic radio sources received at two or more radio observatories 
\cite{112}. 
Generally, geodetic observing sessions run for 24 hours and
observe a number of different radio sources distributed across the sky. 
The observatories can be widely
separated; the sensitivity of the observations to variations in the 
orientation of the Earth increases with the size
of the VLBI network. VLBI is the only technique capable of measuring all 
components of the Earth's orientation accurately and
simultaneously. Currently, VLBI determinations of Earth-rotation variations, 
and of the coordinates of terrestrial sites and
celestial objects are made routinely and regularly with estimated accuracies 
of about +/-0.2 milliarcsecond or
better \cite{112}, \cite{113}. Such a high precision of observations requires
an extremely accurate accounting for different physical effects 
in propagation of light from radio sources
to observer including relativistic gravitational time delay.

There have been many papers dealing with relativistic effects which must be
accounted for in VLBI data processing (see, e.g., \cite{99}, \cite{114},
\cite{115}, and references therein). The common
efforts of many researches in this area have resulted in the creation of
what is commonly believed now to be as a 
`standard'
model of VLBI data processing which is called a consensus model \cite{115}
emerged from a workshop held in 1990 \cite{114}. The accuracy
limit chosen for the consensus VLBI relativistic time delay model is $10^{-12}$
seconds (one picosecond) of differential VLBI delay for baselines less than two
Earth radii in length. As it was stated, in the model all terms of order
$10^{-13}$ seconds or larger were included to ensure that the final result was
accurate at the picosecond level. 
By definition, extragalactic source coordinates derived from the consensus
model should have no apparent motions due to solar system relativistic 
effects at the picosecond level. Our purpose in this section is to
analyze critically this statement and to show that the consensus model is not 
enough
elaborated, at least theoretically, in accounting for relativistic 
effects in propagation of light at the picosecond level.
For this reason, we propose necessary modification of the consensus model
to make it applicable at the level of accuracy approaching to $10^{-13}$
seconds without any restrictions.

In what follows we work for simplicity with the barycentric coordinate 
time of the 
solar system only. Precise definition of
the measuring procedure applied in VLBI requires, however, derivation of 
relativistic
relationship between the proper time of observer and the barycentric 
coordinate time. It is given, e.g. in \cite{116}, and can be added
to the formalism of the present section for adapting it to 
practical applications. A
complete description of such an extended formalism will be given elsewhere.
 
The VLBI time delay (see Fig. \ref{covariant3}) 
to be calculated is the time of arrival of
electromagnetic signal, $t_2$, at station 2 
minus the
time of arrival of the same signal, $t_1$, at station 1. The time of arrival at 
station 1 serves 
as the
time reference for the measurement. In what follows, unless explicitly stated
otherwise, all vectors and scalar
quantities are assumed to be calculated at $t_1$ except for position of
the source of light ${\bf x}_0$ which is always calculated at the 
of time of light emission $t_0$. We use for calculation of the
VLBI time delay equations (\ref{qer}), (\ref{shapd}) referred to the barycentric
coordinate frame of the solar system. The equations give us
\begin{eqnarray}
\label{vlbi1}
t_2-t_1&=&|{\bf x}_2(t_2)-{\bf x}_0|-|{\bf x}_1-{\bf x}_0|+\Delta(t_2,t_0)-
\Delta(t_1,t_0)\;,
\end{eqnarray}
where ${\bf x}_0$ are coordinates of the source of light, ${\bf x}_2(t_2)$ are
coordinates of the station 2 at the moment $t_2$, ${\bf x}_1$ are
coordinates of the station 1 at the moment $t_1$. The differential relativistic 
time delay is given in the form
\begin{eqnarray}
\label{vlbi2}
\Delta(t_2,t_0)-\Delta(t_1,t_0)&=&2\sum_{a=1}^N\left[B_a(s_2,s_0)-B_a(s_1,s_0)
\right]\;,
\end{eqnarray}
where the difference of the $B_a$'s up to the linear with respect to
velocities of the solar system bodies reads (see equation (\ref{47}))
\begin{eqnarray}
\label{vlbi3}\hspace{-1 cm}
B_a(s_2,s_0)-B_a(s_1,s_0)&=&\ln\frac{r_{1a}-{\bf k}_1\cdot{\bf r}_{1a}}
{r_{2a}-{\bf k}_2\cdot{\bf r}_{2a}}-\ln\frac{r_{0a}-{\bf k}_1\cdot{\bf r}_{0a}}
{r_{0a}-{\bf k}_2\cdot{\bf r}_{0a}}\\\nonumber\\\nonumber\mbox{}&&+
{\bf k}_2\cdot{\bf v}_a(s_2)\ln\left(r_{2a}-{\bf k}_2\cdot{\bf r}_{2a}\right)-
{\bf k}_1\cdot{\bf v}_a(s_1)\ln\left(r_{1a}-{\bf k}_1\cdot{\bf r}_{1a}\right)
\\\nonumber\\\nonumber\mbox{}&&+
\int^{s_1}_{s_0}\ln\left[t^{\ast}_1+{\bf k}_1\cdot{\bf
x}_a(\zeta)-\zeta\right]\;{\bf k}_1\cdot{\dot{\bf v}}_a(\zeta)\;d\zeta-
\int^{s_2}_{s_0}\ln\left[t^{\ast}_2+{\bf k}_2\cdot{\bf
x}_a(\zeta)-\zeta\right]\;{\bf k}_2\cdot{\dot{\bf v}}_a(\zeta)\;d\zeta\;.
\end{eqnarray}
Herein $s_1$ and $s_2$ are retarded times determined iteratively from the
equations
\begin{eqnarray}
\label{vlbi4}
s_1&=&t_1-|{\bf x}_1-{\bf x}_a(s_1)|\;,
\\\label{vlbi4a}
s_2&=&t_2-|{\bf x}_2(t_2)-{\bf x}_a(s_2)|\;,
\end{eqnarray}
the quantity ${\bf r}_{1a}={\bf x}_1-{\bf x}_a(s_1)$ 
is the vector from the $a$-th body to the station 1, 
${\bf r}_{2a}={\bf x}_2(t_2)-{\bf x}_a(s_2)$ 
is the vector from the $a$-th body to the station 2, 
$r_{1a}=|{\bf r}_{1a}|$, $r_{2a}=|{\bf r}_{2a}|$
and
\begin{eqnarray}
\label{vlbi5}
t^{\ast}_1&=&t_1-{\bf k}_1\cdot{\bf x}_{1}\;,
\\\label{vlbi5a}
t^{\ast}_2&=&t_2-{\bf k}_2\cdot{\bf x}_2(t_2)\;,
\end{eqnarray}
are the moments of the closest approach of the light rays 1 and 2 to the 
barycenter
of the solar system. It will be also helpful in comparing our approach 
with the
consensus model to use the moments of the closest 
approach of the light rays 1 and 2 to the $a$-th body which we will 
define according to the rule  
\begin{eqnarray}
\label{vlbi5b}
{t^{\ast}_1}_a&=&t_1-{\bf k}_1\cdot{\bf r}_{1a}\;,
\\\label{vlbi5c}
{t^{\ast}_2}_a&=&t_2-{\bf k}_2\cdot{\bf r}_{2a}\;.
\end{eqnarray} 
It is worth emphasizing that our definitions of times ${t^{\ast}_1}_a$ 
and ${t^{\ast}_2}_a$ are slightly
different from the definitions of similar quantities given in the
`standard' consensus
model. It relates to the definition of positions of bodies in the vectors
${\bf r}_{1a}$ and ${\bf r}_{2a}$. In our case we refer the coordinates
of the bodies to the
retarded times $s_1$ and $s_2$ respectively while in the consensus model
they are taken at the times $t_1$ and $t_2$. It introduces some
uncertainty into the notion of the instant of the closest approach of
light ray to to the body or barycenter of the solar system which appears
due to not covariant formulation of the relativistic time delay in the
consensus model. There is no such an uncertainty in our approach which
is fully covariant in the first post-Minkowskian approximation.     

The unit vectors ${\bf k}_1$ and ${\bf k}_2$ are defined as
\begin{equation}
\label{vlbi6}
{\bf k}_1=\frac{{\bf x}_1-{\bf x}_0}{|{\bf x}_1-{\bf
x}_0|}\;,\quad\quad\quad\quad\quad
{\bf k}_2=\frac{{\bf x}_2(t_2)-{\bf x}_0}{|{\bf x}_2(t_2)-{\bf x}_0|}\;,
\end{equation}
which shows that they have slightly different orientations in space.
Let us introduce the barycentric baseline vector at the time of arrival $t_1$
through the definition ${\bf B}={\bf x}_2(t_1)-{\bf x}_1(t_1)$. 
Let us stress
that the baseline vector lies on the hypersurface of constant time
$t_1$. The original version of the relativistic relationship of the barycentric 
baseline 
vector to the 
geocentric one, ${\bf b}$, can be found in \cite{116} or 
later publications \cite{99}, \cite{115}. We shall neglect this relativistic 
difference in the 
expression for the Shapiro time delay because it is inessential in our
present discussion. Thus, we assume
${\bf B}={\bf b}$. The difference between the
vectors ${\bf k}_1$ and ${\bf k}_2$ may be found using the 
expansion with respect
to powers of the small parameter $b/R$ where $R$ is the distance between the 
barycenter of the solar system and source of light. We have
\begin{eqnarray}
\label{vlbi7}
{\bf x}_2-{\bf x}_0&=&{\bf x}_1-{\bf x}_0+{\bf b}+{\bf
v}_2\;(t_2-t_1)+O\left(\frac{v^2}{c^2}\;b\right)\;,\\\nonumber\\\nonumber
|{\bf x}_2-{\bf x}_0|&=&|{\bf x}_1-{\bf x}_0|+{\bf b}\cdot{\bf k}_1+{\bf
v}_2\cdot{\bf k}_1\;(t_2-t_1)+
O\left(\frac{v^2}{c^2}\;b\right)+O\left(\frac{b^2}{R}\right)\;,
\end{eqnarray}
where ${\bf v}_2$ is the velocity of station 2 with respect to the barycenter of
the solar system. These expansions yield
\begin{eqnarray}
\label{vlbi9}
{\bf k}_2&=&{\bf k}_1+\frac{{\bf k}_1\times({\bf b}\times{\bf k}_1)}{R}+
O\left(\frac{v}{c}\frac{b}{R}\right)+O\left(\frac{b^2}{R^2}\right)\;,
\end{eqnarray}
and for the time delay (\ref{vlbi1})
\begin{eqnarray}
\label{vlbi8}
t_2-t_1&=&{\bf k}_1\cdot{\bf b}\left[1+{\bf v}_2\cdot{\bf
k}_1+O\left(\frac{v^2}{c^2}\right)+O\left(\frac{b}{R}\right)\right]+
\Delta(t_2,t_0)-\Delta(t_1,t_0)\;.
\end{eqnarray}
As a consequence of the previous expansions we also have the following
equalities
\begin{eqnarray}
\label{vlbi10a}
t^{\ast}_2-t^{\ast}_1&=&
\frac{({\bf b}\times{\bf k}_1)({\bf k}_1\times{\bf x}_{2})}{R}+
O\left(\frac{b\;v^2}{c^2}\right)+O\left(\frac{b^2}{R}\right)\;,
\end{eqnarray}
\begin{eqnarray}
\label{vlbi10}
{t^{\ast}_2}_a-{t^{\ast}_1}_a&=&
({\bf k}_1\cdot{\bf b})({\bf k}_1\cdot{\bf v}_a)-
(r_{2a}-r_{1a})({\bf k}_1\cdot{\bf v}_a)+
\frac{({\bf b}\times{\bf k}_1)({\bf k}_1\times{\bf r}_{2a})}{R}+
O\left(\frac{b\;v^2}{c^2}\right)+O\left(\frac{b^2}{R}\right)\;,
\end{eqnarray}
which evidently shows that, e.g. for the Jupiter and for the source of light at
infinity, the time 
difference ${t^{\ast}_2}_\odot-{t^{\ast}_1}_\odot$  
is of the order 
$(R_{\oplus}/c)(v_{J}/c)\simeq 75$ nanoseconds, that is,
rather small but still may be important in the analysis of observational errors.
For VLBI observations of the solar system objects the time difference  
${t^{\ast}_2}_\odot-{t^{\ast}_1}_\odot$ can approach the value 
$R_{\oplus}/c\simeq
30$ ms which can not be ignored at all.
The time difference 
$t^{\ast}_2-t^{\ast}_1$ can be considered for extra-solar objects as
negligibly small since it is of the order $(R_{\oplus}/c)$ by the annual 
parallax of
the source of light which makes it much less than $1$ picosecond. In
case of VLBI observations of the solar system objects the time
difference $t^{\ast}_2-t^{\ast}_1$ can not be ignored anymore but we do
not elaborate it here. 
Now we can simplify formula (\ref{vlbi3}).

First of all, taking into account the relationship (\ref{vlbi9}), we obtain 
\begin{eqnarray}
\label{vlbi11}
\ln\frac{r_{0a}-{\bf k}_1\cdot{\bf r}_{0a}}{r_{0a}-{\bf k}_2\cdot{\bf
r}_{0a}}&=&-\frac{({\bf b}\times{\bf k}_1)({\bf k}_1\times{\bf r}_{0a})}
{R(r_{0a}-{\bf k}_1\cdot{\bf r}_{0a})}\simeq O\left(\frac{b}{R}\right)\;,
\end{eqnarray}
which is of the order of the annual parallax of the source of light.
This term can be neglected in the delay formula (\ref{vlbi3}) since it gives
a contribution to the delay for extra-solar system objects much less 
than 1 picosecond \cite{117}. After noting that in the expression for 
the difference of the two integrals in (\ref{vlbi3}) one can equate ${\bf
k}_2={\bf k}_1$ and $t^\ast_2=t^\ast_1$, we state that the difference
reads 
\begin{eqnarray}
{\hspace{-1cm}}\int^{s_1}_{s_2}\ln\left[t^\ast_1+{\bf k}_1\cdot{\bf 
x}_a(\zeta)-\zeta
\right]{\bf k}_1\cdot\dot{\bf v}_a(\zeta)d\zeta&=&
{\bf k}_1\cdot\dot{\bf v}_a(s_1)\biggl\{(r_{2a}-{\bf k}_1\cdot
{\bf r}_{2a})\ln(r_{2a}
-{\bf k}_1\cdot{\bf r}_{2a})
\\\nonumber&&
{\hspace{-1cm}}-(r_{1a}-{\bf k}_1\cdot
{\bf r}_{1a})\ln(r_{1a}-{\bf k}_1\cdot{\bf r}_{1a})
+r_{1a}-r_{2a}-
{\bf k}_1\cdot{\bf r}_{1a}+{\bf k}_1\cdot{\bf r}_{2a}\biggr\}
\;,
\end{eqnarray}
and after multiplication by the factor $2Gm_a/c^3$ is much
less than 1 picoarcsecond. Hence, we neglect those two integrals from the
expression for the VLBI delay $\Delta(t_1,t_2)=\Delta(t_2,t_0)-\Delta(t_1,t_0)$.

Finally, taking into account that ${\bf k}_1=-{\bf K}$, up to the corrections
of order of the annual parallax, we get for the time delay
\begin{eqnarray}
\label{vlbi12}
\Delta(t_1,t_2)&=&2\sum_{a=1}^N m_a\left(1+{\bf K}\cdot{\bf v}_a\right)\;
\ln\frac{r_{1a}+{\bf K}\cdot{\bf r}_{1a}}
{r_{2a}+{\bf K}\cdot{\bf r}_{2a}}\;,
\end{eqnarray} 
where ${\bf v}_a={\bf v}_a(s_1)$, $r_{1a}=|{\bf r}_{1a}|$, 
$r_{2a}=|{\bf r}_{2a}|$, and
\begin{equation}
\label{defi}
{\bf r}_{1a}={\bf x}_{1}(t_1)-{\bf x}_{a}(s_1)\;,\quad\quad\quad\quad
{\bf r}_{2a}={\bf x}_{2}(t_2)-{\bf x}_{a}(s_2)\;.
\end{equation}
We emphasize that our formula (\ref{vlbi12}) includes the first correction for 
the velocity of
the bodies deflecting light rays. Moreover, there is a difference between
the definitions of the vectors ${\bf r}_{1a}$, ${\bf r}_{2a}$ in our model
(\ref{vlbi12}) and the consensus model (see \cite{115}, chapter 12, 
formula (1)). In our case the coordinates of stations ${\bf x}_1$, ${\bf x}_2$ 
are taken at the
instants $t_1$, $t_2$ respectively, and the coordinates of the light-deflecting 
bodies are
calculated at the retarded times $s_1$, $s_2$ defined in
(\ref{vlbi4}), (\ref{vlbi4a}) which is a direct consequence of our rigorous
approach of the integration of equations of light propagation. On the other 
hand,
in the consensus model coordinates of stations are taken also at the 
instants $t_1$, $t_2$ but coordinates of the $a$-th body are
calculated only at the time $t^{\ast}_{1a}$ defined in
(\ref{vlbi5b}). Strictly speaking, this prescription can be justified only for
VLBI observations of the distant, extra-solar objects and is only
marginally correct for VLBI observations of the solar sytem objects.
The prescription to obtain the position of the gravitating body at the
time of closest approach of the ray path to the body was based on an intuitive 
guess (see, for instance, \cite{10}). 
Such a guess gives a rather good approximation but can not be adopted as a
self-consistent theoretical recommendation in doing practically important 
numerical processing of VLBI observations and, especially, in the dedicated
experiments specifically designed to test gravitational deflection of light 
in the solar system \cite{118}. 

If we
denote by $\Delta t_{grav}$ the VLBI delay in the consensus model, 
as it is described in the 
{\it IERS Conventions} (see \cite{115}, formulas (1), (2) of chapter 12, 
or formula (5) from \cite{99}), and put the PPN parameter
$\gamma=1$, 
we get in the framework of General Relativity the following relationship
between the Lorentz-covariant expression for the time delay in our model and 
$\Delta t_{grav}$:
\begin{eqnarray}
\label{vlbi14}
\Delta(t_1,t_2)&=&\Delta t_{grav}
+2\sum_{a=1}^N \frac{G m_a}{c^4}\left({\bf K}\cdot{\bf v}_a\right)
\ln\frac{r_{1a}+{\bf K}\cdot{\bf r}_{1a}}
{r_{2a}+{\bf K}\cdot{\bf r}_{2a}}-2\sum_{a=1}^N \frac{G m_a}{c^4}\frac{({\bf 
v}_a
\times {\bf r}_{1a})({\bf b}\times{\bf r}_{1a})}{r^3_{1a}}+...\;,
\end{eqnarray}   
where  
ellipses denote residual terms, and 
we have restored the universal gravitational constant $G$ and speed of
light $c$ for convenience. One can see that the Shapiro time delay
in the consensus model was not properly defined although it had no 
consequences for practical observations in the recent past. Indeed, 
the third term in the 
right hand side of
(\ref{vlbi14}) is so small that can be neglected for any
observational configuration of the source of light and the deflecting 
body including the Earth. Expansion of the second term in the right hand
side of equation (\ref{vlbi14}) with respect to powers $b/d_a$, where
$d_a$
is the impact parameter of the light ray with respect to the $a$-th
light-deflecting body, gives
\begin{eqnarray}
\label{vlbi15}
2\frac{G m_a}{c^4}\left
({\bf K}\cdot{\bf v}_a\right)\ln\frac{r_{1a}+{\bf K}\cdot{\bf r}_{1a}}
{r_{2a}+{\bf K}\cdot{\bf r}_{2a}}&=&-\;2\;\frac{G m_a}{c^4}\left
({\bf K}\cdot{\bf v}_a\right)\frac{{\bf b}\cdot({\bf
n}_{1a}+{\bf K})}{r_{1a}+{\bf K}\cdot{\bf r}_{1a}}=
-\;4\;\frac{G m_a}{c^4}\left
({\bf K}\cdot{\bf v}_a\right)
\frac{{\bf b}\cdot({\bf n}_{1a}+{\bf K})}{d_a}\frac{r_{1a}}{d_a}\;,
\end{eqnarray}  
where the unit vector ${\bf n}_{1a}={\bf r}_{1a}/r_{1a}$.
For the light ray grazing, for example, the limb of the Sun the term under
consideration can reach a few picoseconds. The effect amounts 1
picosecond for radio source being at the angular distance 10 arcminutes from
the Jupiter if Jupiter is at the distance 5 AU from the Earth. This may will 
have real impact in
near future on the treatment of the gravitational deflection of light by
massive solar system plantes in the specialized high-precision VLBI experiments. 

We would like to note that 
relativistic perturbations of light propagation
caused by velocities of moving gravitating bodies were considered by Klioner
\cite{119} in order to find corresponding corrections to the consensus model of
VLBI data processing. That author approached the problem doing calculations on 
the base of the
post-Newtonian metric tensor. As we have shown in the present paper, such an 
approximation is not exact enough 
to take properly
into account all effects of retardation in the metric which contribute to
velocity dependent terms in the propagation of light. Nevertheless, at
least formally, the
result published in \cite{119}, equation (4.9), coincides with our
equation (\ref{vlbi12}) but the coordinates and velocities of
the light-ray-deflecting bodies are taken at the time of the closest approach
of photon to the $a$-th body. This
produces errors of the order comparable with the last term 
shown in the right hand side of
(\ref{vlbi14}) which are negligibly small. Hence, we conclude that the 
relativistic model of VLBI data
processing proposed in \cite{119} is although theoretically incomplete but 
practically good enough for
applications at the level of accuracy about one picosecond for 
astronomical objects with negligibly small parallaxes.

\subsubsection{Relativistic Space Astrometry}

Space astrometry is a new branch of fundamental astrometry. Ground-based
telescopes may reach the angular resolution not better than $0.01$
arcseconds. This limits our ability to create a fundamental inertial system
on the sky \cite{120} with the accuracy required for a much
better understanding the laws of translational and rotational motions of
celestial bodies both inside and outside of the solar system. The epoch of the
space astrometry began in 1989 when the HIPPARCOS satellite was successfully
launched by Ariane 4 of the European Space Agency on 8 August 1989. 
Despite the failure to
put the satellite on the intended geostationary orbit 
at 36,000 Km from Earth the astrometric
program has been completely fulfilled \cite{121}. 
As a result the new astrometric
catalogue of all stars up to 13-th stellar magnitude was obtained. It 
includes
about 120,000 stars and has a precision of around $0.002$ arcseconds. 
Unfortunately,
such high precision can not be retained longer than 10 years because of
errors in determination of proper motions of stars. For this reason the
second analogous mission having the same or better astrometric accuracy
should be launched in near future.

Rapid industrial development of space technologies allows us to hope that in
the next several years the precision of astrometric satellites will 
reach a few
microarcseconds or even better in the determination of positions, proper 
motions,
and parallaxes of celestial objects. All together, the photometric sensitivity
of measuring devices will be substantially improved. As an example, we refer
to a new space project of the European Space Agency named GAIA (Galactic
Astrometric Interferometer for Astrophysics). In the framework of this 
project \cite{2} positions, proper motion, and 
parallaxes of about 
$1000$ million stars up to $20$ stellar magnitude are to be measured with
accuracy better than $10$ microarcsecond. It means that practically 
almost all 
stars in our Galaxy will be observed and registered.

Such extremely difficult observations can not be processed adequately 
if 
numerous relativistic corrections are not taken into account in a proper
way \cite{99}. Indeed, the
relativistic deflection of light caused by the Sun is not less than $1$
milliarcsecond throughout all of the sky. Major planets produce a 
relativistic
deflection of light about 1 microarcsecond at the angular distances from $1$
to $90$ degrees outside the planet \cite{122}. It
is worth emphasizing that the relativistic deflection of light produced by the
Earth reaches a maximal value of about $550$ microarcseconds and should be
accounted for any position of a star with respect to the Earth. In addition,
the reduction of astrometric observations made on the moving platform will
require an extremely careful consideration of relativistic aberration
\cite{corr.7} and
classic parallax terms in order to reduce the measurements to the solar system
barycenter - the point to which the origin of the fundamental inertial
system is attached. Perhaps, it would be more properly to say that data
processing of observations from modern space astrometric satellites should be 
fully based on
general relativistic conceptions rather than on a classical approach in 
which the
relativistic corrections are considered as additive and are taken into 
account at the very last stage of the 
reduction of observations.

As far as we know, the first attempt to construct such a 
self-consistent theory of astrometric
observations was proposed by Brumberg \& Kopeikin \cite{123} 
and further explored
in \cite{116} and \cite{4}. The main idea of the 
formalism is to exploit to a full extent a
relativistic theory of reference frames in the solar system
developed in papers (\cite{124} - \cite{127}, 
\cite{74}, and references therein). An independent, 
but similar 
approach with more emphasis on mathematical details
was presented in papers \cite{75}, \cite{128}, \cite{129}.
One global and several local reference frames have been constructed by
solving in a specific way the Einstein equations for gravitational field.
The global frame is
the barycentric reference frame of the solar system with origin at the
barycenter. Among the local 
frames the
most important for us is the geocentric frame with origin at the
geocenter and the proper reference frame
of an observer (or the satellite in case of a space mission like HIPPARCOS or 
GAIA). All reference frames are harmonic \cite{130} and were constructed in such 
a way to
reduce to minimum all fictitious coordinate perturbations which may be 
caused by unsophisticated technique in  
using coordinate transformations from one frame to another. We
have discovered and outlined corresponding relativistic transformations between
the frames which generalize the well-known Lorentz transformation in the 
special
theory of relativity and minimize the magnitude of unphysical coordinate
dependent terms. Proceeding in this way we have achieved a
significant progress in describing relativistic aberration, classic parallax,
and proper motion corrections \cite{4}. However, the problem of propagation of 
light 
rays from
distant sources of light to an observer in the non-stationary gravitational 
field
of the solar system was not treated thoroughly enough. This section refines the 
problem and gives its final solution. 

The quantity which we are specifically interested in is the direction 
towards the source of
light (star, quasar) measured by a fictitious observer being at rest at 
the point with
the solar system barycentric
coordinates $(t,{\bf x})$. This direction is given, actually, by the 
equation (\ref{dop}) and can explicitly be written as follows
\vspace{0.3cm}       
\begin{eqnarray}
\label{direct}
s^i(\tau,\hat{\bm{\xi}})&=&K^i+
2\sum_{a=1}^N \frac{m_a}{\sqrt{1-v^2_a}}\;
\frac{\left(1-{\bf k}\cdot {\bf v}_a\right)^2}
{r_a-{\bf k}\cdot{\bf r}_a}\;\frac{P^i_{\;j}\;r_a^j}{r_a-
{\bf v}_a\cdot{\bf r}_a}-2\sum_{a=1}^N\frac{m_a}{\sqrt{1-v^2_a}}
\frac{2-{\bf k}\cdot {\bf v}_a}
{r_a-{\bf v}_a\cdot {\bf r}_a}\;P^i_{\;j}\;v^j_a\\\nonumber\\\mbox{}&&
-\frac{2}{R}\sum_{a=1}^N \frac{m_a}{\sqrt{1-v_a^2}}\frac{1-{\bf k}
\cdot{\bf v}_a}
{r_a-{\bf k}\cdot{\bf r}_a}\;P^i_{\;j} r_a^j-\frac{4}{R}
\sum_{a=1}^N \frac{m_a}{\sqrt{1-v_a^2}}
\ln\left(\frac{r_a-{\bf k}\cdot{\bf r}_a}{2R}\right)\;P^i_{\;j} v_a^j
+...\;,\nonumber
\vspace{0.3cm}
\end{eqnarray}
where positions and velocities of the solar system light-deflecting 
bodies are calculated at the retarded time
$s=t-|{\bf x}-{\bf x}_a|$,  
$R=|{\bf x}-{\bf x}_a|$ is the
distance from the source of light to observer, and ellipses denote 
residual terms depending on
accelerations of the bodies given by the retarded integrals
(\ref{53})-(\ref{54}). We have neglected all terms
depending on accelerations of the bodies because of their insignificant 
numerical
value. Further simplification of equation (\ref{direct}) is possible if we 
remember that the velocities of bodies, ${\bf v}_a$, comprising the solar system 
are small in comparison with the speed of light, and distances, $R$, to stars 
are very large compared to the size of the solar system. 
This makes it possible to omit all terms being quadratic with 
respect to ${\bf v}_a$ as well as 
fifth term in the right hand side 
of (\ref{direct}) being inversely proportional to $R$. It yields
\begin{eqnarray}
\label{sght} 
s^i(\tau,\hat{\bm{\xi}})&=&K^i+
2\sum_{a=1}^N \frac{G m_a}{c^2}\left(1-\frac{2}{c}{\bf k}\cdot{\bf v}_a
+\frac{1}{c}{\bf v}_a\cdot{\bf n}_a-\frac{r_a}{R}\right)
\frac{{\bf k}\times({\bf n}_a\times{\bf k})}{r_a-{\bf k}\cdot{\bf r}_a}-
4\sum_{a=1}^N \frac{G m_a}{c^3}\frac{{\bf k}\times({\bf v}_a\times{\bf 
k})}{r_a}\;,
\end{eqnarray}
where ${\bf n}_a={\bf r}_a/r_a$, the sign ``$\times$" denotes the usual
Euclidean vector product, and we restored the fundamental constants 
$G$ and $c$ for convenience. The equation (\ref{sght}) eliminates
incompleteness in
the derivation of the similar formula given by 
Klioner \cite{53} which 
was obtained using the post-Newtonian expression for the metric tensor and
under the assumption of rectilinear and uniform motion of the
light-deflecting bodies. As we have already noted many times in the 
present
paper, the post-Newtonian approximation for the 
metric tensor does not 
take into account all necessary effects of retardation \cite{131} 
which are essential in
the derivation of the equation (\ref{sght}). Klioner \& Kopeikin \cite{4} have 
simply 
copied the result
of \cite{53} due to the absence at that time of a better theoretical treatment 
of influence of body's velocities on the propagation of light rays. With the
mathematical technique invented in the present paper the equation
(\ref{sght}) gives the correct answer to this question and closes the 
problem.

The leading order term in (\ref{sght}) gives the
well-known expression for the angle of deflection of light rays in the
gravitational field of a static, spherically symmetric body. The velocity 
dependent
terms in (\ref{sght}) describe small corrections which may be important in data
analysis of future space missions. The very last term in the large round
brackets in (\ref{sght}) may slightly change magnitude of 
the angle of gravitational 
deflection for some nearby stars or objects within the solar system if
the impact parameter of the light ray is small and the deflection angle
is expected to be rather large.  Parallactic
corrections to the direction $s^i$ are extracted from the unit vector 
${\bf K}$ by its expansion
in powers of the ratio (the barycentric distance to observer)/(the
barycentric distance to a star). Account for 
aberrational corrections is made by means of relating the  
direction to the star, 
$s^i$, observed by a fixed fictitious observer, to the direction observed by a
moving real observer, with the help of the matrix of relativistic 
transformation
displayed in section VII of the paper [4]. It is worth
emphasizing that the correction for aberration must be done first before
account for parallax. Complete analysis of the relativistic algorithm of
processing observations of celestial objects made from a board of a space
observatory will be given elsewhere.

\subsection{Doppler Tracking of Interplanetary Spacecrafts}

\subsubsection{Approximation Scheme for Calculation of the Doppler shift}

The Doppler tracking of interplanetary spacecrafts \cite{132}, \cite{133}, 
\cite{102} is the only method
presently available to search for gravitational waves in the low
frequency regime ($10^{-5}$ - $1$ Hz). Several experiments have been
carried out so far, for instance, VOYAGER, PIONEER, ULYSSES, 
GALILEO and
MARS-OBSERVER. The space-probe CASSINI represents the next step in such
gravitational wave
Doppler experiments \cite{103}. 
Its primary target is to study the Saturn system.
However, the spacecraft carries on board much improved instrumentation
and will perform three long (40 days each) dedicated data acquisition
runs in 2002, 2003 and 2004 to search for gravitational waves with
expected sensitivity about twenty times better than that achieved so
far. The detection of gravitational waves requires the precise knowledge of the
Doppler frequency shift caused by the solar system's bodies lying near the line
of sight of observer to spacecraft (see Fig. \ref{covariant5}). 

Another important implementation of the Doppler tracking is the Global
Positioning System (GPS) which uses accurate, stable atomic clocks in
satellites and on Earth to provide world-wide position and time determination.
These clocks have relativistic frequency shifts which are so large that,
without accounting for numerous relativistic effects, the system would not
function (\cite{ashby}, and references therein). Quite recently, the Europeian 
Space Agency (ESA) has adopted a new
program aimed at achieving an even better precision in measuring time and frequency
in space-time observations.
The program is called the Atomic Clock Ensemble in Space (ACES) and will be
carried out on board of the International Space
Station (ISS). The principal idea is to use a cold atom clock in absence of 
gravity which will outperform the fountains clock on the ground with the
potential accuracy of $5\times 10^{-17}$ \cite{clock}. 

An adequate treatment of such gravitational wave and time-metrology 
high-precision experiments require advanced theoretical
development of the corresponding analytic algorithm which properly 
accounts for
all terms of order $10^{-16}$ and higher in the
classic Doppler and gravitational shifts between transmitted and received 
electromagentic frequencies caused by the relative motion of the
spacecraft with respect to observer and time-dependent gravitational
field of the solar system bodies. In this paragraph we discuss basic
principles of the Doppler tracking observations and give the most 
important relationships for calculation of the relevant effects. However,
the complete theory involves so many specific details that it would be
unreasonable to give all of them in the present paper.  Therefore, 
only basic elements of the Doppler tracking theory are given here 
and particular details will be published somewhere else. 

Let us assume (see Fig. \ref{covariant5}) 
that an electromagnetic signal is being transmitted from the point with 
barycentric 
coordinates ${\bf x}_0$ located on 
the Earth with
frequency $\nu_0$ at the barycentric time $t_0$. It travels to the 
interplanetary spacecraft, is
received on its board at the point with barycentric 
coordinates ${\bf x}_1$ with frequency $\nu_1$ at the barycentric time
$t_1$, and is transponded back
to the Earth (on exactly the same frequency $\nu_1$) where one observes 
this signal at the point with barycentric 
coordinates ${\bf x}_2$ with frequency $\nu_2$ at the barycentric time
$t_2$. It is
worthwhile to emphasize that because of the motion of the receiver with
respect to the transmitter during the light travel time of the signal 
the observed frequency $\nu_2$ is different from 
the emitted frequency $\nu_0$ even if the signal is transponded 
from the spacecraft being momentarily at rest with respect to the 
barycentric
coordinates of the solar system.

Proper time of the transmitter at the instant of signal's emission is 
denoted by ${\cal T}_0$
and at the instant of the signal's reception  by ${\cal T}_2$. Proper time of
the spacecraft's transponder is denoted as $T_1$. Barycentric time at 
the emission point
is $t_0$, at the point of reception, $t_2$, and at the specacraft's
position, $t_1$. We follow arguments similar to those
used in section VI.C. The spectral 
shift of electromagnetic frequency $\nu_0$ with
respect to $\nu_1$ is given by the equation
\begin{eqnarray}
\label{dt1}
1+z_1&=&\frac{\nu_0}{\nu_1}=\frac{dT_1}{dt_1}\frac{dt_1}{dt_0}
\frac{dt_0}{d{\cal T}_0}\;,
\end{eqnarray}
and the shift of the frequency $\nu_1$ with respect to $\nu_2$ is
described by the similar relationship
\begin{eqnarray}
\label{dt2}
1+z_1&=&\frac{\nu_1}{\nu_2}=\frac{dt_1}{dT_1}\frac{dt_2}{dt_1}\frac{d{\cal T}_2}
{dt_2}\;.
\end{eqnarray}
Here the time derivatives $dT_1/dt_1$ and $dt_1/dT_1$ are calculated at the 
spacecraft's
position, $dt_0/d{\cal T}_0$ is calculated at the point of emission, and 
$d{\cal T}_2/t_2d$ at the point of reception. Time derivatives $dt_1/dt_0$
and $dt_2/dt_1$ are obtained from the solution of equation of propagation of
electromagnetic signal in time-dependent gravitational field of the solar
system (\ref{qer}) establishing theoretical description of the 
transmitter-spacecraft (up-) and spacecraft-receiver (down-) radio links.

In practice, when Doppler tracking observations are made, 
the frequency $\nu_2$ of the
receiver is kept fixed. It relates to the fact that the 
frequency band of the receiver must be rather narrow to decrease the
level of stochastic noise fluctuations and to increase the sensitivity 
of the receiver to detect a very weak  
radio signal transponded to the Earth from the spacecraft. 
On the other hand, technical
limitations on the range of the transmitted frequency are not so restrictive
and it can be changed smoothly in a very broad band according
to a prescribed frequency modulation law. This law of modulation is 
chosen in such a way
to ensure the receiving of the transponded signal from the spececraft 
exactly at
the frequency $\nu_2$. It requires to know precisely the ephemerides
of transmitter, observer, and spacecraft as well as the law of propagation of
electromagnetic signal on its round-trip journey. Hence, one needs to
know the Doppler shift $\delta\nu/\nu_2$ where 
$\delta\nu=\nu_0-\nu_2$.
From equations (\ref{dt1}), (\ref{dt2}) we have
\begin{eqnarray}      
\label{dt3}
\frac{\delta\nu}{\nu_2}&=&\frac{\nu_0}{\nu_2}-1=
\frac{dt_0}{d{\cal T}_0}\frac{dt_1}{dt_0}\frac{dt_2}{dt_1}\frac{d{\cal T}_2}
{dt_2}-1\;.
\end{eqnarray}
As one can see from (\ref{dt3}) there is no need to know explicitly the
transformation between the proper time of the spacecarft, $T_1$, and the 
barycentric time of the solar system, $t_1$. This remark simplifies 
calculations.
 
Accounting for relationship (\ref{60})
and expression (\ref{lw}) for the metric tensor yields at the point of
emission
\begin{eqnarray}
\label{dt4}
\frac{dt_0}{d{\cal T}_0}&=&\left[\left(1-v_0^2\right)\left(1+2\sum^N_{a=1}
\frac{m_a\sqrt{1-v_{a0}^2}}{r_{0a}-{\bf v}_{a0}\cdot{\bf r}_{0a}}\right)-
4\sum^N_{a=1}
\frac{m_a}{\sqrt{1-v_{a0}^2}}\frac{(1-{\bf v}_0\cdot{\bf
v}_{a0})^2}{r_{0a}-{\bf v}_{a0}\cdot{\bf r}_{0a}}\right]^{-1/2}\;,
\end{eqnarray}
where ${\bf v}_0(t_0)$ is the barycentric velocity of emitter, ${\bf
v}_{a0}={\bf v}_a(s_0)$ is the barycentric velocity of the $a$-th gravitating 
body,
$r_{0a}=|{\bf r}_{0a}|$, ${\bf r}_{0a}={\bf x}_0(t_0)-{\bf x}_a(s_0)$, and
$s_0=t_0-r_{0a}$ is the retarded time corresponding to the time of 
emission, $t_0$, of radio signal. 

Similar arguments give
\begin{eqnarray}
\label{dt5}
\frac{d{\cal T}_2}{dt_2}&=&\left[\left(1-v_2^2\right)\left(1+2\sum^N_{a=1}
\frac{m_a\sqrt{1-v_{a2}^2}}{r_{2a}-{\bf v}_{a2}\cdot{\bf r}_{2a}}\right)-
4\sum^N_{a=1}
\frac{m_a}{\sqrt{1-v_{a2}^2}}\frac{(1-{\bf v}_2\cdot{\bf
v}_{a2})^2}{r_{2a}-{\bf v}_{a2}\cdot{\bf r}_{2a}}\right]^{1/2}\;,
\end{eqnarray}
where ${\bf v}_2(t_2)$ is the barycentric velocity of emitter, ${\bf
v}_{a2}={\bf v}_a(s_2)$ is the barycentric velocity of the $a$-th gravitating 
body,
$r_{2a}=|{\bf r}_{2a}|$, ${\bf r}_{2a}={\bf x}_2(t_2)-{\bf x}_a(s_2)$, and
$s_2=t_2-r_{2a}$ is retarded time corresponding to the time, $t_2$, 
of signal's reception. 

For up- and down- radio links the relationship (\ref{61}) 
yields respectively
\begin{eqnarray}
\label{dt6} 
\frac{dt_1}{dt_0}&=&\frac{1+{\bf K}_1\cdot {\bf v}_0+2\displaystyle{
\sum_{a=1}^N\;m_a\left[\frac{\partial s_1}{\partial t_0}
\frac{\partial}{\partial s_1}+\frac{\partial s_0}{\partial t_0}
\frac{\partial}{\partial s_0}+\frac{\partial t_1^{\ast}}{\partial t_0}
\frac{\partial}{\partial t_1^{\ast}}+\frac{\partial k_1^i}{\partial t_0}
\frac{\partial}{\partial k_1^i}\right]B_a(s_1,s_0,t_1^{\ast},{\bf k}_1)}}
{1+{\bf K}_1\cdot {\bf v}_1-2\displaystyle{
\sum_{a=1}^N\;m_a\left[\frac{\partial s_1}{\partial t_1}
\frac{\partial}{\partial s_1}+
\frac{\partial s_0}{\partial t_1}\frac{\partial}{\partial s_0}+
\frac{\partial t_1^{\ast}}{\partial t_1}
\frac{\partial}{\partial t_1^{\ast}}+\frac{\partial k_1^i}{\partial t_1}
\frac{\partial}{\partial k_1^i}\right]B_a(s_1,s_0,t_1^{\ast},{\bf k}_1)}}\;,
\end{eqnarray}
and
\begin{eqnarray}
\label{dt7} 
\frac{dt_2}{dt_1}&=&\frac{1+{\bf K}_2\cdot {\bf v}_1+2\displaystyle{
\sum_{a=1}^N\;m_a\left[\frac{\partial s_2}{\partial t_1}
\frac{\partial}{\partial s_2}+\frac{\partial s_1}{\partial t_1}
\frac{\partial}{\partial s_1}+\frac{\partial t_2^{\ast}}{\partial t_1}
\frac{\partial}{\partial t_2^{\ast}}+\frac{\partial k_2^i}{\partial t_1}
\frac{\partial}{\partial k_2^i}\right]B_a(s_2,s_1,t_2^{\ast},{\bf k}_2)}}
{1+{\bf K}_2\cdot {\bf v}_2-2\displaystyle{
\sum_{a=1}^N\;m_a\left[\frac{\partial s_2}{\partial t_2}
\frac{\partial}{\partial s_2}+
\frac{\partial s_1}{\partial t_2}\frac{\partial}{\partial s_1}+
\frac{\partial t_2^{\ast}}{\partial t_2}
\frac{\partial}{\partial t_2^{\ast}}+\frac{\partial k_2^i}{\partial t_2}
\frac{\partial}{\partial k_2^i}\right]B_a(s_2,s_1,t_2^{\ast},{\bf k}_2)}}\;.
\end{eqnarray}
Here the retarded time $s_1$ comes out from the relation 
$s_1=t_1-|{\bf x}_1-{\bf x}_a(s_1)|$, and
\begin{equation}
\label{dt8}
{\bf k}_1=-{\bf K}_1=\frac{{\bf x}_1(t_1)-{\bf x}_0(t_0)}
{|{\bf x}_1(t_1)-{\bf x}_0(t_0)|}\;,\quad\quad\quad\quad
{\bf k}_2=-{\bf K}_2=\frac{{\bf x}_2(t_2)-{\bf x}_1(t_1)}
{|{\bf x}_2(t_2)-{\bf x}_1(t_1)|}\;,
\end{equation}
are the unit vectors which define direction of propagation of 
transmitted and transponded radio signals respectively, and
\begin{equation}
\label{dt9}
t_1^{\ast}=t_0-{\bf k}_1\cdot{\bf x}_0\;,\quad\quad\quad\quad
t_2^{\ast}=t_1-{\bf k}_2\cdot{\bf x}_1\;.
\end{equation}

\subsubsection{Auxilary Partial Derivatives}

The relationships (\ref{add1})-(\ref{add4}) allow to write down corresponding
expressions for the retarded times $s_0$, $s_1$, and $s_2$. One has to
carefully dinstinguish between derivatives for the up- and down-radio 
links. For the transmitter-spacecraft up-radio link we have
\begin{eqnarray}
\label{dt10}
\frac{\partial s_1}{\partial t_1}&=&\frac{r_{1a}-{\bf k}_1\cdot{\bf
r}_{1a}}
{r_{1a}-{\bf v}_{a1}\cdot{\bf r}_{1a}}-\frac{({\bf k}_1\times{\bf v}_1)
\cdot({\bf k}_1\times{\bf
r}_{1a})}{r_{1a}-{\bf v}_{a1}\cdot{\bf r}_{1a}}\;,\\\nonumber\\
\label{dt11}
\frac{\partial s_1}{\partial t_0}&=&\frac{(1-{\bf k}_1\cdot{\bf v}_0)
({\bf k}_1\cdot{\bf
r}_{1a})}{r_{1a}-{\bf v}_{a1}\cdot{\bf r}_{1a}}\;,\\\nonumber\\
\label{dt12}
\frac{\partial s_0}{\partial t_0}&=&\frac{r_{0a}-{\bf v}_0\cdot{\bf r}_{0a}}
{r_{0a}-{\bf v}_{a0}\cdot{\bf r}_{0a}}\;,\\\nonumber\\
\label{dt13}
\frac{\partial s_0}{\partial t_1}&=&0\;.
\end{eqnarray}
These formulas must be used in equation (\ref{dt6}).
For the spacecraft-receiver down-radio link we obtain
\begin{eqnarray}
\label{dt14}
\frac{\partial s_2}{\partial t_2}&=&\frac{r_{2a}-{\bf k}_2\cdot{\bf
r}_{2a}}
{r_{2a}-{\bf v}_{a2}\cdot{\bf r}_{2a}}-\frac{({\bf k}_2\times{\bf v}_2)
\cdot({\bf k}_2\times{\bf
r}_{2a})}{r_{2a}-{\bf v}_{a2}\cdot{\bf r}_{2a}}\;,\\\nonumber\\
\label{dt15}
\frac{\partial s_2}{\partial t_0}&=&\frac{(1-{\bf k}_2\cdot{\bf v}_1)
({\bf k}_2\cdot{\bf
r}_{2a})}{r_{2a}-{\bf v}_{a2}\cdot{\bf r}_{2a}}\;,\\\nonumber\\
\label{dt16}
\frac{\partial s_1}{\partial t_1}&=&\frac{r_{1a}-{\bf v}_1\cdot{\bf r}_{1a}}
{r_{1a}-{\bf v}_{a1}\cdot{\bf r}_{1a}}\;,\\\nonumber\\
\label{dt17}
\frac{\partial s_1}{\partial t_2}&=&0\;.
\end{eqnarray}
These formulas must be used in equation (\ref{dt7}). We point out that
the meaning of the time derivative (\ref{dt10}) is completely different 
from that of the time derivative (\ref{dt16}) although they are 
calculated at one
and the same point of transponding of the radio signal. At the first
sight it may look surprising. However, if one remembers that the
derivative (\ref{dt10}) is calculated along the transmitter-spacecraft 
light path and that (\ref{dt16}) along the spacecraft-receiver 
light path, which have opposite directions and 
different parameterizations, the difference becomes
evident.

The other set of time derivatives required in subsequent calculations
reads as follows,    
\begin{equation}
\label{dt18}
\frac{\partial k_1^i}{\partial t_1}=\frac{({\bf k}_1\times({\bf v}_1
\times{\bf k}_1))^i}{R_{01}}\;,\quad\quad\quad\quad
\frac{\partial k_1^i}{\partial t_0}=-\frac{({\bf k}_1\times({\bf v}_0
\times{\bf k}_1))^i}{R_{01}}\;,
\end{equation}
\begin{equation}
\label{dt19}
\frac{\partial k_2^i}{\partial t_2}=\frac{({\bf k}_2\times({\bf v}_2
\times{\bf k}_2))^i}{R_{21}}\;,\quad\quad\quad\quad
\frac{\partial k_2^i}{\partial t_1}=-\frac{({\bf k}_2\times({\bf v}_1
\times{\bf k}_2))^i}{R_{21}}\;,
\end{equation}
\begin{equation}
\label{dt20}
\frac{\partial t_1^\ast}{\partial t_0}=1-{\bf k}_1\cdot{\bf
v}_0+\frac{{\bf v}_0\cdot{\bm{\xi}}_1}{R_{01}}\;,
\quad\quad\quad\quad\frac{\partial t_1^\ast}{\partial t_1}=-
\frac{{\bf v}_1\cdot{\bm{\xi}}_1}{R_{01}}\;,
\end{equation}
\begin{equation}
\label{dt21}
\frac{\partial t_2^\ast}{\partial t_1}=1-{\bf k}_2\cdot{\bf
v}_1+\frac{{\bf v}_1\cdot{\bm{\xi}}_2}{R_{21}}\;,
\quad\quad\quad\quad\frac{\partial t_2^\ast}{\partial t_2}=-
\frac{{\bf v}_2\cdot{\bm{\xi}}_2}{R_{21}}\;,
\end{equation}
where $R_{01}=|{\bf x}_0-{\bf x}_1|$ is the radial distance between
emitter on the Earth and spacecraft, $R_{21}=|{\bf x}_2-{\bf x}_1|$ is 
the radial distance between receiver on the Earth and spacecraft, and for the
impact parameters hold ${\bm{\xi}}_1={\bf k}_1\times({\bf x}_1\times{\bf
k}_1)$, and ${\bm{\xi}}_2={\bf k}_2\times({\bf x}_1\times{\bf k}_2)$.

Partial derivatives of functions $B_a(s_1,s_0,t_1^\ast,k_1^i)$ and 
$B_a(s_2,s_1,t_2^\ast,k_2^i)$ can be found by making use of
relationships (\ref{62})-(\ref{pe}). This yields
\begin{eqnarray}
\label{dt22}
\frac{\partial B_a(s_1,s_0,t_1^\ast,k_1^i)}{\partial s_1}&=&
\frac{1}{\sqrt{1-v^2_{a1}}}\;
\frac{(1-{\bf k}_1\cdot {\bf v}_{a1})^2}{r_{1a}-{\bf k}_1\cdot {\bf
r}_{1a}}\;,
\\\nonumber\\
\label{dt23}
\frac{\partial B_a(s_1,s_0,t_1^\ast,k_1^i)}{\partial s_0}&=&
-\frac{1}{\sqrt{1-v^2_{a0}}}\;
\frac{(1-{\bf k}_1\cdot {\bf v}_{a0})^2}{r_{0a}-{\bf k}_1\cdot{\bf r}_{0a}}\;,
\\\nonumber\\
\label{dt24}
\frac{\partial B_a(s_1,s_0,t_1^\ast,k_1^i)}{\partial t_1^\ast}&=&
-\frac{1}{\sqrt{1-v^2_{a1}}}
\frac{1-{\bf k}_1\cdot {\bf v}_{a1}}{r_{1a}-{\bf k}_1\cdot {\bf r}_{1a}}+
\frac{1}{\sqrt{1-v^2_{a0}}}
\frac{1-{\bf k}_1\cdot {\bf v}_{a0}}{r_{0a}-{\bf k}_1\cdot {\bf
r}_{0a}}\;,\\\nonumber\\
\label{dt25}
\frac{\partial B_a(s_1,s_0,t_1^\ast,k_1^i)}{\partial k_1^i}&=&-
\frac{1-{\bf k}_1\cdot {\bf v}_{a1}}{\sqrt{1-v^2_{a1}}}
\frac{x_a^j(s_1)}{r_{1a}-{\bf k}_1\cdot {\bf r}_{1a}}+
\frac{1-{\bf k}_1\cdot {\bf v}_{a0}}{\sqrt{1-v^2_{a0}}}
\frac{x_a^j(s_0)}{r_{0a}-{\bf k}_1\cdot {\bf r}_{0a}}\\\nonumber
\\\nonumber&&
+\frac{2v_{a1}^j}{\sqrt{1-v_{a1}^2}}\ln(r_{1a}-{\bf k}_1\cdot {\bf r}_{1a})-
\frac{2v_{a0}^j}{\sqrt{1-v_{a0}^2}}\ln(r_{0a}-{\bf k}_1\cdot {\bf
r}_{0a})\;,
\end{eqnarray}
and
\begin{eqnarray}
\label{dt26}
\frac{\partial B_a(s_2,s_1,t_2^\ast,k_2^i)}{\partial s_2}&=&
\frac{1}{\sqrt{1-v^2_{a2}}}\;
\frac{(1-{\bf k}_2\cdot {\bf v}_{a2})^2}{r_{2a}-{\bf k}_2\cdot {\bf
r}_{2a}}\;,
\\\nonumber\\
\label{dt27}
\frac{\partial B_a(s_2,s_1,t_2^\ast,k_2^i)}{\partial s_1}&=&
-\frac{1}{\sqrt{1-v^2_{a1}}}\;
\frac{(1-{\bf k}_2\cdot {\bf v}_{a1})^2}{r_{1a}-{\bf k}_2\cdot{\bf r}_{1a}}\;,
\\\nonumber\\
\label{dt28}
\frac{\partial B_a(s_2,s_1,t_2^\ast,k_2^i)}{\partial t_2^\ast}&=&
-\frac{1}{\sqrt{1-v^2_{a2}}}
\frac{1-{\bf k}_2\cdot {\bf v}_{a2}}{r_{2a}-{\bf k}_2\cdot {\bf r}_{2a}}+
\frac{1}{\sqrt{1-v^2_{a1}}}
\frac{1-{\bf k}_2\cdot {\bf v}_{a1}}{r_{1a}-{\bf k}_2\cdot {\bf
r}_{1a}}\;,\\\nonumber\\
\label{dt29}
\frac{\partial B_a(s_2,s_1,t_2^\ast,k_2^i)}{\partial k_2^i}&=&-
\frac{1-{\bf k}_2\cdot {\bf v}_{a2}}{\sqrt{1-v^2_{a2}}}
\frac{x_a^j(s_2)}{r_{2a}-{\bf k}_2\cdot {\bf r}_{2a}}+
\frac{1-{\bf k}_2\cdot {\bf v}_{a1}}{\sqrt{1-v^2_{a1}}}
\frac{x_a^j(s_1)}{r_{1a}-{\bf k}_2\cdot {\bf r}_{1a}}\\\nonumber
\\\nonumber&&
+\frac{2v_{a2}^j}{\sqrt{1-v_{a2}^2}}\ln(r_{2a}-{\bf k}_2\cdot {\bf r}_{2a})-
\frac{2v_{a1}^j}{\sqrt{1-v_{a1}^2}}\ln(r_{1a}-{\bf k}_2\cdot {\bf
r}_{1a})\;.
\end{eqnarray}
We have neglected in formulas (\ref{dt24}),(\ref{dt25}) and (\ref{dt28}),
(\ref{dt29}) all terms depending on accelerations of the solar system
bodies.

\subsubsection{Relativistic Effect for Doppler Measurement near Solar and
Planetary Conjunctions}

The relationships (\ref{dt3}) - (\ref{dt29}) constitute the basic elements 
of the post-Minkowskian (Lorentz-covariant) Doppler 
tracking theory. They are sufficient to calculate the Doppler response for
any conceivable relative configuartion of transmitter, spacecraft, and
the solar system bodies. We shall consider in this section only 
the case when the spacecraft is
beyond a massive solar system body like Sun, Jupiter or Saturn when the 
impact parameters of up- and down-radio links are small compared with
distances from the body to transmitter, receiver, and spacecraft. 
We shall also rectrict ourselves
to the consideration of gravitational shift of frequency only. Actually, this
case is similar to gravitational lens. Thus, we neglect all terms of
order $m_a/r_{0a}$, $m_a/r_{1a}$, $m_a/r_{2a}$, $m_a/R_{01}$, 
$m_a/R_{21}$ as well as terms being quadratic with respect to the
velocity $v_a$. It is worthwhile to point out that the round trip 
travel time of the transmitted radio signal is much shorter than orbital
period of any of the solar system body. For this reason, all functions with 
the retarded time argument
entering the equations can be expanded around the time of transmission of
the signal which is precisely determined by atomic clocks. Taking into
account these remarks and making use of relationship (\ref{rewr1}) we obtain
\begin{equation}
\label{dopshift}
\left(\frac{\delta\nu}{\nu_2}\right)_{gr}^{obs}=\frac{2}{c}({\bf v}_a-
\frac{r_1}{R}{\bf v}_0-\frac{r_0}{R}{\bf
v}_1)\cdot{\bm{\alpha}}({\bm{\xi}}_a)\;,\quad\quad\quad\quad\quad
\alpha^i({\bm{\xi}}_a)=\frac{4G m_a}{c^2 d_a^2}\;\xi_a^i\;,
\end{equation}
where ${\bf v}_0$ is velocity of transmitter, ${\bf v}_1$ is velocity of 
spacecraft, ${\bf v}_a$ is velocity of
the $a$-th gravitating body deflecting trajectory of 
the emitted radio signal at the
angle $\alpha^i$, $d_a=|{\bm{\xi}}_a|$ is the length of the impact
parameter of the light ray with respect to the $a$-th body, $r_1$ is the
distance between the transmitter and the light-deflecting body, $r_0$ is the
distance between the spacecraft and light-deflecting body, and $R=|{\bf
x}_0-{\bf x}-1|\simeq r_0+r_1$. Formula (\ref{dopshift}) for Doppler shift by
gravitational lensing depends on velocities of transmitter, spacecraft, and the
ligh-deflecting body and generalizes that obtained independently by Bertotti \&
Giampieri \cite{corr.4} who considered only static gravitational lens. In case,
of Doppler tracking observations of spacecraft in the field of Sun the
difference between the two formulas is negligible, but the motion of the lensing
body may be important in the case of Doppler observations of spacecrafts in the
field of giant planets like Jupiter or Saturn. 

Approximate 
value of the Doppler shift is determined by the
expression $\delta\nu/\nu_2=2\alpha(v_\oplus/c)\cos\varphi$, where 
$\alpha$ is the
deflection angle of the light ray, $v_\oplus$ is velocity of the Earth, and
$\varphi$ is the angle between $v_\oplus$ and the impact parameter.
For the Sun the deflection angle over the whole sky is not
less than 1 milliarcsecond or $\simeq 4.85\cdot 10^{-9}$ radians. 
The relative velocity of
the Earth with respect to speed of light is about $10^{-4}$. These simple
estimates applied to the Doppler shift's formula elucidate that the
gravitational shift of frequency in Doppler tracking of interplanetary
spacecraft caused by the Sun is not less than $\simeq 4.85\cdot 10^{-13}$ 
for any 
location of the spacecraft in the sky. If the path of the radio link
grazes the Sun's surface the Doppler shift will be about $8.47\cdot
10^{-10}$ - a quantity which can be measured rather easily. The same
kind of estimates gives for radio signals grazing the Jupiter and Saturn
the Doppler shifts about $
7.76\cdot 10^{-12}$ and $2.91\cdot 10^{-12}$, respectively, which
can be also measured in practice. 
    
Our formalism for derivation of corresponding relationships for the
description of high-precise Doppler tracking of interplanetary
spacecrafts can be compared with approaches based on the post-Newtonian
approximation scheme (see, e.g., \cite{corr.4}, \cite{116}). The advantage of 
the post-Minkowskian approach used
in this paper is that it automatically accounts for all effects
related to velocities of gravitating bodies through the expressions of
the {\it Li\'enard-Wiechert} potentials. The post-Newtonian scheme makes
calculations much longer and not so evident. 

\subsubsection{Comparision of Two Mathematical Techniques for Calculation of
the Doppler effect} 

It is worthy from the methodological point of view to compare calculation of
the {\it Doppler effect in terms of frequency}, used throughout the present 
paper, with that {\it in terms of energy} 
(see \cite{corr.2} for definition) used, e.g. by Bertotti \& Giampieri
\cite{corr.4}. Let us introduce definitions of the 4-velocity of observer 
$u^\alpha=u^0(1,v^i)$, the 4-velocity of source of light
$u_0^\alpha=u_0^0(1,v_0^i)$, the 4-momenta of photon at the point of emission
${\cal K}_0^\alpha={\cal K}_0^0(1,\dot{x}^i(t_0))$ and the point of observation
${\cal K}^\alpha={\cal K}^0(1,\dot{x}^i(t))$, where $u^0=dt/d{\cal T}$, $u_0^0=dt_0/d{\cal
T}_0$, ${\cal K}_0^0=dt_0/d\lambda_0$ and 
${\cal K}^0=dt/d\lambda$ with $\lambda$ and $\lambda_0$ being values of the
affine parameter along the light geodesic at the points of emission and
observation. Then, using the definition of the {\it Dopppler
effect in terms of energy} (\ref{enrg})  
it is not difficult to show that equation (\ref{enrg}) can be recast 
into the form 
\begin{eqnarray}
\label{asd}
\frac{\nu}{\nu_0}&=&\frac{u^0 {\cal K}^0\biggl\{g_{00}(t,{\bf x})+g_{0i}(t,{\bf x})
\left[\dot{x}^i(t)+v^i\right]+g_{ij}(t,{\bf x})\dot{x}^i(t)v^i\biggr\}}
{u_0^0 {\cal K}_0^0\biggl\{g_{00}(t_0,{\bf x}_0)+g_{0i}(t_0,{\bf
x}_0)\left[\dot{x}^i(t_0)+v_0^i
\right]+g_{ij}(t_0,{\bf x}_0)\dot{x}^i(t_0)v_0^i\biggr\}}\;.   
\end{eqnarray}
Calculation of time component ${\cal K}^0$ of the 4-momentum of photon in (\ref{asd}) 
can be done if
one knows the relationship of the affine parameter $\lambda$ along the light
geodesic and coordinate time $t$. This is found by solution of the time
component of the equation for the light geodesic ($G=c=1$) 
\begin{eqnarray}
\label{geod}
\frac{d^2 t}{d\lambda^2}&=&-\left(\Gamma^0_{00}+2\Gamma^0_{0i}\dot{x}^i+ 
\Gamma^0_{ij}\dot{x}^i\dot{x}^j\right)\left(\frac{d t}{d\lambda}\right)^2\;.
\end{eqnarray}
Using decomposition (\ref{2}) of the metric tensor and parametrization 
(\ref{11}) along the unperturbed light ray, equation (\ref{geod}) may be
written
\begin{eqnarray}
\label{geod1}
\frac{d^2 t}{d\lambda^2}&=&-\left[\frac{1}{2}k^\alpha k^\beta {\partial}_t
h_{\alpha\beta}-k^\alpha\hat{\partial}_\tau 
h_{0\alpha}\right]\left(\frac{d t}{d\lambda}\right)^2\;,
\end{eqnarray}
where the constant vector $k^\alpha=(1,k^i)=(1,{\bf k})$, and the substitution for the unperturbed 
trajectory of light ray in $h_{\alpha\beta}$ is done
after taking a partial derivative with respect to coordinate time $t$. Solution
of equation (\ref{geod1} can be found by iterations using expansion
\begin{eqnarray}
\label{aff}
\lambda&=&E^{-1}\left[t+{\cal F}(t)\right]\;,
\end{eqnarray}
where $E$ is the constant photon's energy at past null infinity measured by
a fictitious observer being at rest, and 
the function ${\cal F}(t)$ is of the order $O(h_{\alpha\beta})$. It is
obtained by solution of the equation
\begin{eqnarray}
\label{iter}
\frac{d^2{\cal F}}{d\tau^2}=\frac{1}{2}k^\alpha k^\beta {\partial}_t
h_{\alpha\beta}-k^\alpha\hat{\partial}_\tau h_{0\alpha}\;.  
\end{eqnarray}
Solving the differential equation (\ref{iter}) one finds 
\begin{equation}
\label{ko}
{\cal K}^0(\tau)=E^{-1}\left[1-\dot{\cal F}(\tau)\right]\;,\quad\quad\quad\quad
{\cal K}_0^0\equiv {\cal K}^0(\tau_0)=E^{-1}\left[1-\dot{\cal F}(\tau_0)\right]\;,
\end{equation}
and
\begin{eqnarray}
\label{kob}
\dot{\cal F}(\tau)&=&
\frac{1}{2}k^\alpha k^\beta \displaystyle{\int_{-\infty}^\tau}\left[
\frac{\partial h_{\alpha\beta}(t, {\bf x})}{\partial t}
\right]_{t=\sigma+t^\ast;\;{\bf x}={\bf k}\sigma+\hat{\bs{\xi}}}\;  
d\sigma-k^\alpha h_{0\alpha}(\tau)\;,
\end{eqnarray}
\begin{eqnarray}
\label{koba}
\dot{\cal F}(\tau_0)&=&
\frac{1}{2}k^\alpha k^\beta \displaystyle{\int_{-\infty}^{\tau_0}}\left[
\frac{\partial h_{\alpha\beta}(t, {\bf x})}{\partial t}
\right]_{t=\sigma+t^\ast;\;{\bf x}={\bf k}\sigma+\hat{\bs{\xi}}}\; 
d\sigma-k^\alpha h_{0\alpha}(\tau_0)\;.
\end{eqnarray}
After examination of structure of integrands in the integrals of the
expressions (\ref{kob}), (\ref{koba}) one notes that 
\begin{eqnarray}
\label{note}
\left[\frac{\partial h_{\alpha\beta}(t, {\bf x})}{\partial t}
\right]_{t=\sigma+t^\ast;\;{\bf x}={\bf k}\sigma+\hat{\bs{\xi}}}&=&
\frac{\partial h_{\alpha\beta}(\sigma+t^\ast, {\bf k}\sigma+\hat{\bm{\xi}})}
{\partial t^\ast}\;.
\end{eqnarray}
A remarkable property of equality (\ref{note}) is that the parameter $t^\ast$ is
independent from the argument $\sigma$ of the integrand in (\ref{kob}), (\ref{koba})
and, for this reason, the derivative with respect to $t^\ast$ can be taken out
of the sign of the integrals. It allows to transform, e.g., the integral (\ref{kob})
into the form
\begin{eqnarray}
\label{qaw}
\displaystyle{\int_{-\infty}^{\tau}}\left[
\frac{\partial h_{\alpha\beta}(t, {\bf x})}{\partial t}
\right]_{t=\sigma+t^\ast;\;{\bf x}={\bf k}\sigma+\hat{\bs{\xi}}}\;d\sigma&=&
\frac{\partial}{\partial t^\ast}\displaystyle{\int_{-\infty}^{\tau}}
h_{\alpha\beta}(\sigma+t^\ast, {\bf
k}\sigma+\hat{\bm{\xi}})d\sigma\;.
\end{eqnarray}
Using solution (\ref{lw}) for $h_{\alpha\beta}$ and relationship 
(\ref{newvar}) relating total differentials of coordinate time $\sigma$ and
retarded time $\zeta$ one obtains
\begin{eqnarray}
\label{lim}
\frac{\partial}{\partial t^\ast}\displaystyle{\int_{-\infty}^{\tau}}
h_{\alpha\beta}(\sigma+t^\ast, {\bf
k}\sigma+\hat{\bm{\xi}})d\sigma&=&4\sum_{a=1}^N\left[
\frac{\partial}{\partial t^\ast}\displaystyle{\int_{-\infty}^{s(\tau,t^\ast)}}
\frac{\hat{T}_{\alpha\beta}^a(\zeta)-
\frac{1}{2}\eta_{\alpha\beta}\hat{T}^\lambda_{a\lambda}(\zeta)}{t^\ast+{\bf
k}\cdot{\bf x}_a(\zeta)-\zeta}d\zeta\right]\;,
\end{eqnarray}
where the upper limit $s(\tau, t^\ast)$ of integral in the right hand side
is calculated by means of solution of equation, $s+|{\bf
k}\sigma+\hat{\bm{\xi}}-{\bf x}_a(s)|=\tau+t^\ast$, and 
depends on time $\tau$ and instant of the closest approach $t^\ast$ considered
as a parameter \cite{133a}. For the upper limit depends on $t^\ast$ the
derivative $\partial/\partial t^\ast$ of the integral in square brackets 
is taken both from the
integrand of the integral and its upper limit. It is possible to eliminate
dependence of the upper limit of the integral on the parameter $t^\ast$. It
will be achieved if one takes time $t$ as independent variable instead of $\tau$
and finds the upper limit of the integral from the equation (\ref{ret}) as we have
done previously while calculating the {\it Doppler effect in terms of frequency.}
Such a procedure gives us   
\begin{eqnarray}
\label{limx}
4\sum_{a=1}^N\left[\frac{\partial}{\partial t^\ast}
\displaystyle{\int_{-\infty}^{s(\tau,t^\ast)}}
\frac{\hat{T}_{\alpha\beta}^a(\zeta)-
\frac{1}{2}\eta_{\alpha\beta}\hat{T}^\lambda_{a\lambda}(\zeta)}{t^\ast+{\bf
k}\cdot{\bf x}_a(\zeta)-\zeta}d\zeta\right]&=&4\sum_{a=1}^N\left[
\frac{\partial}{\partial t^\ast}\displaystyle{\int_{-\infty}^{s(t)}}
\frac{\hat{T}_{\alpha\beta}^a(\zeta)-
\frac{1}{2}\eta_{\alpha\beta}\hat{T}^\lambda_{a\lambda}(\zeta)}{t^\ast+{\bf
k}\cdot{\bf x}_a(\zeta)-\zeta}d\zeta\right]\;\\\nonumber\\\nonumber&+&
4\;\sum_{a=1}^N\frac{\hat{T}_{\alpha\beta}^a(s)-
\frac{1}{2}\eta_{\alpha\beta}\hat{T}^\lambda_{a\lambda}(s)}{r_a(s)-{\bf
k}\cdot{\bf r}_a(s)}\frac{r_a}{r_a(s)-{\bf v}_a(s)\cdot{\bf r}_a(s)}\;,
\end{eqnarray}
where the second term in the right hand side is a 
partial derivative of the upper limit of integral in (\ref{lim}) with respect
to $t^\ast$ \cite{133c} and the upper limit $s(t)$ 
of the last integral is treated as independent from $t^\ast$ \cite{133b}. 
Finally, one has
\begin{eqnarray}
\label{fina}
\frac{1}{2}k^\alpha k^\beta\displaystyle{\int_{-\infty}^{\tau}}\left[
\frac{\partial h_{\alpha\beta}(t, {\bf x})}{\partial t}
\right]_{t=\sigma+t^\ast;\;{\bf x}={\bf k}\sigma+\hat{\bs{\xi}}}\;d\sigma
&=&\frac{1}{2}k^\alpha k^\beta
h_{\alpha\beta}(\tau)\\\nonumber\\\nonumber&
-&2\sum_{a=1}^N \left[m_a C_a(s)-\frac{m_a}{\sqrt{1-v^2_a}}
\frac{(1-{\bf k}\cdot{\bf v}_a)^2}{r_a-{\bf k}\cdot{\bf r}_a}
\frac{{\bf k}\cdot{\bf r}_a}{r_a-{\bf v}_a\cdot{\bf r}_a}
\right]\;,
\end{eqnarray}
where the function $C_a(s)$ is displayed in (\ref{50}).
Similar arguments give
\begin{eqnarray}
\label{qaws}
\frac{1}{2}k^\alpha k^\beta\displaystyle{\int_{-\infty}^{\tau_0}}\left[
\frac{\partial h_{\alpha\beta}(t, {\bf x})}{\partial t}
\right]_{t=\sigma+t^\ast;\;{\bf x}={\bf k}\sigma+\hat{\bs{\xi}}}\;d\sigma
&=&\frac{1}{2}k^\alpha k^\beta
h_{\alpha\beta}(\tau_0)\\\nonumber\\\nonumber&
-&2\sum_{a=1}^N \left[m_a C_a(s_0)-\frac{m_a}{\sqrt{1-v^2_{a0}}}
\frac{(1-{\bf k}\cdot{\bf v}_{a0})^2}{r_{0a}-{\bf k}\cdot{\bf r}_{0a}}
\frac{{\bf k}\cdot{\bf r}_{0a}}{r_{0a}-{\bf v}_{a0}\cdot{\bf r}_{0a}}
\right]\;.
\end{eqnarray}
Going back to the formula (\ref{asd}) of the {\it Doppler effect in terms of 
energy} one can see that it can be factorized in three terms
\begin{eqnarray}
\label{asd1}
\frac{\nu}{\nu_0}&=&{\cal S}_1\cdot{\cal S}_2\cdot{\cal S}_3\;,
\end{eqnarray}
where
\begin{eqnarray}
\label{fac2}
{\cal S}_1&\equiv&\frac{u^0}{u^0_0}=\frac{
1-v_0^2-h_{00}(t,{\bf x}_0)-2h_{0i}(t_0,{\bf
x}_0)v_0^i-h_{ij}(t_0,{\bf x}_0)v_0^i v_0^j}
{1-v^2-h_{00}(t,{\bf x})-2h_{0i}(t,{\bf
x})v^i-h_{ij}(t,{\bf x})v^i v^j}\;,\\
\end{eqnarray}
\begin{eqnarray}
\label{fac1}
{\cal S}_2&\equiv&\frac{{\cal K}^0}{{\cal K}^0_0}=\frac{1-\dot{\cal F}(\tau)}{1-\dot{\cal
F}(\tau_0)}\;,
\end{eqnarray}
\begin{eqnarray}
\label{fac3}
{\cal S}_3&=&\frac{1-{\bf k}\cdot{\bf v}-{\bf
v}\cdot\dot{\bm{\Xi}}(\tau)-k^\alpha h_{0\alpha}(t,{\bf x})-
k^\alpha v^j h_{\alpha j}(t,{\bf x}) }
{1-{\bf k}\cdot{\bf v}_0-{\bf
v}_0\cdot\dot{\bm{\Xi}}(\tau_0)-k^\alpha h_{0\alpha}(t_0,{\bf x}_0)-
k^\alpha v_0^j h_{\alpha j}(t_0,{\bf x}_0)}\;.
\end{eqnarray}
Here ${\bm{\Xi}}(\tau)$ is given in (\ref{27}) and ${\bm{\Xi}}(\tau_0)$ is
obtained from (\ref{27}) by means of calculation of all functions involved at
the instant $\tau_0$. 

On the other hand, our previous result for calculation of the {\it
Doppler shift in terms of frequency} obtained in section VI.C had the
following form
\begin{eqnarray}
\label{ghj}
\frac{\nu}{\nu_0}&=&{\cal S}_1\frac{dt}{dt_0}\;.
\end{eqnarray}
Thus, in order to have an agreement with calcualtion of the {\it Doppler shift
in terms of energy} one must prove that
\begin{eqnarray}
\label{ghy}
\frac{dt}{dt_0}&=&{\cal S}_2\cdot{\cal S}_3\;.
\end{eqnarray}
One can recast the product on the right hand side of (\ref{ghy}) 
accounting for equations (\ref{kob}), (\ref{koba}),
(\ref{fina}), (\ref{qaws}) into the form
\begin{eqnarray}
\label{pttt}
{\cal S}_2\cdot{\cal S}_3&=&\frac{{\cal A}(\tau)}{{\cal A}(\tau_0)}
\frac{{\cal B}(\tau)}{{\cal B}(\tau_0)}\;,
\end{eqnarray}
where
\begin{eqnarray}
\label{ptr}
{\cal A}(\tau)&=&1-{\bf k}\cdot{\bf
v}-\frac{1}{2}v^i k^\alpha k^\beta\hat{\partial}_i B_{\alpha\beta}(\tau)-
\frac{1}{2}k^\alpha k^\beta h_{\alpha\beta}(\tau)\;,\\\nonumber\\
\label{ptr1}
{\cal A}(\tau_0)&=&1-{\bf k}\cdot{\bf
v}_0-\frac{1}{2}v^i_0 k^\alpha k^\beta\hat{\partial}_i B_{\alpha\beta}(\tau_0)-
\frac{1}{2}k^\alpha k^\beta h_{\alpha\beta}(\tau_0)\;,
\\\nonumber\\
\label{ptr2}
{\cal B}(\tau)&=&1+2\sum_{a=1}^N \left[m_a C_a(s)-\frac{m_a}{\sqrt{1-v^2_a}}
\frac{(1-{\bf k}\cdot{\bf v}_a)^2}{r_a-{\bf k}\cdot{\bf r}_a}
\frac{{\bf k}\cdot{\bf r}_a}{r_a-{\bf v}_a\cdot{\bf r}_a}
\right]\;,
\\\nonumber\\
\label{ptr3}
{\cal B}(\tau_0)&=&1+2\sum_{a=1}^N \left[m_a C_a(s_0)-\frac{m_a}{\sqrt{1-v^2_{a0}}}
\frac{(1-{\bf k}\cdot{\bf v}_{a0})^2}{r_{0a}-{\bf k}\cdot{\bf r}_{0a}}
\frac{{\bf k}\cdot{\bf r}_{0a}}{r_{0a}-{\bf v}_{a0}\cdot{\bf r}_{0a}}
\right]\;,
\end{eqnarray}
where the partial derivatives $\hat{\partial}_i B_{\alpha\beta}(\tau)$ and 
$\hat{\partial}_i B_{\alpha\beta}(\tau_0)$ are calculated on the ground of
equation (\ref{21}). With equations (\ref{pttt}) - (\ref{ptr3}) it is 
straightforward to confirm the validity of equation (\ref{ghy}) if one notes that
up to the second order of the post-Minkowskian approximation scheme it holds
\begin{eqnarray}
\label{cvb} 
{\cal B}^{-1}(\tau)&=&1-2\sum_{a=1}^N \left[m_a C_a(s)-
\frac{m_a}{\sqrt{1-v^2_a}}
\frac{(1-{\bf k}\cdot{\bf v}_a)^2}{r_a-{\bf k}\cdot{\bf r}_a}
\frac{{\bf k}\cdot{\bf r}_a}{r_a-{\bf v}_a\cdot{\bf r}_a}
\right]\;,
\end{eqnarray}
so that equation (\ref{pttt}) can be re-written as follows
\begin{eqnarray}
\label{ghyu}
{\cal S}_2\cdot{\cal S}_3&=&\frac{{\cal A}(\tau)}{{\cal A}(\tau_0)
{\cal B}^{-1}(\tau){\cal B}(\tau_0)}\;.
\end{eqnarray}
It is easy to confirm that the numerator and denominator of (\ref{ghyu}) coincide
exactly with those of equation (\ref{61}) used for calculation of the {\it
Doppler shift in terms of frequency} and, for this reason, equation 
(\ref{ghy}) is valid. 
This finalizes the proof of equivalence of using two different 
mathematichal techniques for calculation of the Doppler effect.

In conclusion of this section we would like to point out that the method of
calculation of integrals in formulas (\ref{kob}), (\ref{koba}) exposed in the
sequence of equations (\ref{note}) - (\ref{fina}) significantly simplifies and 
reduces the amount of calculations which have been performed, e.g., in \cite{134} for
studying anisotropies of CMB radiation due to cosmic strings, where 
rather complicated transformations of variables were used
for performing of the integrals under discussion. As we have shown in the
present section such transformations are actually unnecessary.  

\subsubsection{The Explicit Doppler Tracking Formula}

In view of practical applications it is useful to give the explicit formula for
Doppler tracking of satellites. We shall derive it in the present section 
for one-way 
propagation of electromagnetic signals emitted from the point ${\bf x}_0$ at
time $t_0$ and received at the point ${\bf x}$ at time $t$. The Doppler shift
of the observed frequency $\nu$ with respect to the emitted frequency $\nu_0$ is given
by equation (\ref{asd1}) which is to be transformed to separate the special
relativistic Doppler effect from general relativistic corrections. Thus, we
have
\begin{eqnarray}
\label{aa1}
\frac{\nu}{\nu_0}&=&\frac{1-{\bf k}\cdot{\bf v}}{1-{\bf k}\cdot{\bf v}_0}\;
\left[\frac{1-v_0^2}{1-v^2}\right]^{1/2}\;
\left[\frac{{\rm a}(\tau_0)}{{\rm a}(\tau)}\right]^{1/2}\;
\frac{{\rm b}(\tau)}{{\rm b}(\tau_0)}\;,
\end{eqnarray}
where the first two factors out of four describe the 
special relativistic Doppler effect, and the
next terms are general relativistic corrections.
The unit vector ${\bf k}$ given at past null infinity relates to the unit
vector ${\bf K}$ (see (\ref{vvv}) for its
definition) of the boundary value problem through the relationship (\ref{29}) 
which, for the particular case under discussion, reads 
\begin{eqnarray}
\label{aa3}
k^i&=&-K^i+\frac{2}{R}\sum^N_{a=1}m_a\left[\frac{1-{\bf k}\cdot{\bf v}_a}
{\sqrt{1-v_a^2}}\frac{r_a^i-k^i({\bf k}\cdot{\bf r}_a)}
{r_a-{\bf k}\cdot{\bf r}_a}-
\frac{1-{\bf k}\cdot{\bf v}_{a0}}
{\sqrt{1-v_{a0}^2}}\frac{r_{0a}^i-k^i({\bf k}\cdot{\bf r}_{0a})}
{r_{0a}-{\bf k}\cdot{\bf r}_{0a}}\right]\\\nonumber\\\nonumber&+&
\frac{4}{R}\sum^N_{a=1}\left[\frac{m_a}{\sqrt{1-v_a^2}}(v_a^i-k^i({\bf k}\cdot{\bf
v}_a))\ln(r_a-{\bf k}\cdot{\bf r}_a)-
\frac{m_a}{\sqrt{1-v_{a0}^2}}(v_{a0}^i-k^i({\bf k}\cdot{\bf
v}_{a0}))\ln(r_{0a}-{\bf k}\cdot{\bf r}_{0a})\right]\;,
\end{eqnarray}
where $R=|{\bf x}-{\bf x}_0|$. 

The explicit formulas for the functions ${\rm a}(\tau)$ and ${\rm a}(\tau_0)$ are 
derived using (\ref{dt4}) which leads to
\begin{eqnarray}
\label{aa1a}
{\rm a}(\tau)&=&1+2\sum^N_{a=1}
\frac{m_a\sqrt{1-v_{a}^2}}{r_{a}-{\bf v}_{a}\cdot{\bf r}_{a}}-
\frac{4}{1-v^2}\sum^N_{a=1}
\frac{m_a}{\sqrt{1-v_{a}^2}}\frac{(1-{\bf v}\cdot{\bf
v}_{a})^2}{r_{a}-{\bf v}_{a}\cdot{\bf r}_{a}}\;,
\end{eqnarray}
\begin{eqnarray}
\label{aa1b}
{\rm a}(\tau_0)&=&
1+2\sum^N_{a=1}
\frac{m_a\sqrt{1-v_{a0}^2}}{r_{0a}-{\bf v}_{a0}\cdot{\bf r}_{0a}}-
\frac{4}{1-v_0^2}\sum^N_{a=1}
\frac{m_a}{\sqrt{1-v_{a0}^2}}\frac{(1-{\bf v}_0\cdot{\bf
v}_{a0})^2}{r_{0a}-{\bf v}_{a0}\cdot{\bf r}_{0a}}\;.
\end{eqnarray} 
We recall that ${\bf v}_0(t_0)$ is the barycentric velocity of emitter, ${\bf
v}_{a0}={\bf v}_a(s_0)$ is the barycentric velocity of the $a$-th gravitating 
body at the instant $s_0$, $r_{0a}=|{\bf r}_{0a}|$, ${\bf r}_{0a}={\bf x}_0(t_0)-{\bf x}_a(s_0)$, 
and $s_0=t_0-r_{0a}$ is the retarded time corresponding to the time of 
emission, $t_0$, of the radio signal. Besides this,
${\bf v}(t)$ is the barycentric velocity of receiver, ${\bf
v}_{a}={\bf v}_a(s)$ is the barycentric velocity of the $a$-th gravitating 
body at the instant $s$, $r_{a}=|{\bf r}_{a}|$, ${\bf r}_{a}={\bf x}(t)-
{\bf x}_a(s)$, and $s=t-r_{a}$ is the retarded time corresponding to the time 
of reception, $t$, of the radio signal.

Omitting all terms in
equation (\ref{53}) for the integral $C_a$
depending on accelerations of the bodies' center-of-mass, 
and reducing similar terms, we 
obtain for the functions in the last factor of the basic relationship 
(\ref{aa1}) the following explicit result
\begin{eqnarray}
\label{aa2a}
{\rm b}(\tau)&=&1+2\sum^N_{a=1}\frac{m_a}{\sqrt{1-v_a^2}}
\frac{1-{\bf k}\cdot{\bf v}_a}{r_a-{\bf v}_a\cdot{\bf r}_a}
\left[\frac{(1-{\bf k}\cdot{\bf v}_a)({\bf k}
\times{\bf v})\cdot({\bf k}\times{\bf r}_a)}{r_a-{\bf k}\cdot{\bf r}_a}
-\frac{({\bf k}\times{\bf v}_a)\cdot({\bf k}\times{\bf r}_a)}
{r_a-{\bf k}\cdot{\bf r}_a}+{\bf k}\cdot{\bf v}_a\right]\;,
\end{eqnarray}
\begin{eqnarray}
\label{aa2b}
{\rm b}(\tau_0)&=&
1+2\sum^N_{a=1}\frac{m_a}{\sqrt{1-v_{a0}^2}}
\frac{1-{\bf k}\cdot{\bf v}_{a0}}{r_{0a}-{\bf v}_{a0}\cdot{\bf r}_{0a}}
\left[\frac{(1-{\bf k}\cdot{\bf v}_{a0})({\bf k}\times{\bf v}_0)\cdot({\bf k}
\times{\bf r}_{0a})}{r_{0a}-{\bf k}\cdot{\bf r}_{0a}}
-\frac{({\bf k}\times{\bf v}_{a0})\cdot({\bf k}\times{\bf r}_{0a})}
{r_{0a}-{\bf k}\cdot{\bf r}_{0a}}+{\bf k}\cdot{\bf v}_{a0}\right].
\end{eqnarray}

The formulas (\ref{aa1}) -(\ref{aa2b}) describe the Doppler shift of the radio signal
transmitted from observer to spacecraft. The Doppler shift of the radio signal
transponded back to the observer is described by a similar set of equations with
corresponding attachment of all quantities to the instant of the signal's
reflection from the spacecraft and to the one of the signal's reception. In case of
light grazing a gravitating body, the formula (\ref{aa1}) gives, of course, the
result shown already in (\ref{dopshift}). 

\section{Discussion}

\subsection{Basic Results}

The long-standing problem of 
relativistic astrophysics and astrometry concerning
propagation of electromagnetic signals in the weak but arbitrarily fast
changing, time-dependent gravitational
field of an astronomical N-body system is analytically solved in the 
present paper in the first post-Minkowskian
approximation of General Relativity . The gravitational field, described by
the perturbation $h_{\alpha\beta}$ of the Minkowski metric tensor 
$\eta_{\alpha\beta}$ of the flat space-time, is presented
in the form of the {\it Li\'enard-Wiechert} potentials and depends on
coordinates, ${\bf x}_a$, ($a=1,2,...,N$), and velocities, ${\bf v}_a$, 
of the bodies taken at the retarded instants of time. There is no any
restriction on the motion of the bodies except for that ${\bf v}_a < c$ (speed 
of light). The relativistic
equations of light propagation are integrated in the field of the 
{\it Li\'enard-Wiechert} potentials and their solution are found 
in algebraically closed form. Exact analytic expressions for the 
integrated time delay, the angle of light
deflection, and the gravitational shift of electromagnetic frequency caused
by the gravitational fields of arbitrary moving bodies are derived and
all possible residual terms are shown explicitly. One can compare theoretical 
elegancy
and completeness of the Lorentz-covariant formalism of the present paper with 
various 
approaches of other
authors to the same problem of light propagation in time-dependent gravitational
fields (see, for example, \cite{134} - \cite{137}).

The applications of the Lorentz-covariant theory of light propagation,
developed in the present paper, to
relativistic astrophysics and astrometry are as follows:
\begin{itemize}
\item general theory of the Shapiro time delay in binary pulsars is
developed and all corrections with respect to velocities of pulsar and 
its companion to the standard logarithmic expression of the time delay in
static gravitational field are found. Particular attention is paid to 
the terms being linear in velocities which generalize the formula for the
Shapiro time delay which existed in the parameterized post-Keplerian formalism
discussed by Damour \& Taylor \cite{68}.
Lorentz-covariant post-Minkowskian approach to the time delay
calculations 
is compared with the post-Newtonian approach the enigmatic efficiency of which 
remained puzzling for a long time, is fully explained both in
terms of the analytic mathematical technique and in the visual
language of Minkowski diagrams;       
\item equation of gravitational lens, moving arbitrarily fast and possessing
spin-dipole and quadrupole components, is derived. Gravitational shift 
of spectral lines of the lensed source of light is worked out and
its influency on the anisotropy of cosmic microwave background radiation 
is discussed;  
\item the expression for the Shapiro time delay, caused by the solar
system bodies, is re-analyzed to improve accuracy of pulsar timing data
processing programs and of the consensus model of very long baseline
interferometry;   
\item relativistic deflection of light in the solar system gravitational
field is obtained with accounting for all velocity-dependent terms in the
first post-Minkowskian approximation. This result will be important in
future space astrometric missions like GAIA (ESA), SIM (NASA), etc.;
\item theoretical formulation of the Doppler tracking of interplanetary 
spacecrafts is achieved at the level of residual terms of order 
$10^{-16}$.   
\end{itemize}  
We could not elaborate in the present paper all possible aspects of the
Lorentz-covariant approach to the problem of propagation of light rays in
time-dependent gravitational fields of isolated astronomical systems. Some of
the most important theoretical developments which can be done in future are 
outlined in the following section. 

\subsection{Future Prospects}

The clear mathematical formulation of the Lorentz-covariant theory of
light propagation in gravitational fields of arbitrary-moving bodies and
the elegant method for solving related problems coming up in 
this framework and being based on the proper account for all retardation
effects, delivers new fascinating opportunities for much deeper exploration 
of the following open problems of modern relativistic astrophysics and
astrometry:
\begin{itemize}
\item propagation of light rays in the field of arbitrary-moving bodies
endowed with spin-dipole and quadrupole moments. This requires the
knowledge of the expression for the singular tensor of energy-momentum
of point-like particles with spin and quadrupole moments. Spin
contribution to the tensor can be found, for example, in \cite{46} 
but the structure of the tensor with the quadrupole (and
higher) multipole seems to be unknown. Solution of the given problem will
admit a precise mathematical treatment of timing observations of a pulsar
orbiting a Kerr black hole, as well as a unique interpretation of those X-ray
and gamma-ray sources which are assumed to have a Kerr black hole at the
center of their accretion disks related;  
\item extension of the Lorentz-covariant theory presented in this paper
to the event of strong gravitational fields. It will require finding
solutions of the equations of light propagation in the second post-Minkowskian
approximation of general relativity or another alternative theory of
gravity. Here one can expect to find differences between predictions of
two gravity theories which may be used for suggesting new observational tests of
the theories. It is also interesting to note \cite{30} that if the 
light-deflecting body or/and observer 
move too fast in a specific direction even the weak and
hence linear gravitational field can become strong in a chosen coordinate frame.
In such a case the linear post-Minkowskian approximation is not enough to give
unambigious observational predictions of relativistic gravitational effects in
propagation of light rays and a second iteration of the Einstein equations 
is required;
\item elaboration of the formalism of the present paper on the case of 
polarized electromagnetic wave to calculate the rotation angle 
of the plane of polarization along the 
null geodesic path of the wave - the Skrotskii effect \cite{138}, see also
\cite{134};
\item inclusion in the formalism of the given paper of relativistic effects
of gravitational waves from localized sources like supernova explosion,
massive binary black holes in nuclei of active galaxies, cataclysmic and
ordinary binary stars in our galaxy, etc. The first decisive step towards
the adequate interpretation of these gravitational wave effects has been
done in our paper \cite{1}. However, a more involved
technique is required to take into account motion of the sources of
gravitational waves with respect to observer. We expect that new
interesting effects may be found along this line;
\item calculation of response of space gravitational wave interferometers
like LISA to the signals emitted by gravitationally induced 
oscillations of the Sun (so-called g-modes). The combined technique of
this and our previous paper \cite{1} is undoubtedly
enough for getting the answer to that problem;
\item application of the formalism of the present paper to the case of
small-angle scattering problem of fast-moving self-gravitating bodies and
calculation of gravitational waveforms (cf., e.g., \cite{29}); 
\item development of physically adequate, high-precision algorithms for 
data processing of observations of space astrometric satellites and 
navigation systems like GPS 
as well as very long baseline interferometry. Practical
necessity in such algorithms is strongly felt already today and will 
permanently grow following achievements in the rapid development of advanced 
space technologies.
\end{itemize}  

We could continue the list of subjects for future work. For example, we
did not touch cosmological applications of the formalism of the present
paper. This will require some modifications of equations of light
propagation to account for cosmological expansion of the universe. No
doubt, the interpretation of observations of anisotropy of cosmic microwave
background radiation induced by, e.g. cosmic strings \cite{134}, 
can be made more theoretically adequate in the framework
of the presented new scheme.  

\acknowledgments
{We are greateful to N. Wex, A. Jessner and G. Giampieri for 
valuable discussions as well as to K. Nordtvedt, B. Mashhoon, 
A.F. Zakharov, and S.A. Klioner for important remarks. 
S.M. Kopeikin is pleased to acknowledge the hospitality 
of G. Neugebauer (TPI, FSU Jena) and R. Wielebinski
(MPIfR, Bonn). This work was partially 
supported by the Th\"uringer Ministerium f\"ur Wissenschaft, Forschung und 
Kultur Grant No B501-96060 (SMK) and by the Max-Planck-Gesellschaft Grant
No 02160-361-TG74 (GS).           

\newpage
\begin{figure*}
\centerline{\psfig{figure=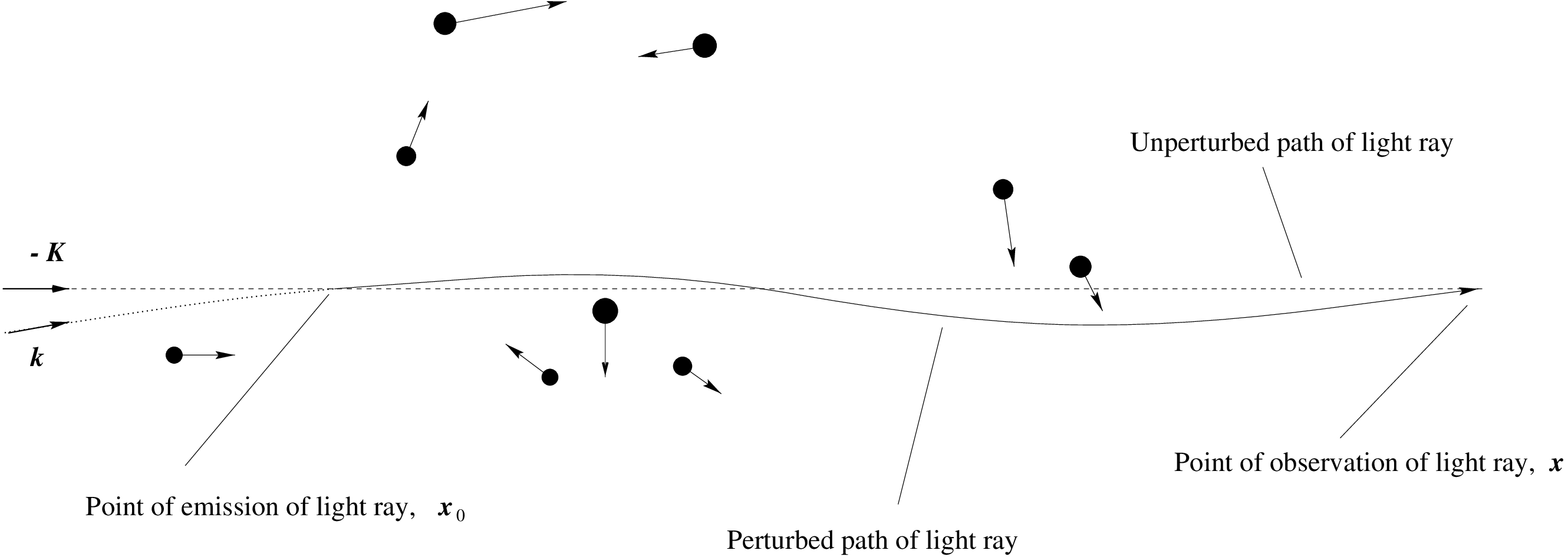,angle=270,height=6 cm,width=19 cm}}
\vspace{2 cm}
\caption{Illustration of the light-ray's propagation history. The light ray is emitted at the instant of time $t_0$ at the point ${\bf
x}_0$ and arrives at the point of observation ${\bf x}$ at the instant of 
time $t$. Light-deflecting bodies moves along accelerated world lines during 
the time of propagation of the light ray; their velocities at some 
intermediate instant of time are shown by black arrows. In the absence of the
light-ray-deflecting bodies the light ray would propagate along an unperturbed 
path (dashed line) which is a straight line passing through the points of 
emission, ${\bf x}_0$, and observation, ${\bf x}$. Direction of the 
unperturbed path is determined by the unit vector ${\bf K}=
-({\bf x}-{\bf x}_0)/|{\bf x}-{\bf x}_0|$. In the presence of the 
light-ray-deflecting bodies the light ray propagates along the perturbed 
path (solid line). The perturbed trajectory of the light ray is bent and 
twisted due to the gravitoelectric (mass-induced) and gravitomagnetic 
(velocity-induced) fields of the bodies. The initial boundary condition for 
the equation of light propagation is determined by the unit vector ${\bf k}$ 
defined at past null infinity by means of a dynamical backward-in-time 
prolongation (dotted line) of the perturbed trajectory of light from the point 
of emission ${\bf x}_0$ in such a way that the tangent vector of the 
prolongated trajectory coincides with that of the perturbed light-ray's 
trajectory at the point of emission. Relationship between unit vectors 
${\bf k}$ and ${\bf K}$ includes relativistic bending of light and is given 
in the text by equation (\ref{29}). }     
\label{covariant1}
\end{figure*}
\newpage
\begin{figure*}
\centerline{\psfig{figure=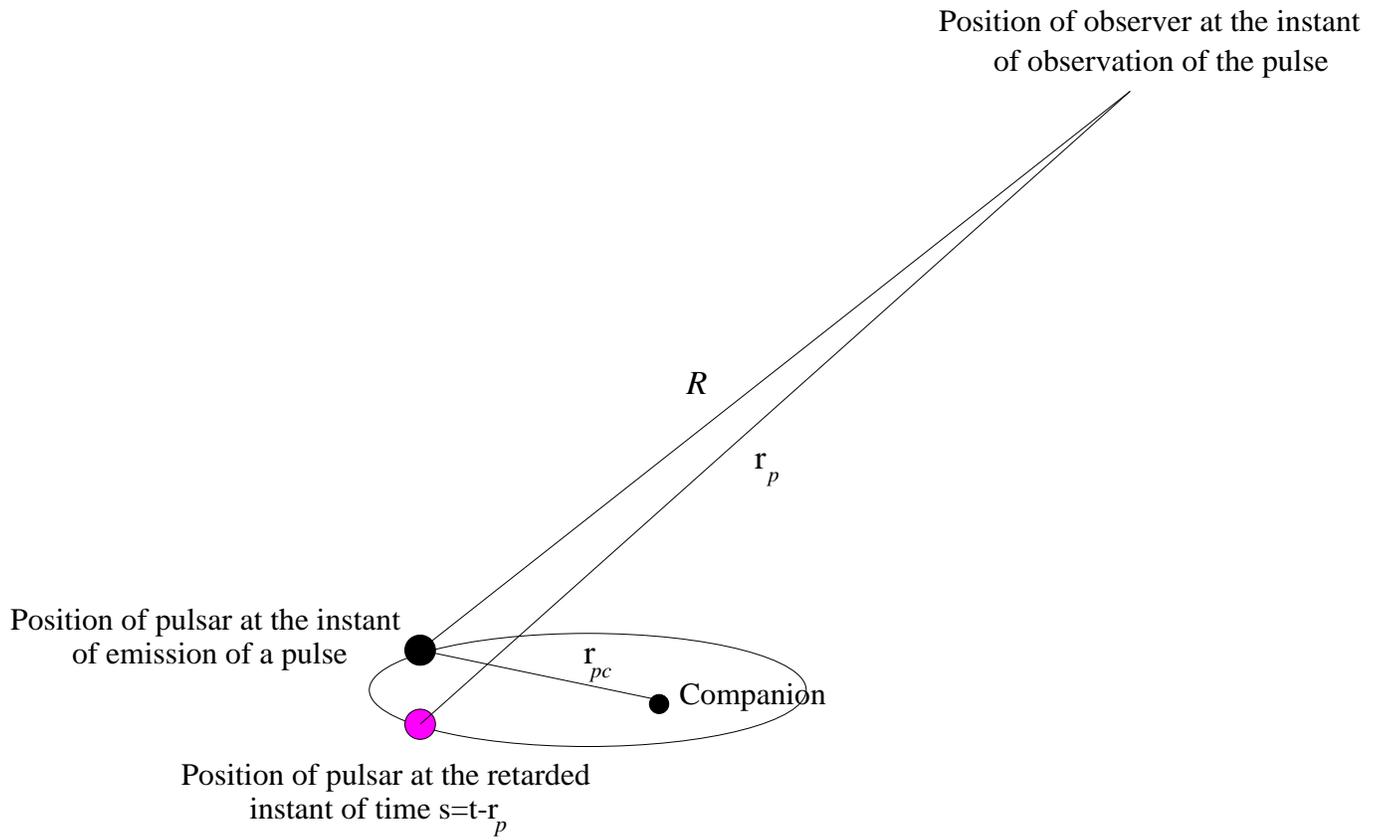,angle=270,height=11 cm,width=18 cm}}
\vspace{2 cm}
\caption{Schematic illustration of the Shapiro time delay in a binary
pulsar. The pulsar emits radio signal at the time $t_0$ which reaches the
observer at the time $t$. For calculation of the Shapiro time delay 
positions of the pulsar and its companion must be taken at the retarded
instants of time $s_0$ and $s$ corresponding to those $t_0$ and $t$. See
also Fig. \ref{covariant4} for further explanations. }
\label{shapiro}
\end{figure*}
\newpage
\begin{figure*}
\centerline{\psfig{figure=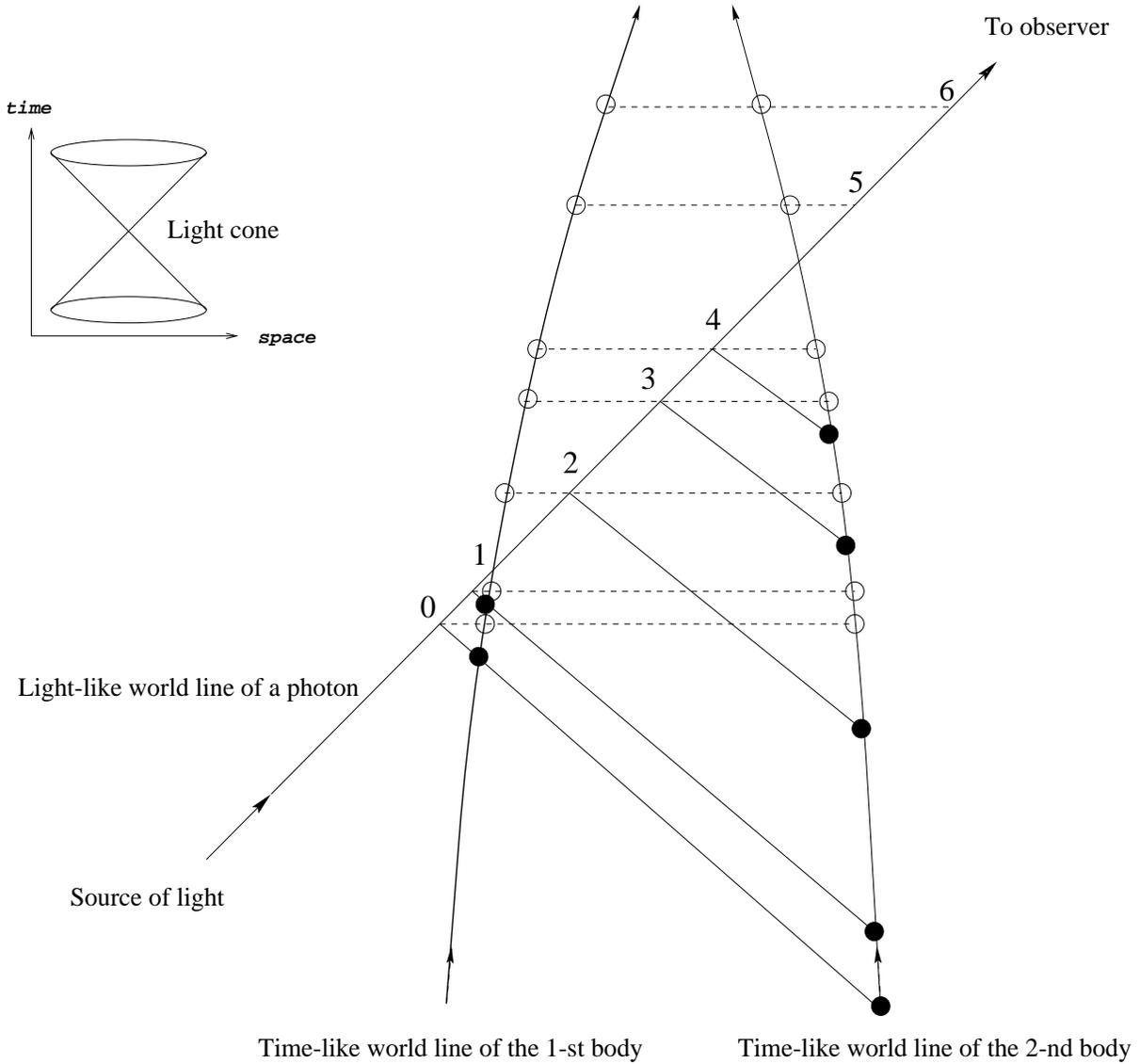,angle=270,height=15 cm,width=16 cm}}
\vspace{2 cm}
\caption{Schematic space-time diagram showing the relationship between
positions of a photon taken at different instants of time 
$t_0$, $t_1$,..., $t_6$ (events
0,1,...,6 on the photon's null world line) and positions of the
light-ray-deflecting bodies (marked by the black circles) 
taken at the instants of the
retarded time corresponding to the instants $t_0$, $t_1$,...,$t_6$. For
simplicity only the two-body system is considered.
The photon is deflected by the retarded
gravitational field of the bodies expressed through the 
{\it Li\'enard-Wiechert} potentials. Also shown are positions of the 
bodies (marked by the unfilled circles)
taken on the space-like hypersurfaces (dashed lines) of the time instants 
$t_0$, $t_1$,..., $t_6$. As the photon approaches towards the system
(events 0,1) 
it moves in the variable gravitational field of two bodies. After
crossing the system (events 5,6) the gravitational field at the photon's 
position is
"frozen" since the photon moves along the same light cone as 
the gravitational field propagates. The "freezing" of gravitational field
takes place during propagation of the photon inside the system (events
2,3,4). Spatial positions of gravitating bodies taken at the retarded instants 
of
time are very close to those taken at the hypersurfaces of constant time when
photon moves near or inside the system. It explains why the
post-Newtonian solution for the metric tensor can be applied in this 
situation as well as the post-Minkowskian one for calculation of the
photon's propagation. Retarded and instanteneous spatial 
positions of the gravitating bodies are drastically different when
photon is at large distance from the system (far outside the near zone). In
this case only the post-Minkowskin retarded solution for the metric tensor can 
be
applied for an adequate description of the gravitational perturbations of 
the photon's trajectory. }
\label{covariant4}
\end{figure*}
\newpage
\begin{figure*}
\centerline{\psfig{figure=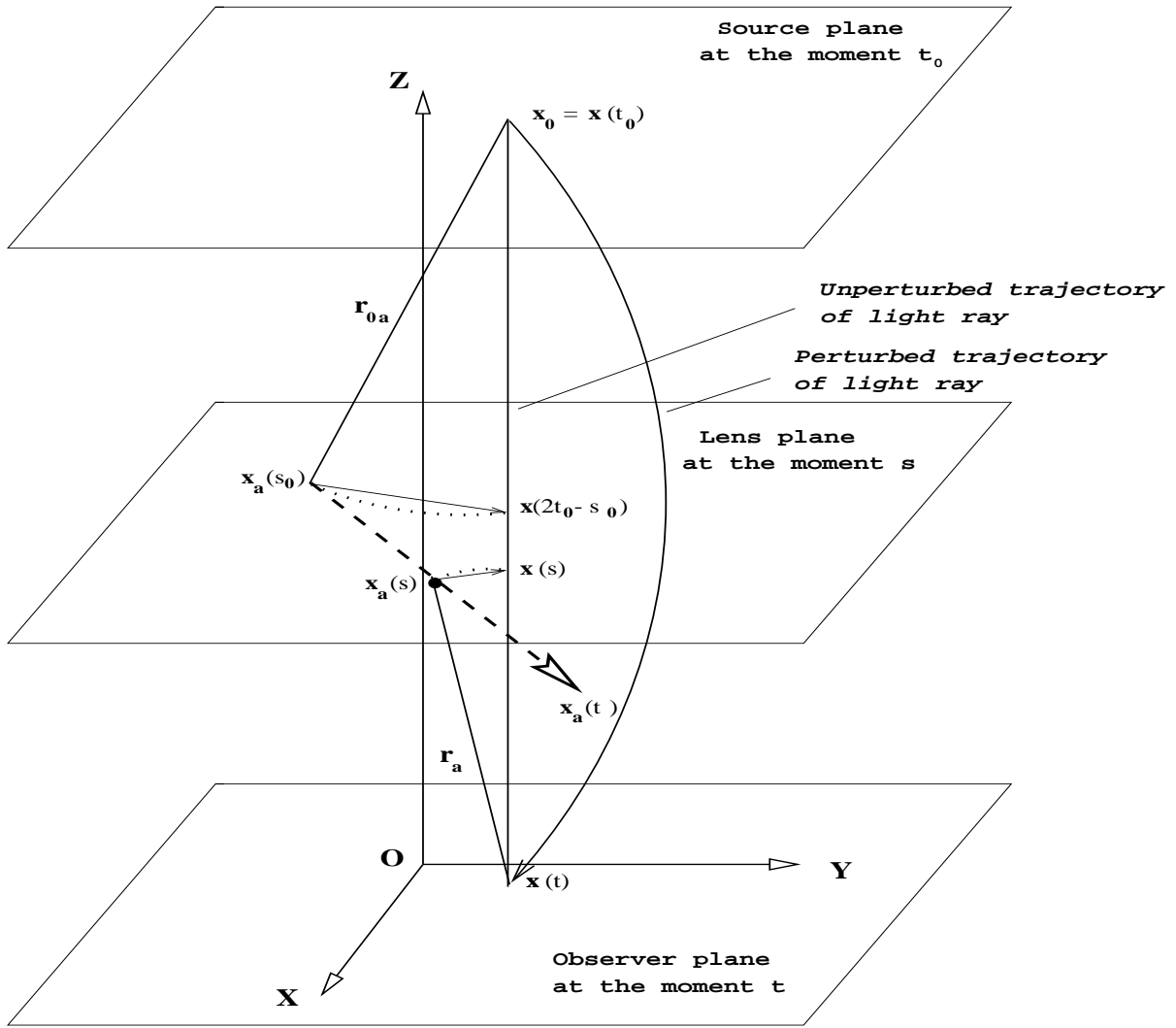,angle=270,height=14 cm,width=16 cm}}
\vspace{2 cm}
\caption{Relative configuration of observer, source of light
and a moving gravitational lens deflecting light rays which are
emitted at the moment $t_0$ at the point ${\bf x}_0$ and received at the 
moment $t$ at the point ${\bf x}$. The lens moves along straight line with
constant velocity from the retarded position ${\bf x}_a(s_0)$ through that 
${\bf x}_a(s)$ and arrives to the point ${\bf x}_a(t)$ at the moment of
observation. Characteristic time of the process
corresponds to the time of propagation of light from the point of emission up
to the point of observation.}
\label{movingGL}
\end{figure*}
\newpage
\begin{figure*}
\centerline{\psfig{figure=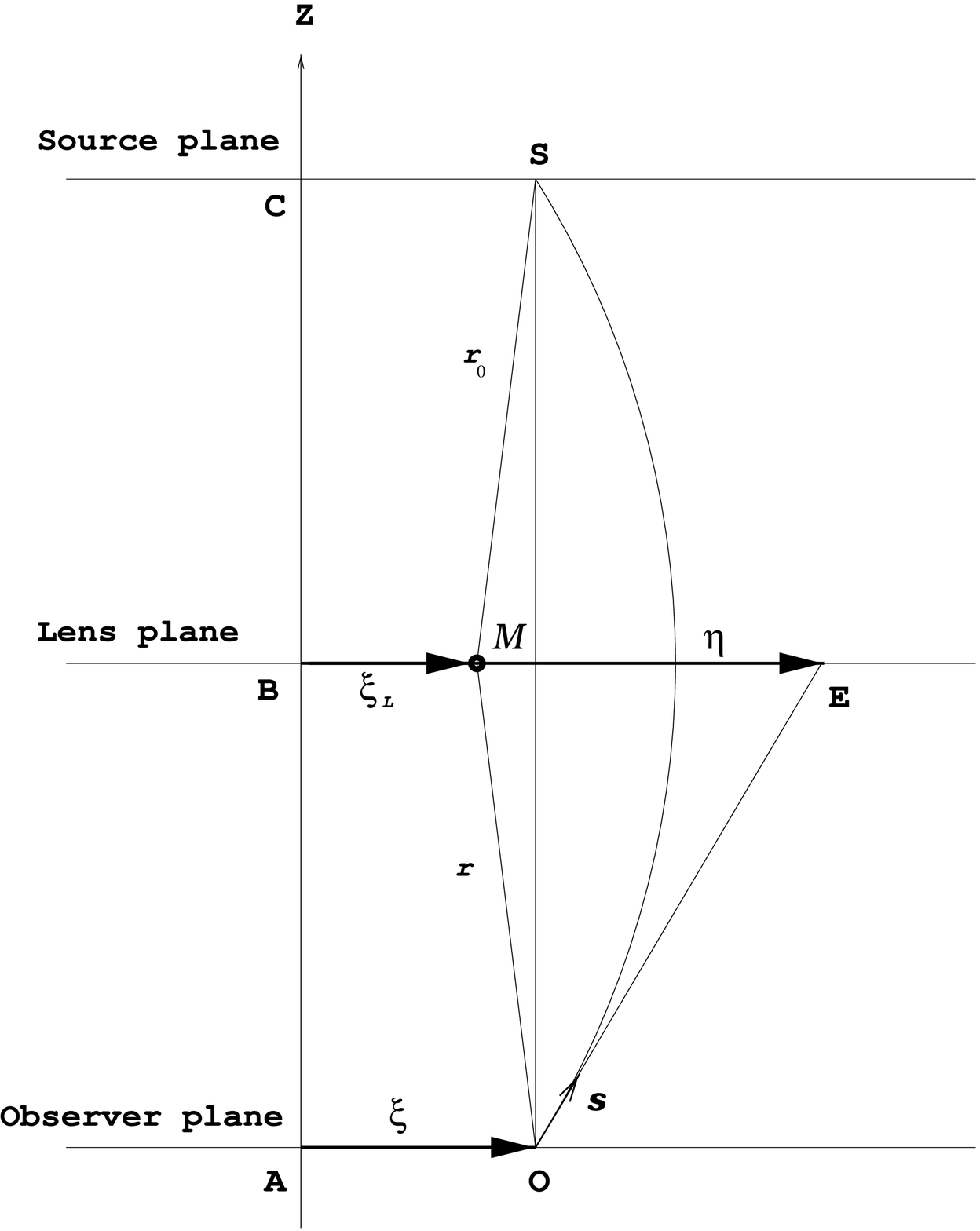,angle=0,height=16 cm,width=14 cm}}
\vspace{2 cm}
\caption{The gravitational lens geometry for a moving lens ${\cal M}
=\sum^N_{a=1} m_a$
being at the distance $r$ from the point of observation O with
coordinates $x^i(t)$. 
A source of light S with coordinates $x_0(t_0)$ is at the
distance $R$ from O. Vector ${\bm{\xi}}$ is the impact parameter of the
unperturbed path of photon in the observer plane. 
Vector ${\bm{\xi}}_L$ denotes position of the center of mass of the 
lensing object in the
lens plane.
Vector ${\bm{\eta}}=\vec{BE}$ is the observed image position of the background 
source of
light S shifted in the lens plane from its true position by the gravitational 
field of the lens to the point E. Coordinates of the lens are 
$X^i(\lambda^\ast)={\cal M}^{-1}\sum^N_{a=1} m_a x_a^i(\lambda^\ast)$,
and coordinates of the point E are
$x^i(\lambda^\ast)=x^i(t)+s^i(\lambda^\ast-t)$.}
\label{movingGL2}
\end{figure*}
\newpage
\begin{figure*}
\centerline{\psfig{figure=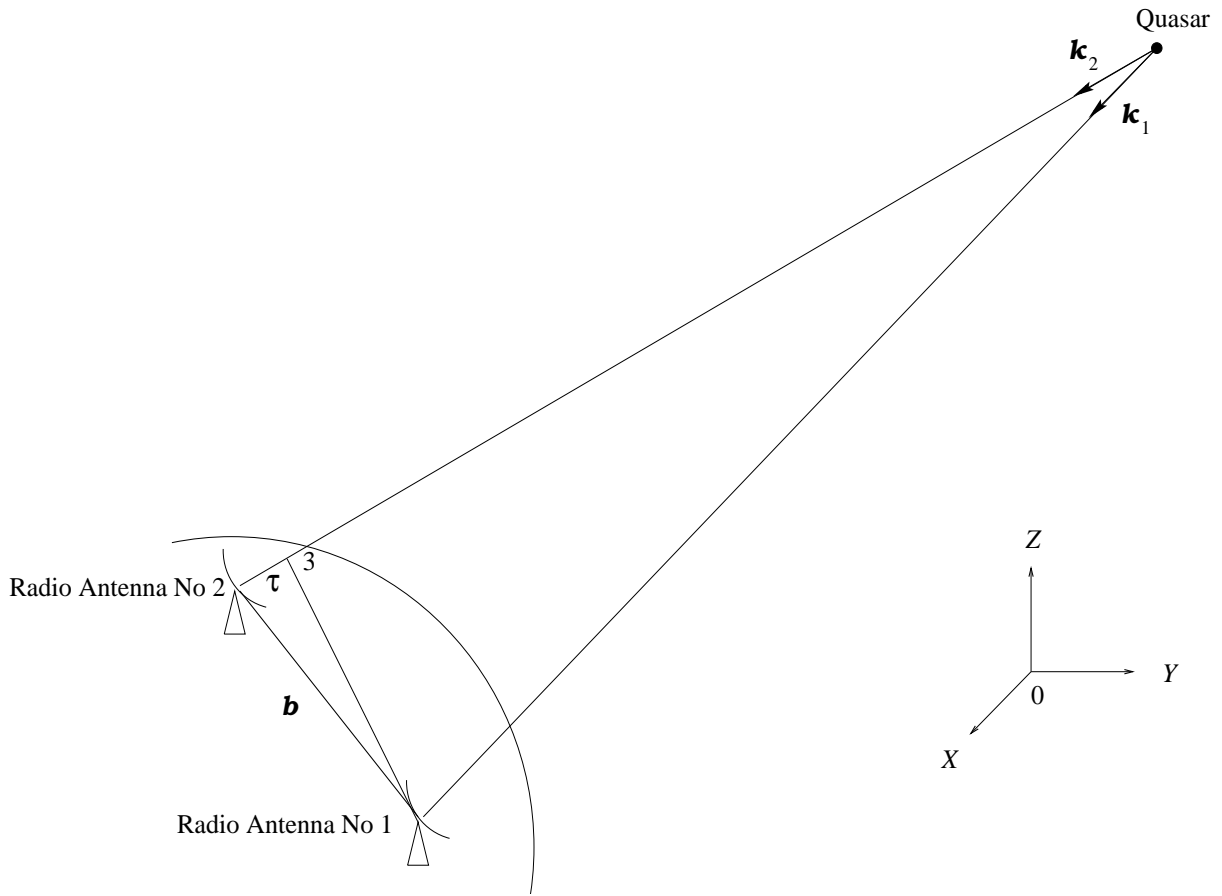,angle=270,height=12 cm,width=16 cm}}
\vspace{2 cm}
\caption{Geodetic very long baseline interferometry measures delay $\tau$ 
(the light travel time between points 2 and 3) in times of
arrival of radio signal from a quasar at the first and second radio 
antennas, $\tau=\tau_2-\tau_1$, located on the Earth' surface. 
Diurnal rotation and orbital motion of the Earth makes 
the delay to be dependent on time. This allows to determine the baseline
${\bf b}$ between two antennas, astrometric coordinates of the quasar,
motion of the Earth's pole, parameters of precession and nutations, and
many others. Modern data processing of VLBI observations is fully based
on the relativistic conceptions and was supposed to be accurate up to 1
picosecond.}
\label{covariant3}
\end{figure*}
\newpage
\begin{figure*}
\centerline{\psfig{figure=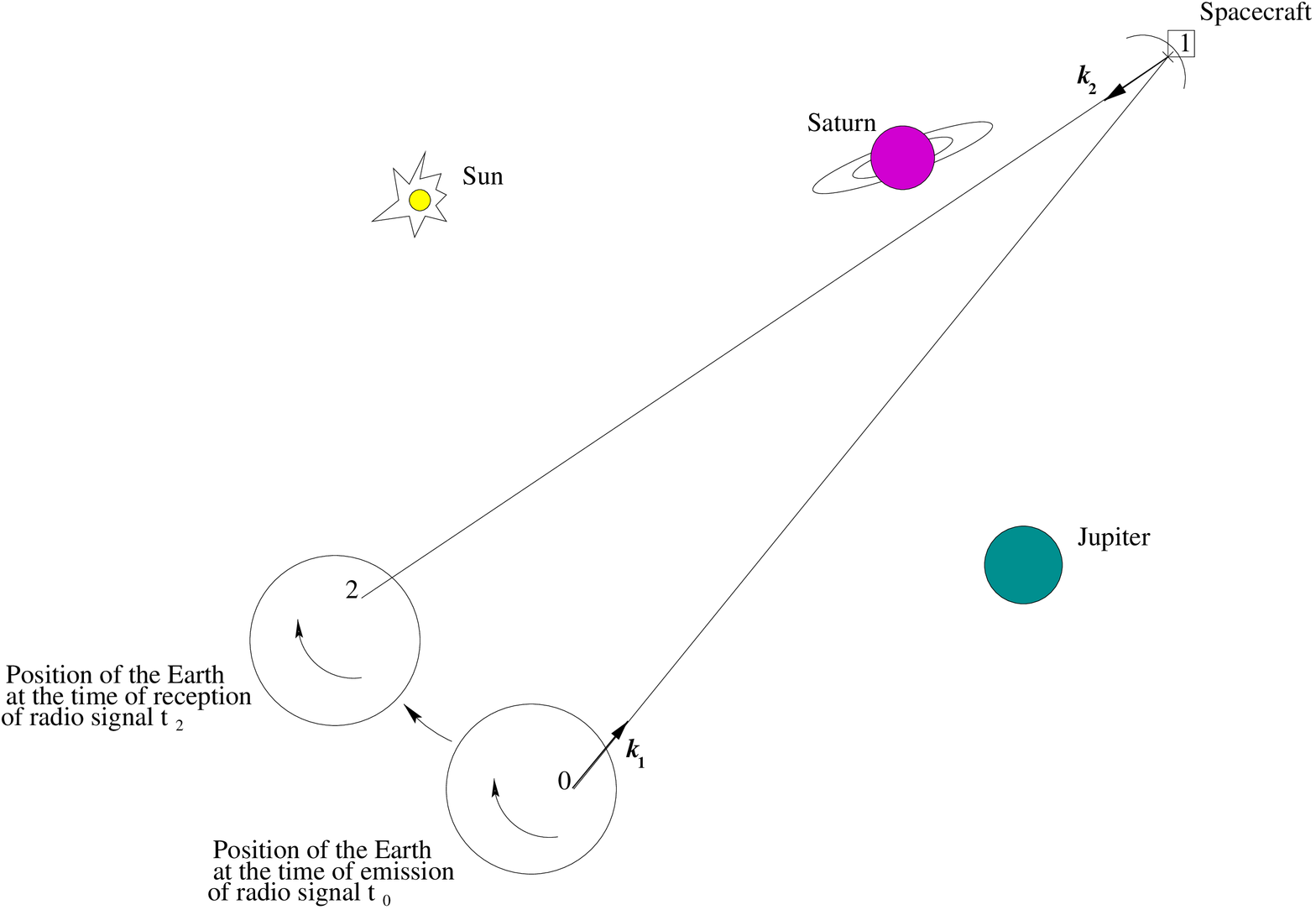,angle=270,height=11 cm,width=16 cm}}
\vspace{2 cm}
\caption{Spacecraft Doppler tracking experiment in deep space. Radio signal is 
transmitted at the
time $t_0$ and at the point 0 on the Earth along the unit vector 
${\bf k}_1$. The radio signal reaches the spacecraft at the moment $t_1$
and at the point 1 somewhere in the solar system and responds
back to the Earth exactly at the same time $t_1$ along the unit vector
${\bf k}_2$ which has a different orientation from ${\bf k}_1$. The
responded signal arrives at the reception point 2 on the Earth 
at the time $t_2$. During the round-trip time of the radio signal the Earth
rotates around its own axis and moves along the orbit. Hence, the
barycentric position and velocity of the transmitter is different from the 
barycentric position and velocity of the receiver despite of that their
topocentric positions on the Earth can coincide. When the impact parameter of 
the radio signal's trajectory is small the gravitational Doppler shift of the 
transmitted 
frequency with respect to the received frequency is estimated approximately 
as $\delta\nu/\nu_2=2\alpha(v_\odot/c)\cos\varphi$, where $\alpha$ is the
deflection angle of the light ray, $v_\odot$ is velocity of the Earth, and
$\varphi$ is the angle between $v_\odot$ and the impact parameter.}
\label{covariant5}
\end{figure*}


\begin{references}
\bibitem{1} Kopeikin, S.M., Sch\"afer G., Gwinn, C.R., \& Eubanks, T.M., 
1999, Phys. Rev. D., 59, 084023
\bibitem{2} Lindegren, L., and Perryman, M.A.C., 1996, A\&AS, 116, 579
\bibitem{3} Peterson, D. \& Shao, M. 1997, In: {\it Proc. of the ESA 
Symposium 
"HIPPARCOS, Venice 97"}, ed. Battrick, B.,  ESA Publications Division,
Noordwijk: The Netherlands,  p. 789
\bibitem{4} Klioner, S.A. \& Kopeikin, S.M., 1992, Astron. J., 104, 897; the
idea of the method used by Klioner \& Kopeikin was put forward in \cite{116}.
\bibitem{5} Misner, C.W., Thorne, K.S. \& Wheeler, J.A., 1973, {\it 
Gravitation},
W. H. Freeman \& Company: New York
\bibitem{6} Fock, V.A., 1959, {\it The Theory of Space, Time and 
Gravitation},
Pergamon Press: London
\bibitem{7} Soffel, M.H., 1989, {\it Relativity in Astrometry, Celestial 
Mechanics and
Geodesy}, Springer-Verlag: Berlin 
\bibitem{8} Brumberg, V.A., 1991, {\it Essential Relativistic Celestial
Mechanics}, Adam Hilger: Bristol
\bibitem{9} Will, C.M., 1993, {\it Theory and Experiment in Gravitational 
Physics}, 
Cambridge University Press: Cambridge 
\bibitem{10} Hellings, R.W., 1986, Astron. J., 91, 650; 92, 1446; see also
Sovers, O.J. \& Fanselow, J.L., 1987, {\it Observation model and Parameter
Partials for the JPL VLBI Parameter Estimation Software 'Masterfit"-1987}, JPL
Pub. 83-89, Rev. 3, and \cite{112}, discussion on pages 1406-7 after formula
(3.23)
\bibitem{11} Kopeikin, S.M., 1997, J. Math. Phys., 38, 2587
\bibitem{12} In what follows we shall often use geometrical units, e.g. 
see \cite{5} in which $G=c=1$.
\bibitem{13} The difference between 
the parameters $\gamma^\ast$ and $\gamma$ includes terms of the post-Newtonian 
order of
magnitude. In the weak-field limit the numerical value of $\gamma^\ast$
coincides with that of
$\gamma$ which is one of the parameters of  
the standard parameterized post-Newtonian (PPN) formalism \cite{9}. 
For more details see \cite{14}, \cite{15}, and references therein.
\bibitem{14} Nordtvedt, K., 1985, \apj, 297, 390
\bibitem{15} Damour, T. \& Esposito-Far\`ese, G., 1996, Phys. Rev. D, 53, 5541
\bibitem{16} Damour, T. \& Esposito-Far\`ese, G., 1998, Phys. Rev. D, 58, 
044003
\bibitem{17} Landau, L.D. \& Lifshitz, E.M., 1971, {\it The Classical Theory of
Fields}, Pergamon: Oxford
\bibitem{18} Space-time indices $\alpha, \beta,...,$ etc. run
from 0 to 3 and are 
raised and lowered by means of $\eta_{\alpha\beta}$. Spatial indices 
$i,j,k,...,$
etc. run from 1 to 3 and are raised and lowered by means of the Kronecker symbol
$\delta_{ij}$, so that, actually, the upper and lower case spatial indices
are not distinguished. Repeated greek and latin indices preassume summation 
from 0 to 3 and from 1 to 3 respectively.
\bibitem{19} The harmonic gauge \cite{6} is fixed in the first post-Minkowskian
approximation by the four differential 
conditions 
\begin{eqnarray}
\label{harm}
h_{\alpha}^{\;\beta},_{\beta}-\frac{1}{2} h^{\beta}_{\;\beta},_{\alpha}=0\;,
\end{eqnarray}
where
comma denotes a partial derivative with respect to a corresponding coordinate.
\bibitem{20} Weinberg, S., 1972, {\it Gravitation and Cosmology}, 
John Wiley \& Sons: New York
\bibitem{21} Jackson, J.D., 1974, {\it Classical Electrodynamics}, 
John Wiley \& Sons: New York
\bibitem{22} We take only the retarded Green 
function and abandoned the advanced one as we assume that the N-body system 
under consideration is absolutely isolated from possible external gravitational
enviroment. It is equivalent to the assumption that there is no gravitational 
radiation impinging onto the system in the first post-Minkowskian 
approximation. It is interesting to emphasize that in higher post-Minkowskian
approximations existence of the tail gravitational radiation effects
\cite{23} - \cite{25} brings 
about a small fraction of incoming radiation as being backscattered on
the static part of the curvature generated by the monopole component in
multipole expansion of the metric tensor. However, although the backscattered 
radiation is incoming, it does not come in from past null infinity 
("scri minus"). Therefore, it has nothing to do with advanced Green function. 
Its again purely outgoing radiation at future null infinity ("scri plus"). 
We omit such backscatter terms in what follows for they appear only in the 
higher orders of the post-Minkowskian approximation scheme.
\bibitem{23} Blanchet, L. \& Damour, T., 1988, Phys. Rev. D, 37, 1410
\bibitem{24} Sch\"afer, G., 1990, Astron. Nachrichten, 311, 213
\bibitem{25} Blanchet, L. \& Sch\"afer, 1993, Class. Quant. Grav., 10, 2699
\bibitem{26} It is more appropriate
to denote the retarded time for the $a$-th body as $s_a$ which would have 
reflected
the
dependence of the retarded time on the number of the body. However, it would
make notations and the presentation of subsequent formulas more cumbersome. For
this reason we use $s$ instead of $s_a$ keeping in mind that if it
is not
stated otherwise, coordinates, velocity, and acceleration of the $a$-th body are 
taken at the
corresponding retarded time $s_a$. This remark is crucial, e.g., in the
discussion regarding the definition of the center of mass of the $N$ body system
(for more details see the explanations after equation (\ref{g11bb})).
\bibitem{27} In addition, we emphasize that the
condition of weak-field approximation, that is $|h^{\alpha\beta}\ll 1|$, leads
to general restriction on the velocity of a moving body $(G m_a)/(c^2 r_a)\ll
(1-v_a/c)^{1/2}$. In the particular case, where the velocity ${\bf v}_a$ of the 
$a$-th body is almost parallel to ${\bf r}_a$ one gets the stronger restriction 
$(G m_a)/(c^2 r_a)\ll (1-v_a/c)^{3/2}$ (\cite{28} - \cite{30}).
\bibitem{28} Kov{\'a}cs, S.J. \& Thorne, K.S., 1978, \apj, 224, 62
\bibitem{29} Westpfahl, K., 1985, Fortschritte der Physik, 33, 417
\bibitem{30} Mashhoon, B., 1992, Phys. Lett. A, 163, 7
\bibitem{31} Thorne, K.S. \& Kov{\'a}cs, S.J., 1975, \apj, 200, 245
\bibitem{32} Crowley, R.J. \& Thorne, K.S., 1977, \apj, 215, 624
\bibitem{33} Kov{\'a}cs, S.J. \& Thorne, K.S., 1977, \apj, 217, 252
\bibitem{34} It is worth noting that this statement is true only 
if the origin of the
coordinate system is in between the source of light and observer. Under
other circumstances the variable $\tau$ may be always either positive or
negative.
\bibitem{35} At the given stage of our study we
do not fix the freedom in choosing the origin of the coordinate system by
assuming, for example, that the origin coincides with the center of mass
of the $N$-body system. Specific choices of the origin will be done later on.
\bibitem{35a} The principal differential identity used for derivation of 
equation (\ref{14}) and applied to any smooth function $F(t,{\bf x})$ 
reads as follows
\begin{eqnarray}
\label{pr}
\left[\frac{\partial F(t,{\bf x})}{\partial x^i}+
k_i\;\frac{\partial F(t,{\bf x})}{\partial t}\right]_{{\bf x}={\bf
k}\;(t-t_0)+{\bf x}_0}&=&\frac{\partial F(\tau+t^\ast,{\bf k}\tau+{\bm{\xi}})}
{\partial \xi^i}+
k_i\;\frac{\partial F(\tau+t^\ast,{\bf k}\tau+{\bm{\xi}})}{\partial \tau}\;,
\end{eqnarray}
and allows to change the order of operations of partial differentiation and 
substitution for the unperturbed light ray trajectory in the equation for light
geodesics \cite{1}.  
\bibitem{36} More precisely, this kind of equation is
known in the literature as "retarded-functional differential system" because of
the dependence of gravitational potentials on the retarded time argument. Such
an equation belongs to the framework of "predictive relativistic mechanics" 
\cite{37} - \cite{43}.
\bibitem{37} Currie, D.G., 1966, Phys. Rev., 142, 817
\bibitem{38} Hill, R.N., 1967, J. Math. Phys., 8, 201
\bibitem{39} Bel, L., 1970, Ann. Inst. H. Poincar{\'e}, 12, 307
\bibitem{40} Bel, L., Salas, A. \& Sanchez, J.M., 1973, Phys. Rev. D., 7, 1099
\bibitem{41} Bel, L. \& Fustero, X., 1976, Ann. Inst. H. Poincar{\'e}, A25, 411
\bibitem{42} Bel, L., Damour, T., Deruelle, N., Iba{\~n}ez, J. \& Martin,
J., 1981, Gen. Rel. Grav., 13, 963
\bibitem{43} Damour, T., 1983, In: {\it Gravitational Radiation}, Les
Houches 1982, eds. Deruelle, N. \& Piran, T., North-Holland: Amsterdam
\bibitem{44} We again emphasize that the new parameter $\zeta$ depends on the
index of each body. For this reason it would be better to denote it as 
$\zeta_a$.
We do not use this notation to avoid the appearance of a large number of 
subindices.
\bibitem{44a} Calculation of the derivative $\hat{\partial}_i B^{\alpha\beta}(s_0)$ at
the point of light emission is obtained from the formula (\ref{21}) where all
quantities involved are to be taken at the retarded time $s_0$.
\bibitem{45} We emphasize that our formalism admits to work with world lines
of arbitrary moving bodies without restricting them to straight lines 
only. More precisely, in the harmonic gauge (\ref{harm})
the equations of motion of the bodies result from the harmonic coordinate 
conditions (\ref{harm}). In the first post-Minkowskian approximation 
these conditions allow motion of bodies only along straight lines with
constant speeds. However, if in finding the metric tensor the non-linear 
terms in the Einstein equations are taken 
into account, the bodies may show accelerated motion without structurally
changing the linearized form of the {\it Li\'enard-Wiechert} solution 
for the metric tensor which is used for integration of equations of motion of a
photon.
\bibitem{46} Iba{\~n}ez, J., Martin, J. \& Ruiz, E., 1984, Gen. Rel. Grav.,
16, 225
\bibitem{47} Sch\"afer, G., 1985, Annals Phys., 161, 81
\bibitem{48} Kopeikin, S.M., 1985, Sov. Astron., 29, 516
\bibitem{49} Damour, T., 1987, In: {\it 300 Years of Gravitation}, eds.
Hawking, S.W. \& Israel, W., Cambridge University Press: Cambridge, p. 128
\bibitem{50} Shapiro, I.I., 1964, Phys. Rev. Lett., 13, 789
\bibitem{51} Again
it would be better to denote the retarded time as $s_{0a}$ emphasizing its
dependence on the number index of the body under consideration. We keep
in mind this remark but do 
not use such a
notation to avoid confusing mixture of indices.
\bibitem{52} Nordtvedt, K., 1980, unpublished work (private communication)
\bibitem{53} Klioner, S.A., 1989, {\it Propagation of light in the barycentric
reference system with the motion of the gravitating masses taken into 
account},
preprint of the Institute of Applied Astronomy No 6, Leningrad (in Russian)
\bibitem{54} Wex, N., 1995, {\it Relativistische Effekte in der Beobachtung und
Bewegung von Pulsaren}, PhD thesis, TPI, FSU of Jena, Germany
\bibitem{55} The case of an
observer moving with respect to the harmonic coordinate system with velocity 
$v^i$ may
be considered after completing the additional Lorentz transformation 
described by the matrix $L^{\alpha}_{\;\beta}$ 
with components (e.g., see \cite{5}, formula 2.44)
\vspace{0.3 cm}
\begin{equation}
\label{lorentz}
L^0_{\;0}=\gamma\equiv\frac{1}{\sqrt{1-{\beta}^2}}\;,\quad\quad\quad\quad
L^0_{\;i}=L^i_{\;0}=-\beta\gamma n^i\;,\quad\quad\quad\quad 
L^i_{\;j}=\delta^{ij}+(\gamma-1)n^i n^j\;,
\end{equation}
where $\beta=v/c$, and $n^i=v^i/v$ is the unit vector in the direction of
motion of the observer. 
\bibitem{56} Note that in the
relativistic terms of any formula of the present paper 
we are allowed to use
the substitution $\delta_{ij}-p_i p_j=\delta_{ij}-k_i k_j=P_{ij}$.
\bibitem{57} Valenti, J.A., Butler, R.P. \& Marcy, G.W., 1995, PASP, 107, 966
\bibitem{58} Cochran, W., 1996, In: {\it Proc. Workshop of High Resolution
Data Processing}, eds. Iye, M., Takata, T. and Wampler. J., SUBARU
Telescope Technical Report, NAOJ, No 55, p. 30
\bibitem{59} Kopeikin, S.M., \& Ozernoy, L. M., 1998, \apj, to be published; 
preprint astro-ph/9812446 
\bibitem{corr.3} Synge, J.L., 1971, {\it Relativity: The General Theory},
North-Holland: Amsterdam
\bibitem{60} Brumberg, V.A., 1972, {\it Relativistic Celestial Mechanics}, 
Nauka: Moskow (in Russian)
\bibitem{corr.2} Synge  calls the relationship (\ref{58}) as the
{\it Doppler effect in terms of frequency} (see \cite{corr.3}, page 122). It is
fully consistent with definition of the {\it Doppler shift in terms of energy}
(see \cite{corr.3}, page 231) when one compares the energy of photon at the
points of emission and observation of light. The {\it Doppler shift in terms 
of energy} is given by
\begin{eqnarray}
\label{enrg}
1+z&=&\frac{\nu}{\nu_0}=\frac{u^\alpha {\cal K}_\alpha}
{u_0^\alpha {\cal K}_{0\;\alpha}}\;,
\end{eqnarray}
where $u_0^\alpha$, $u^\alpha$ and ${\cal K}_{0\;\alpha}$, ${\cal K}_\alpha$ 
are 
the 4-velocities of source of light and observer and the 4-momenta of photon at
the points of emission and observation respectively. It is quite easy 
to see that both mentioned formulations of the Doppler shift
effect are equaivalent. Indeed, taking into account that
$u^\alpha=dx^\alpha/d{\cal T}$ and ${\cal K}_\alpha=
{\partial \varphi}/{\partial x^\alpha}$, where $\varphi$ is the phase of the
electromagnetic wave, we obtain  $u^\alpha {\cal K}_\alpha=d\varphi/d{\cal T}$.
Thus,
\begin{eqnarray}
\label{volk}
1+z&=&\frac{d\varphi}{d\varphi_0}\frac{d{\cal T}_0}{d{\cal T}}\;.
\end{eqnarray}
The
phase of electromagnetic wave remains constant along the light ray trajectory. 
For this reason, $d\varphi/d\varphi_0=1$, and equation (\ref{58}) holds. 
\bibitem{60a} Taking times $\tau$ and $\tau_0$ as primary quantities 
instead of $t$ and $t_0$ brings in the retarded times $s$ and $s_0$ dependence
on the time of the closest approach $t^\ast$, that is either $s=s(t,t_0)$,
$s_0=s_0(t_0)$ or $s=s(\tau,\tau_0,t^\ast)$, $s_0=s_0(\tau_0,t^\ast)$. It
introduces partial derivatives of $s$ and $s_0$ with respect to $t^\ast$ 
and modifies formula (\ref{61}) as well.
\bibitem{60b} If one uses times $\tau$ and $\tau_0$ as time variables the
equalities (\ref{ret}) assume the form
\begin{eqnarray}
\label{rtuo}
s+|{\bf x}(\tau_0)+{\bf k}(\tau,\tau_0)(\tau-\tau_0)-{\bf
x}_a(s)|&=&\tau+t^\ast\;,\\
s_0+|{\bf x}(\tau_0)-{\bf x}_a(s)|&=&\tau_0+t^\ast\;,
\end{eqnarray}
from which and (\ref{ret}) it follows that
\begin{equation}
\label{rtxo}
\frac{\partial s(t,t_0)}{\partial t}=\frac{\partial s(\tau,\tau_0,t^\ast)}{\partial
t^\ast}\;,\quad\quad\quad \frac{\partial s_0(t_0)}{\partial t_0}=
\frac{\partial s_0(\tau_0,t^\ast)}{\partial t^\ast}\;. 
\end{equation} 
\bibitem{62} This statement may not be valid in the case of Doppler tracking 
observations of spacecrafts in the solar system.
\bibitem{63} Damour, T., 1984, Found. of Phys., 14, 987
\bibitem{64} Damour, T. \& Sch\"afer, G., 1991, Phys. Rev. Lett., 66, 2549
\bibitem{65} Damour, T. \& Esposito-Far\`ese, G., 1992, Phys. Rev. D, 46, 4128
\bibitem{66} Damour, T. \& Esposito-Far\`ese, G., 1998a, Phys. Rev. D, 58,
042001
\bibitem{67} Zaglauer, H.W., 1992, \apj, 393, 685
\bibitem{68} Damour, T. \& Taylor, J.H., 1992, Phys. Rev. D, 45, 1840. 
Strictly speaking, some of the PK parameters depend, actually, on four
unknown quantities - masses of pulsar and its companion and two angles of
orientation of the pulsar's angular velocity of proper rotation. However, the PK
parameters presently measured in most binary pulsar systems depend on masses of
the stars only. 
\bibitem{69} Kaspi, V.M., Taylor, J.H., \& Ryba, M.F., 1994, \apj, 428, 713
\bibitem{70} Stairs, I.H., Arzoumanian, Z., Camilo, F., Lyne, A.G., Nice, D.J.,
Taylor, J.H., Thorsett, S.E. \& Wolszczan, A., 1998, \apj, 505, 352
\bibitem{71} Taylor, J.H., 1994, Rev. Mod. Phys., 66, 711
\bibitem{72} Blandford, R. \& Teukolsky, S.A., 1976, \apj, 205, 580
\bibitem{73} More precisely, the
coordinates of the point of emission are constant in the pulsar proper
reference frame. The relativistic transformation from the proper reference frame
of the pulsar to the binary pulsar barycentric coordinate system 
\cite{74},
\cite{75} reveals that if the pulsar moves along the elliptic
orbit the barycentric vector ${\bf X}$ actually depends on time. 
However, this periodic relativistic perturbation of the vector is
of order $(|{\bf X}|/c)(v^2_p/c^2)$ where $v_p$ is a characteristic
velocity of the pulsar with respect to the barycenter of the binary system. For
a typical distance $|{\bf X}|\simeq 50\div 100$ km this is too small for
being measurable. Another reason for the time dependence of the
barycentric vector ${\bf X}$ on
time arises due to the effects of aberration \cite{76}, 
the orbital pulsar parallax 
\cite{77}, and the bending delay \cite{78}. These
effects are also small and can be neglected in the formula for the 
Shapiro time delay.
\bibitem{74} Brumberg, V.A. \& Kopeikin, S.M., 1989, Nuovo Cim., 103B, 63
\bibitem{75} Damour, T., Soffel, M. \& Xu, Ch., 1991, Phys. Rev. D, 43, 3273
\bibitem{76} Smarr, L.L. \& Blandford, R., 1976, \apj, 207, 574
\bibitem{77} Kopeikin, S.M., 1995, \apj, 439, L5 
\bibitem{78} Doroshenko, O.V. \& Kopeikin, S.M., 1995, Mon. Not. R. Astron.
Soc., 274, 1029
\bibitem{79} Compare, for instance, 
with formula (3) in \cite{80}.
\bibitem{80} Damour, T. \& Deruelle, N., 1986, Ann. Inst. H. Poincar{\'e}, A44,
263
\bibitem{81} Let us note once again that 
the post-Newtonian scheme can be applied
without restriction only if the length of the light ray trajectory is 
small compared with the size of the gravitating system. This situation is
realized in the observations of the solar system objects. We analyse this
case in section 7C (see also Fig. \ref{covariant4}) in more details.
\bibitem{82} For comparison with other phenomenological timing models worked
out by other researchers see, e.g., \cite{83}.
\bibitem{83} Klioner, S.A. \& Kopeikin, S.M., 1994, \apj, 427, 951
\bibitem{83a} Kopeikin, S.M., 1994, \apj, 434, L67
\bibitem{83b} To be more precise, the post-Newtonian scheme may
give inconsistent results for light propagation in those terms which are proportional
to the product of mass of the light-deflecting body and square of its velocity,
that is, $Gm_a v_a^2$. At the same time
the post-Minkowskian approach of the present paper allows to treat all such
terms without ambiguity. Nevertheless, these terms 
are not enough for complete description of the light-ray 
trajectory
because the first post-Minkowskian approximation does not include terms being
quadratic with respect to gravitational constant $G$ which may be comparable in
self-gravitating systems with terms of order $Gm_a v_a^2$.  
\bibitem{84} We neglect the proper motion of the pulsar in the
sky which brings about the small secular change in coordinates of the
vector ${\bf k}$. The error of the approximation is about $\frac{G
m_c}{c^3}\frac{\mu T_{span}}{1-\sin i}$, where $\mu$ is the proper
motion of the pulsar and $T_{span}$ is the total span of observation.
This error is much smaller than 1 $\mu$s being presently unmeasurable.
\bibitem{85} Birkinshaw, M., 1989, In: {\it Lecture Notes in Physics 330, 
Gravitational Lenses}, eds. Moran, J.M., Hewitt, J.N. and Lo, K.L.,  
Springer-Verlag: Berlin, p. 59
\bibitem{86} The integrals in (\ref{53})-(\ref{54}) are identically zero 
because of our assumption that the velocities of the gravitating bodies 
are constant.
\bibitem{87} Schneider, P., Ehlers, J. \& Falco, E.E., 1992, {\it 
Gravitational Lenses}, Springer-Verlag: Berlin
\bibitem{88} Zakharov, A.F. \& Sazhin, M.V., 1998, {\it Gravitational
Microlensing}, Physics-Uspekhi, 41, 945
\bibitem{89} If we suppose that the dipole moment of the lens
${\cal I}^i$ is
not equal to zero, then the expression for the gravitational lens potential
$\psi$ assumes the form
\begin{eqnarray}
\label{dipole}
\psi&=&\left[{\cal M}-{\bf k}\cdot\dot{\bm{\cal I}}(t^\ast)-
{\cal I}^i(t^{\ast})\;{\hat{\partial}}_i+
\left({\bf k}\times{\bm{\cal S}}\right)^i{\hat{\partial}}_i+
\frac{1}{2}\;{\cal I}^{ij}(t^{\ast})\;{\hat{\partial}}_{ij}
\right]\ln |{\bm{\xi}}|\;,
\end{eqnarray}
where the impact parameter ${\bm{\xi}}$ is the distance from the origin
of the coordinate system to the point of the closest approach of light
ray to the lens. The 
scrutiny examination of the multipole structure of the shape of the 
curves of 
constant value of $\psi$ in cosmological gravitational lenses \cite{90}, 
\cite{91} may reveal the
presence of dark matter in the lens and identify the position of its center of 
mass, velocity and density distribution which can be compared with analogous
characteristics of luminous matter in the lens. In case of the transparent
gravitational lens the expression for the
gravitational lens potential in terms of transverse-traceless (TT) 
internal and external multipole
moments can be found in \cite{11}. Discussion of observational effects
produced by the spin of the lens is given in \cite{92}.
\bibitem{90} Kaiser, N., \& Squires, G., 1993, \apj, 404, 441
\bibitem{91} Bartelmann, M., 1998, 
In: {\it Proc.
18-th Texas Symp. on Relativistic Astrophysics and Cosmology}, eds. 
Olinto, A.V.,
Frieman, J.A. and Schramm, D.N.  World Scientific: Singapore, p. 533
\bibitem{92} Dymnikova, I.G., 1986, Sov. Phys. Usp., 29, 215
\bibitem{93} We emphasize that in the linear with respect to ${\bf v}_a/c$ 
approximation the gravitational shift of frequency depends only on  
transverse component of relative motion of lens and observer.  
Dependence of the gravitational shift of frequency on longitudinal 
motion of lens (radial velocity) appears only if one takes quadratic and
higher order powers in ${\bf v}_a/c$.
\bibitem{corr.1} It is worthwhile to point out that in the expression 
${\bm{\xi}}-{\bm{\xi}}_a$ 
for the impact parameter the term ${\bm{\xi}}$ must be treated as 
${\bm{\xi}}={\bf k}\times({\bf x}_0\times{\bf k})$ where ${\bf x}_0$ are 
coordinates of the
source of light. The unique interpretation of meaning of 
the impact parameter ${\bm{\xi}}$ 
in the last term of (\ref{vic}) as well as in the expression (\ref{damour}) 
for the gravitational lens potential $\psi$ is achieved immediately after 
solving relativistic equations of light geodesic in which the unperturbed 
trajectory of light ray is used everywhere. This eliminates ambiguity in
the definition of ${\bm{\xi}}$.  
\bibitem{94} Remember that $\partial t^\ast/\partial t_0\simeq 1$ and 
$\partial t^\ast/\partial t\simeq 0$ as a consequence of (\ref{tstar}).
\bibitem{95} Birkinshaw, M. \& Gull, S.F., 1983, Nature, 302, 315
\bibitem{96} Critics of the results of the paper \cite{95} by Gurvits \&
Mitrofanov \cite{97} is not rigorously justified. A discrepancy by a factor of 2
between amplitudes of the perturbation of the background cosmic radiation in
\cite{95} and \cite{97} has an algebraic origin rather than a physical one.
\bibitem{97} Gurvits, L.I. \& Mitrofanov, I.G., 1986, Nature, 324, 349 
\bibitem{98} Aghanim, N., Prunet, S., Forni, O., \& Bouchet, F.R., 1998, 
A\&A, 334, 409
\bibitem{corr.4} Bertotti, B. \& Giampieri, G., 1992, Class. Quantum Grav., 9, 
777
\bibitem{corr.5} Iess, L., Giampieri, G., Anderson, J.D. \& Bertotti, B., 1999,
Class. Quantum Grav., 16, 1487
\bibitem{99} Kaplan, G.H., 1998, Astron. J., 115, 361 
\bibitem{100} Reasenberg, R.D., Shapiro, I.I., Macneil, P.E., Goldstein, R.B., 
Breidenthal, J.C., Brenkle, J.P., Cain, D.L., Kaufman, T.M., Komarek, T.A. \&
Zygielbaum, A.I., 1979, \apj, 234, L219
\bibitem{101} Williams, J.G., Newhall, X.X. \& Dickey, J.O., 1996, Phys. Rev. D, 
53, 6730
\bibitem{102} Bertotti, B. \& Iess, L., 1985, Gen. Rel. Grav., 17, 1043 
\bibitem{103} Bertotti, B., Vecchio, A. \& Iess, L., 1999
Phys. Rev. D, 59,  082001
\bibitem{corr.6} Krisher, T.P., Morabito, D.D. \& Anderson, J.D., 1993, Phys.
Rev. Lett., 70, 2213
\bibitem{104} Ohta, T., Okamura, H., Kimura, T. \& Hiida, K., 1973, Prog. Theor.
Phys., 68, 2191
\bibitem{105} Anderson, J.L. \& Decanio, T.C., 1975, Gen. Rel. Grav., 6, 197
\bibitem{106} Damour, T. \& Sch\"afer, G., 1985, Gen. Rel. Grav., 17, 879 
\bibitem{107} More
details about how to integrate the equations of light propagation accounting 
for (static) high-order multipoles can be found in the papers \cite{4}, 
\cite{11}.
\bibitem{108} Ehlers, J., Rosenblum, A., Goldberg, J.N. \& Havas, P., 1976, 
\apj, 
208, L77
\bibitem{109} If one tries to perform a global integration of
(\ref{inst1}) using the Taylor time series expansion of the bodies'
coordinates, the correct logarithmic behavior of the integral 
takes place only if the
first two terms in the expansion are taken into account 
which is physically equivalent to the
case of bodies moving uniformly along straight lines. The logarithm
diverges if limits of integration in (\ref{inst1}) goes to $+\infty$ and
$-\infty$ respectively. Account for the 
third term in the expansion (accelerated motion of the bodies) 
supresses the logarithmic behavior of the integral for large intervals
of integration comparable with the characteristic Keplerian period of the
system, and brings about incorrect prediction for the Shapiro
time delay and the total angle of deflection of light. As bodies in
self-gravitating systems always move with acceleration,  
one evidently has a mathematical inconsistency 
in the Taylor time series expansion for finding a numerical value of the
integral (\ref{inst1}) in the case where the photon goes beyond the limits 
of the near zone of the system.
\bibitem{110} Doroshenko, O.V. \& Kopeikin, S.M., 1990, Sov. Astron., 34, 496
\bibitem{111} Taylor, J.H., Weisberg, J.M., 1989, \apj, 345, 434
\bibitem{112} Sovers, O.J, Fanselow, J.L. \& Jacobs, C.S., 1998, Rev. 
Mod. Phys., 70, 1393
\bibitem{113} Petrov, L., 1999, {\it Absolute methods for determination of 
reference system from VLBI observations}. In: Proceedings 
of the 13-th Working Meeting on European VLBI for Geodesy and Astrometry, 
eds. W.Schl\"ter and H. Hase, BKG, Wettzell, Germany, pp. 138-143 
\bibitem{114} {\it Proceedings of the U. S. Naval
Observatory Workshop on Relativistic Models for Use in Space Geodesy}, U. S.
Naval Observatory: Washington, D. C., ed. Eubanks, T.M., 1991 
\bibitem{115} {\it IERS
Conventions}, IERS Technical Note 21., Obs. de Paris, ed. McCarthy, D.D., 1996 
\bibitem{116} Kopeikin, S.M., 1990, Sov. Astron., 34, 5
\bibitem{117} For VLBI observations of spacecrafts in the
solar system the term under discussion can be important and deserves a
more careful study.
\bibitem{118} Treuhaft, R.N. \& Lowe, S.T., 1991, Astron. J., 102, 1879
\bibitem{119} Klioner, S.A. 1991, 
In: {\it Proc. of AGU Chapman
Conference on Geodetic VLBI: Monitoring Global Change}, NOAA Technical
Report NOS 137 NGS 49, ed. Carter, W.E., American Geophysical Union: 
Washington D.C., p. 188
\bibitem{120}  Lieske, J.H. \& Abalakin, V.K., 1990, {\it Inertial Coordinate
System on the Sky}, IAU Symp. 141, Kluwer: Dordrecht
\bibitem{121} European Space Agency Information
Note No 41-97, Paris, 8 December 1997; http://www.esa.int/Info/97/info41.html
\bibitem{122}  Brumberg, V.A., Klioner, S.A. \& Kopeikin, S.M., 1990, In:
{\it Inertial Coordinate System on the Sky}, eds. Lieske, J. H. and Abalakin, V.
K., Kluwer: Dordrecht, p. 229
\bibitem{corr.7} Krivov, A.V., 1994, Astronomy Reports, 38, 691
\bibitem{123} Brumberg, V.A. \& Kopeikin, S.M., 1989, In: {\it Reference Frames 
in
Astronomy and Geophysics}, eds. Kovalevsky, J., Mueller, I.I., and Kolaczek, B., 
Kluwer: Dordrecht, p. 115
\bibitem{124} Kopeikin, S.M., 1987, Trans. Sternberg State Astron. Inst., 59, 53
(in Russian)
\bibitem{125} Kopeikin, S.M., 1988, Cel. Mech., 44, 87
\bibitem{126} Kopeikin, S.M., 1989, Sov. Astron., 33, 550 
\bibitem{127} Kopeikin, S.M., 1989, Sov. Astron., 33, 665
\bibitem{128} Damour, T., Soffel, M. \& Xu, Ch., 1992, Phys. Rev. D, 45, 1017 
\bibitem{129} Damour, T., Soffel, M. \& Xu, Ch., 1993, Phys. Rev. D, 47, 3124
\bibitem{130} That is, the metric
tensor obeys the special four differential conditions (\ref{harm}) 
which single out the
class of the harmonic coordinates from the infinite number of arbitrary 
coordinate systems
on the space-time manifold.
\bibitem{131} Complete analysis of
differences between two relativistic formulations of light 
deflection - the
post-Newtonian and post-Minkowskian - will be given
elsewhere.
\bibitem{132} Estabrook, F.B. \& Wahlquist, H.R., 1975, Gen. Rel. Grav., 6, 439
\bibitem{133} Smarr, L.L., Vessot, R.F.C., Lundquist, C.A., Decher, R. \&
Piran, T., 1983, Gen. Rel. Grav., 15, 129
\bibitem{ashby} Ashby, N., 1998, In: {\it Gravitation and Relativity : At the turn 
of the Millennium}, 
Proceedings of the GR-15 Conference, IUCAA, Pune, India, Dec. 16-21, 1997,
eds. Dadhich, N. and Narlikar, J., 
InterUniversity Centre for Astronomy and Astrophysics: Pune, p. 231
\bibitem{clock} Laurent, Ph., Lemonde, P., Simon, E., Santarelli, G., Clairon,
A., Dimarcq, N., Petit, P., Audoin, C., \& Salomon, C., 1998, Eur. Phys. J.,
D3, 201 
\bibitem{133a} We again emphasize that among three variables $t$, $\tau$, and
$t^\ast$ only two can be considered as independent because of relationship
$\tau=t-t^\ast$ derived in (\ref{10}). The same is valid for the set $t_0$,
$\tau_0$, and $t^\ast$.
\bibitem{133c} See formula (\ref{diff}) for calculation of partial derivative
of retarded time $s$ with respect to $t^\ast$. 
\bibitem{133b} Hence, the upper limit of the integral is not differentiated with
respect to $t^\ast$ as it was in equation (\ref{limx}).
\bibitem{134} Stebbins, A., 1988, \apj, 327, 584
\bibitem{135} Plebanski, J., 1960, Phys. Rev., 118, 1396
\bibitem{136} Kovner, I., 1990, \apj, 351, 114
\bibitem{137} Pyne, T. \& Birkinshaw, M., 1993, \apj, 415
\bibitem{138} Skrotskii, G.V., 1958, Sov. Phys. Dokl., 2, 226
88, 529
\end{references}
\end{document}